\documentclass[a4paper,11pt]{article}

\usepackage[utf8]{inputenc}
\usepackage[T1]{fontenc}
\usepackage{amsmath,amssymb,amsthm}
\usepackage{mathrsfs}
\usepackage{mathtools}
\usepackage{bm}
\usepackage{jheppub}
\hypersetup{colorlinks=true,citecolor=blue,linkcolor=blue,urlcolor=blue}

\newtheorem{theorem}{Theorem}[section]

\newtheorem{remark}[theorem]{Remark}

\makeatletter
\renewcommand*\l@section{\@dottedtocline{1}{1.5em}{4.4em}}
\renewcommand*\l@subsection{\@dottedtocline{2}{5.9em}{5.2em}}
\makeatother

\newcommand{\Iplus}{\mathcal I^+}
\newcommand{\Q}{\mathcal Q}
\newcommand{\K}{\mathcal K}
\newcommand{\Dop}{\mathcal D}
\newcommand{\Ddual}{\widetilde{\mathcal D}}
\newcommand{\News}{\widehat N}
\newcommand{\Lie}{\mathcal L}

\newcommand{\dd}{\mathrm d}
\newcommand{\R}{\mathbb R}

\title{Kerr Soft Dressing and the
\texorpdfstring{$w_{1+\infty}$}{w1+infinity} Frame Algebra at Null
Infinity}

\author{Gabriel Menezes}
\affiliation{Departamento de F\'isica Matem\'atica - Instituto de F\'isica, Universidade de S\~ao Paulo,\\R. do Mat\~ao, 1371, 05508-090 S\~ao Paulo -- SP, Brasil}
\emailAdd{gsm@if.usp.br}

\abstract{
We construct the charge-generated
intrinsic/canonical frame dictionary associated with the Kerr-selected
soft dressing.  Starting from the VV supertranslation,
we formulate the higher-spin problem as an inverse problem at null
infinity: the soft kernel
\(\K^{(s,0)}_{AB}[t]\), built from the parity-adapted maximally
longitudinal scalar \(\chi^{(s)}_t\),
is matched to the Kerr-selected exponentiating projection of the universal
soft contribution, thereby determining one parity component of the generator
\(t^{A_1\cdots A_s}\).  Helicity conjugation fixes which one: the
exponentiating source obeys
\(\overline{S^{(s)}_{+,{\rm exp}}}=(-1)^sS^{(s)}_{-,{\rm exp}}\), so the
tower fixes the electric projection of the source at even levels and the
magnetic projection at odd ones, matching the alternation of the Kerr mass
and current moments; for aligned spin the projection is exhaustive and we
solve the tower in closed form.  The prescription reproduces the VV
supertranslation at leading order and fixes the curl, not the divergence,
of a smooth generalized-BMS vector at subleading order.  The reason for the
matching is physical: the same exponentiating soft factor is the classical
limit of the Guevara--Ochirov--Vines spinning three-point operator and
generates the Kerr multipole tower.  We explain the corresponding hard
flux charges and show how their external-state action gives the Ward
representation of the soft theorem.  The polynomial Poisson algebra on
\(T^\ast S^2\), with local \(w_{1+\infty}\)-type reductions, then acts on
these frame-changing generators; it does not close on the Kerr-selected
data alone.  This gives the physical role of
the \(w_{1+\infty}\)-like structure in Kerr black-hole scattering: it
moves the intrinsic/canonical dictionary.  Its observable imprint begins
with displacement memory at \(s=0\) and spin memory at \(s=1\), followed
by higher electric and magnetic memory moments.
}


\begin{document}

\maketitle
\flushbottom

\section{Introduction}

The starting point of this work is a simple but surprisingly rigid fact
about gravitational scattering at null infinity.  A scattering process may
be described in a canonical Bondi frame, adapted to the standard
radiative phase space, or in an intrinsic frame, adapted to the mechanical
center-of-mass description of the hard bodies.  Veneziano and Vilkovisky
showed that these two descriptions are not related by a small gauge choice:
already at order \(G\), they differ by a BMS supertranslation
\cite{Veneziano:2022zwh}.  In their analysis the canonical frame is the
one in which the shear at the past boundary of future null infinity
\(\Iplus_-\) is set to zero, while the intrinsic frame is the frame
naturally obtained from the boosted linearized Schwarzschild fields of the
scattering bodies.  This large-gauge transformation resolves the apparent
discrepancy between the angular-momentum loss inferred in the intrinsic
mechanical frame and the canonical Bondi-frame answer
\cite{Damour:2020tta}.

The second input is the infrared interpretation of the same
supertranslation.  In gravity, hard particles are accompanied by coherent
clouds of soft gravitons.  The leading Faddeev--Kulish dressing can be
written as a soft charge whose parameter is precisely the
Veneziano--Vilkovisky supertranslation
\cite{FaddeevKulish1970,Ware:2013zja,Choi:2019rlz,Elkhidir:2024ward}.
Thus the leading VV dictionary ties together three objects that are often
discussed separately: a BMS frame change, a zero-frequency graviton
operator, and the long-range field carried by the asymptotic state.  This
paper asks whether the same logic persists beyond the leading soft
factor, and if so which part of the higher soft expansion has a clean
spacetime meaning.

The answer we develop is deliberately restricted.  We do not claim that
the complete gravitational soft expansion is determined by Kerr.  Starting
at sub-subleading order the soft theorem contains non-universal terms and
depends on the theory and on the hard data.  Instead, we isolate a
\emph{Kerr-selected exponentiating sector} of the universal soft
contribution.  This is the part of the
soft expansion obtained by repeatedly inserting the spin operator that
appears in the massive three-point amplitude.  It is also the part that
packages the Kerr multipole relation
\begin{equation}
M_\ell+iS_\ell=M(ia)^\ell
\end{equation}
in the classical spinning limit.  Here \(M=M_0=m\) is the black-hole
mass, \(S^\mu\) is the spin vector, and
\(a^\mu=S^\mu/m\) is the Kerr specific-spin, or ring-radius, vector.
In the axisymmetric scalar formula \(a\) denotes the signed magnitude of
this specific-spin vector along the spin axis.  The main claim is that
this sector is the asymptotic charge organization of Kerr multipole soft
dressing.

The concrete problem solved in the paper is therefore the following.
Given the Kerr-selected exponentiating projection at order \(s\), find the
asymptotic generator \(t^{A_1\cdots A_s}\) whose soft charge produces that
field-independent shear displacement.  In formulas, the charge provides a
kernel \(\K^{(s,0)}_{AB}[t]\), and the Kerr soft dressing provides the real,
parity-adapted source \(\mathscr S^{(s)}_{AB,{\rm exp}}\) defined below.
In the VV frame normalization the dictionary is obtained by solving
\begin{equation}
\K^{(s,0)}_{AB}[t]=2G\,\mathscr S^{(s)}_{AB,{\rm exp}}.
\end{equation}
Here \(G\) is Newton's constant.
At leading order this inverse problem is exactly the VV equation for the
supertranslation.  At subleading order the source is magnetic, so it
determines only \(\epsilon^{AB}D_AY_B\), which is why the answer belongs
to generalized BMS and fixes the magnetic potential of the vector.  The
parity alternation is one of the main results of the paper: it is the soft
image of the Kerr no-hair relation, whose mass moments are even in
\(\ell\) and whose current moments are odd.  The algebra is the second:
after the generators have been found, their Poisson/symbol bracket acts on
the intrinsic/canonical frame shifts.  This is the advertised physical
interpretation of the \(w_{1+\infty}\)-like structure.  The memory
discussion then identifies the observables measured by these same shifts:
displacement memory at \(s=0\), spin memory at \(s=1\), and their higher
parity-alternating moments.

This claim sits at the intersection of several developments.  First, the
modern amplitude program has made compact-binary dynamics a problem in
which on-shell data, classical limits, and waveform observables can be
computed together.  Scattering amplitudes have been used to derive
post-Minkowskian potentials and Hamiltonians, impulses, radiated momentum,
radiation kernels and waveform information
\cite{Cheung:2018wkq,Kosower:2018adc,Bern:2019nnu,Bern:2019crd,Cristofoli:2021vyo,Herrmann:2021tct,Buonanno:2022pgc}.  This amplitude
program is now tightly connected to effective-field-theory methods,
self-force theory, EOB dynamics, numerical relativity and waveform
modeling \cite{Donoghue:1994dn,Goldberger:2004jt,Porto:2016pyg,Levi:2018nxp,Barack:2018yvs,Buonanno:1998gg,Buonanno:2000ef,Pretorius:2005gq,Campanelli:2005dd,Baker:2005vv,Blanchet:2013haa}.  Recent work on
radiation, tails, memory and waveform observables makes the zero-frequency
limit especially important \cite{Georgoudis:2023waveform,Alessio:2024soft,Bini:2024waveform,Sen:2024tails,Georgoudis:2025memory,Fucito:2024waveforms}.  The present paper concerns the large-gauge part of
this amplitude-to-spacetime map: the part of the scattering data that
survives as a long-range metric field and as memory at null infinity.
The amplitude side of this story rests on a broad set of on-shell tools:
twistor-inspired tree amplitudes, recursion relations, generalized
unitarity and modern amplitude reviews
\cite{Parke:1986gb,Witten:2003nn,Roiban:2004yf,Gukov:2004ei,Cachazo:2004kj,Britto:2005fq,Bern:1994zx,Fusing,TripleCuteeJets,BCFUnitarity,BDKUniarityReview,Elvang:2015rqa,JJHenrikReview,BernHuangReview}.  Its gravitational applications also draw on the KLT
relations, color-kinematics duality, the double copy and black-hole double
copies \cite{KLT,BCJ,BCJLoop,Monteiro2014cda,BCCJRReview,doublecopyWhitePaper2022}.  These developments provide the technical
background for treating the Kerr three-point amplitude not merely as an
isolated on-shell object, but as a generator of classical long-distance
data.

The same bridge between amplitudes and waveforms is continuous with older
and parallel approaches to conservative and radiative two-body dynamics.
The conservative side includes post-Newtonian methods, post-Minkowskian
methods and EOB theory, including the important lesson that classical
long-range information can arise from loop amplitudes.  The radiative side
includes self-force methods, radiation reaction, and eikonal descriptions
of classical scattering \cite{Holstein:2004dn,Einstein:1938yz,Einstein:1940mt,Ohta:1973je,Jaranowski:1997ky,Damour:1999cr,Blanchet:2000nv,Damour:2001bu,Damour:2014jta,Jaranowski:2015lha,Mino:1996nk,Quinn:1996am,Bertotti:1956pxu,Kerr:1959zlt,Bertotti:1960wuq,Portilla:1979xx,Westpfahl:1979gu,Portilla:1980uz,Bel:1981be,Westpfahl:1985tsl,Damour:2016gwp,Damour:2017zjx,Cheung:2018wkq,Damour:2019lcq,Damgaard:2021rnk,Amati:1987wq,tHooft:1987vrq,Muzinich:1987in,Amati:1987uf,DiVecchia:2022piu,Cristofoli:2021jas,Britto:2021pud,Georgoudis:2023waveform,Cho:2018upo,Kalin:2019rwq,Kalin:2019inp,Cho:2021arx,Bern:2019nnu,Bern:2019crd,Blumlein:2020znm,Manohar:2022dea}.  From this
viewpoint, the soft sector studied here is not a replacement for those
methods.  It is the infrared boundary layer that any amplitude-derived
observable must pass through before it becomes a Bondi-frame statement.

Second, the observational motivation is no longer abstract.  The LIGO,
Virgo and KAGRA network has turned compact binary coalescences into
precision gravitational physics
\cite{LIGOScientific:2014pky,VIRGO:2014yos,KAGRA:2020agh,Abbott:2016blz,
TheLIGOScientific:2017qsa,LIGOScientific:2021qlt,LIGOScientific:2018mvr,
LIGOScientific:2020ibl,LIGOScientific:2021gwtc3}, and next-generation
detectors and waveform models will only sharpen the requirement faced here
\cite{Foucart:2022iwu,LISA:2017pwj,Saleem:2021iwi,Reitze:2019iox,
Punturo:2010zz,Sathyaprakash:2019yqt,Maggiore:2019uih,Kalogera:2021bya,
Berti:2022wzk,Regge:1957td,Zerilli:1970se,Vishveshwara:1970cc,
Teukolsky:1973ha,Schafer:2018kuf,Antonelli:2019ytb,Damour:2014afa,
Ossokine:2017dge,Favata:2013rwa,Samajdar:2018dcx,Purrer:2019jcp,
Huang:2020pba,Gamba:2020wgg}: long-range fields, memory, spin multipoles
and frame conventions must be handled coherently if amplitudes are to be
translated into gauge-invariant observables.

Third, asymptotic-symmetry and celestial-amplitude analyses have revealed
large algebras acting on the soft sector of gravity.  The usual infrared
triangle relates soft theorems, asymptotic charges and memory
\cite{Weinberg1965,Strominger2017,Strominger:2014pwa,Kapec:2016jld,Freidel:2021ytz,Geiller:2024subleading,Cresto:2024wedge,Cresto:2024noether}.  Celestial treatments identify
\(w_{1+\infty}\)-type symmetries acting on soft graviton modes and on
massive scattering states \cite{Guevara:2021abz,Strominger:2021mtt,Hamada:2018vrw,Li:2018gnc,Himwich:2023njb,Pate:2019mfs,Adamo:2021wjo,Adamo:2021lrv,Ball:2021tmb,Donnay:2024carrollianLw,Kmec:2024twistor,Ruzziconi:2025horizons,Agrawal:2024sw,Freidel:2026oneform}.  Our
construction gives a four-dimensional, null-infinity interpretation of
one classical part of this story.  Closest to it, Ref.~\cite{Guevara:2024color}
identifies the same spinning three-point exponential as the structure
constant of a celestial colour symmetry, obtains the Kerr Weyl scalar from
it by a Penrose transform, and finds that the lowest generator acts on
Kerr as a translation, leaving the action of the higher generators open.
What the tower below supplies is precisely that higher action, in the
Lorentzian Bondi-frame language rather than the chiral twistor one.  The global algebra we find is the real
Poisson algebra of polynomial functions on \(T^\ast S^2\); the chiral
celestial \(w_{1+\infty}\) algebra appears only after choosing a local
complex patch and a holomorphic polarization.  This distinction matters:
the subleading transformation relevant for the Kerr soft dressing is a
smooth generator of generalized BMS, not in general a holomorphic
superrotation.
The asymptotic side of the story begins with conformal infinity, the BMS
group and the Ashtekar--Streubel radiative phase space
\cite{Penrose:1964ge,Bondi1962,Sachs1962,Newman:1966ub,Ashtekar:1979iaf,Barnich:2010eb,Ashtekar:2019rpv}.  Modern work has made
the angular-momentum, flux and memory subtleties of supertranslations
especially explicit
\cite{DiVecchia:2022owy,Bonga:2018gzr,Chen:2021kug,Chen:2021szm,Javadinezhad:2022ldc,Javadinezhad:2022hhl,Mao:2023evc,Javadinezhad:2023mtp,Mao:2024ryq,Heissenberg:2024umh}.  The soft theorem
side of the infrared triangle includes the leading and subleading
gravitational soft theorems and their charge interpretations
\cite{Cachazo2014,Bern:2014vva,Strominger2014,He:2014laa,Strominger:2014pwa,Strominger:2016wns}.  Recent higher-spin,
Carrollian and celestial analyses then provide the natural language for
the \(w_{1+\infty}\)-like structures that appear below
\cite{Pate:2019lpp,Mason:2024carrollian,Cresto:2025yangmills,Cresto:2025eym,Cresto:2025thesis,Ahn:2025n8}.  Finally, the multipolar
and radiative language used to interpret the Kerr sector is continuous
with the classic treatment of source multipoles and tails
\cite{Thorne:1980ru,Blanchet:1987wq,Blanchet:1993ec,Donoghue:2022wrw}.

The paper is organized around this chain of ideas, rather than around a
catalog of possible higher-spin constructions.  We first review the VV
dictionary and show that the leading frame equation is sourced by the
leading soft factor.  We then define the soft-charge kernels
at null infinity, including the explicit differential operator that maps
the parameter \(t\) to the soft kernel.  The identification of this kernel
with the exponentiating soft source is motivated by the
Guevara--Ochirov--Vines three-point amplitude for spinning black holes:
the amplitude-level expression is an operator acting on massive spin
states, and its classical limit is the Kerr source exponential.  We then
construct the corresponding hard charges and derive the Ward representation of the soft theorem.
Only after the charge construction is fixed do we discuss the algebra,
memory and celestial reductions.  The algebra is not optional in this
story: it is the composition law of the frame dictionary and the bridge to
the celestial \(w_{1+\infty}\)-like symmetry.  The memory analysis records
which observable shear moments are fixed by that same tower.

Section \ref{sec:vv-main} reviews the VV frames and solves the leading
frame equation.  Section \ref{sec:soft-kerr-main} defines the soft
charges, the universal contribution to the higher soft expansion, and the Kerr
exponent.  Section \ref{sec:leading-subleading-main} derives the leading,
subleading and higher-rank frame equations.  Section \ref{sec:hard-main}
gives the hard charges and Ward identity in detail.  Section
\ref{sec:algebra-main} proves the principal-symbol algebra and relates it
to celestial \(w_{1+\infty}\), including the low-spin checks, the
Hamiltonian origin of the total bracket, the hard-sector Kerr check and
the assumptions behind the result.  Section \ref{sec:dictionary-main}
revisits the intrinsic/canonical dictionary. Section
\ref{sec:memory-main} discusses the all-orders memory probes.

The derivations are placed where the reader needs them.  The VV
section contains the sphere identities and the explicit VV potential; the
soft-charge section contains the Faddeev--Kulish and mode-space matching;
the frame-equation section contains the Bondi frame data and low-spin soft
matching; the Ward section contains the hard charge and mixed soft--hard
checks; the algebra section contains the functional-derivative,
principal-symbol, trace-descendant and ordered-homogeneous-action checks;
the dictionary and memory sections contain the Kerr source and memory
cocycle analyses.  Section \ref{sec:quantization-ordering-main} is then
reserved for the separate quantum-ordering question.  Appendix
\ref{app:spherical-harmonic-basis} gives the harmonic basis translation,
while Appendix \ref{sec:component-mode-boundary-checks} collects
additional component, mode and boundary checks.

\section{Veneziano--Vilkovisky Frames}
\label{sec:vv-main}

\paragraph{Conventions and sphere identities.}
The following conventions and elementary sphere identities are used
throughout the paper.  They are placed here because the VV frame equation
is the first place where the soft polarization, the trace-free Hessian and
the celestial direction \(q^\mu(z)\) enter together.

\subsection{Conventions at Null Infinity}

\subsubsection{Sphere geometry}

The unit sphere metric is denoted by $\gamma_{AB}$, its volume form by
\(\epsilon_{AB}\), and the associated Levi-Civita derivative by \(D_A\);
we choose \(\epsilon_{AB}\epsilon^{AB}=2\).  Curly sphere indices denote
the symmetric trace-free projection.  For a scalar $f$ the trace-free Hessian is
\begin{equation}
\Dop_{AB}f
=
D_A D_B f-\frac12\gamma_{AB}D^2 f.
\end{equation}
For a vector field $Y^A$,
\begin{equation}
D\cdot Y:=D_A Y^A,
\qquad
(\mathrm{curl}\,Y):=\epsilon^{AB}D_A Y_B.
\end{equation}
The electric part of a vector field is determined by $D\cdot Y$ and the
magnetic part by $\mathrm{curl}\,Y$.  In the exponentiating Kerr sector
the equations fix the electric projection at even levels and the magnetic
projection at odd levels.  The opposite-parity component at each level is
not determined by that sector alone.

The null vector associated with a point on the sphere is
\begin{equation}
q^\mu(z)=(1,\Omega^i(z)),
\qquad
q^2=0.
\end{equation}
Here \(\Omega^i(z)\) is the unit spatial direction labelled by \(z\), so
\(q^\mu\) is dimensionless.
If $p^\mu$ is a fixed timelike vector, then
\begin{equation}
x(z):=p\cdot q(z)
\end{equation}
is a linear combination of the $\ell=0$ and $\ell=1$ scalar harmonics.
Therefore
\begin{equation}
\Dop_{AB}x=0.
\label{eq:appendix-DABx-zero}
\end{equation}
This elementary identity is the key to the Veneziano--Vilkovisky solution.

\subsubsection{Polarization tensors}

The trace-free tensor
\begin{equation}
\varepsilon_{AB}^{\mu\nu}(q)
:=
D_{\{A}q^\mu D_{B\}}q^\nu
=
D_Aq^\mu D_Bq^\nu
-\frac12\gamma_{AB}D_Cq^\mu D^Cq^\nu
\end{equation}
is the sphere-index version of a soft-graviton polarization tensor.  With
this notation the leading soft tensor sourced by a massive particle of
momentum $p^\mu$ is
\begin{equation}
S^{(0)}_{AB}(p;q)
=
\frac{p_\mu p_\nu}{p\cdot q}\,
\varepsilon_{AB}^{\mu\nu}(q).
\label{eq:appendix-leading-soft-tensor}
\end{equation}
Using $D_Ax=p\cdot D_Aq$, this can be rewritten as
\begin{equation}
S^{(0)}_{AB}
=
\frac{1}{x}
\left(D_AxD_Bx-\frac12\gamma_{AB}D_CxD^Cx\right).
\label{eq:appendix-leading-soft-x}
\end{equation}
Equation \eqref{eq:appendix-leading-soft-x} makes the solution of the
leading frame equation immediate.

\subsection{Canonical and intrinsic Bondi frames}

Near future null infinity we use retarded Bondi coordinates
\((u,r,z^A)\).  The radiative shear is \(C_{AB}(u,z)\), symmetric and
trace-free with respect to the round metric \(\gamma_{AB}\).  The canonical
Bondi frame is the frame in which the early-time shear at \(\Iplus_-\) is
set to zero.  The intrinsic frame is the frame naturally associated with
the center-of-mass scattering description.  Veneziano and Vilkovisky found
that, already at linearized order in \(G\), the intrinsic frame contains a
nonzero shear.

For a boosted Schwarzschild source with momentum \(p^\mu\), the VV shear is
\begin{equation}
C^{\rm VV}_{AB}
=
-4G\,\frac{p^\mu p^\nu}{p\cdot q}
D_{\{A}q_\mu D_{B\}}q_\nu
\equiv
-4G\,S^{(0)}_{AB}(p;q),
\label{eq:vv-shear-companion-main}
\end{equation}
where \(q^\mu(z)\) is the null direction on the celestial sphere and
\(D_{\{A}q_\mu D_{B\}}q_\nu\) is the trace-free polarization tensor.  A
supertranslation acts on the shear by
\begin{equation}
\delta_T C_{AB}=-2\Dop_{AB}T,
\qquad
\Dop_{AB}T=D_AD_BT-\frac12\gamma_{AB}D^2T.
\label{eq:supertranslation-shear-main}
\end{equation}
The supertranslation mapping the canonical frame to the intrinsic frame therefore satisfies
\begin{equation}
\Dop_{AB}T_{\rm VV}=2G\,S^{(0)}_{AB}.
\label{eq:vv-frame-equation-main}
\end{equation}

\subsection{The Veneziano--Vilkovisky Supertranslation}

As above, let \(x=p\cdot q\).  By Lorentz covariance the solution has the form
\(T=f(x)\), up to ordinary translations.  Since \(x\) is an \(\ell=0,1\)
spherical harmonic combination, from
\eqref{eq:appendix-DABx-zero},
\begin{align}
\Dop_{AB}f(x)
&=
f''(x)
\left(D_AxD_Bx-\frac12\gamma_{AB}D_CxD^Cx\right)
\nonumber\\
&\quad
+
f'(x)\Dop_{AB}x
\nonumber\\
&=
f''(x)
\left(D_AxD_Bx-\frac12\gamma_{AB}D_CxD^Cx\right).
\end{align}
Comparing with \eqref{eq:appendix-leading-soft-x}, the differential
equation reduces to
\begin{equation}
f''(x)=\frac{2G}{x}.
\end{equation}
Therefore
\begin{equation}
f(x)=2G\,x\log x+a+b\,x.
\end{equation}
The terms $a+b\,x$ are in the kernel of $\Dop_{AB}$ and correspond to
ordinary translations.  Dimensionally, \(\log x\) is shorthand for
\(\log(x/\mu)\); a change of the arbitrary reference scale \(\mu\) only
shifts the coefficient \(b\).  We may therefore set the homogeneous terms
to zero and write~\cite{Veneziano:2022zwh}
\begin{equation}
T_{\rm VV}=2G\,(p\cdot q)\log(p\cdot q).
\label{eq:vv-solution-main}
\end{equation}
This agrees with the amplitude derivation of Ref.~\cite{Elkhidir:2024ward},
where the induced retarded-time shift is
\(u\to u+2Gm(v\cdot n)\log(v\cdot n)\) with \(p^\mu=mv^\mu\) and
\(q^\mu=n^\mu\): the two expressions differ by \(2Gm\log m\,(v\cdot n)\),
which is linear in \(p\cdot q\) and therefore an ordinary translation in
the kernel of \(\Dop_{AB}\).  The normalisation \(2G\) used throughout is
fixed by this comparison.
This calculation is the
prototype for the higher construction: the inhomogeneous soft-charge kernel
is chosen so that it reproduces the relevant soft factor.
The same result is derived in stereographic coordinates, and the sphere
identities behind the trace-free Hessian calculation are checked
component by component, in Appendix
\ref{sec:component-mode-boundary-checks}.

This derivation is useful beyond the leading result because it identifies
the ingredients that will reappear at higher spin.  The
canonical frame condition is a boundary condition on the shear; the
intrinsic frame is read from the long-range field naturally associated
with the hard scattering data; and the differential equation relating the
two is sourced by a soft factor projected onto a trace-free sphere tensor.
In this sense the VV supertranslation is not merely a convenient solution
of a Bondi-gauge problem.  It is the first example in this paper of a soft
theorem being used as a boundary-condition-changing operator on the
gravitational phase space.

The physical interpretation of the preceding computation is the following.
The canonical frame is the frame in which the shear vanishes at the past
boundary of future null infinity.  The intrinsic frame is the one naturally
adapted to the mechanical center-of-mass description of the scattering.
The two are related by a large diffeomorphism.  The point of the present
paper is that this large diffeomorphism is also selected by the leading
soft theorem: the inhomogeneous kernel of the soft charge is precisely the
leading soft factor.

This reformulation is what makes the higher-spin extension possible.  We
do not need a closed-form bulk spacetime metric for each higher-spin frame.
We instead use the soft factor itself as the source for the corresponding
asymptotic frame equation.

\section{Soft Charges and the Kerr Exponent}
\label{sec:soft-kerr-main}


This section gives the calculation-level version of the construction used
above.  The aim is to make explicit which sphere identities,
zero-frequency limits and parity projections enter the frame dictionary.
The discussion is deliberately pedestrian: the point is to show how the
VV supertranslation and its higher-spin generalization arise from ordinary
Bondi data and soft operators, rather than to hide the result behind
notation.

There is a second reason to proceed slowly.  The form of the charges is
not meant to be a decorative ansatz.  At \(s=0\) the soft kernel
\(\Dop_{AB}T\) is the standard electric supertranslation kernel.  At
\(s=1\) the Kerr current dipole selects instead its parity dual,
\(\Ddual_{AB}(\epsilon^{CD}D_CY_D)\), where
\(\Ddual_{AB}:=\epsilon_{C(A}\Dop_{B)}{}^{C}\).  This is the magnetic
projection of the generalized-BMS subleading Ward charge: it fixes the
curl of a smooth sphere vector, not its divergence.  It must not be
confused with the complete residual action on a fixed round-sphere Bondi
shear.  The latter belongs to the expanded generalized-BMS phase space,
also transforms the celestial metric, and contains the
divergence-dependent \(u\)-term displayed in
Section~\ref{sec:leading-subleading-main}.  At higher \(s\) we continue the
same parity-adapted, maximally longitudinal construction.  For even \(s\)
one takes \(D_{A_1}\cdots D_{A_s}t^{A_1\cdots A_s}\) and applies
\(\Dop_{AB}\); for odd \(s\) one takes the corresponding curl scalar and
applies \(\Ddual_{AB}\).  The differential operator is therefore selected
by the scalar-potential part of the appropriate parity projection, not by
assuming a new exact nonlinear symmetry at the outset.

This also clarifies the relation to existing BMS and celestial
constructions.  The usual supertranslation charge is recovered at the
first level.  At the second level our kernel is the magnetic projection of
the generalized-BMS Ward charge, with the full spacetime completion kept
distinct.  Celestial multipole and metric-reconstruction analyses show
how asymptotic charges encode multipole data in a celestial basis
\cite{Compere:2022cpm}.  Flux-balance and memory analyses of
gravitational scattering show why angular momentum, memory and
supertranslation frame choices must be treated together
\cite{Compere:2019gft,Riva:2023xxm}.  Our tower should be read as a
Kerr-selected, higher-moment extension of these familiar configurations: it is
the Kerr-selected subsector of the universal soft contribution whose hard
action is fixed by the soft theorem and whose source is the exponentiating
Kerr three-point operator.
It is therefore narrower than a complete nonlinear BMS-flux algebra, but
more structured than an arbitrary soft smearing.

The power \(u^s\) in the charge has the equally concrete origin of a soft
frequency derivative.  A retarded-time moment of the news is the
time-domain version of
\(\partial_\omega^s[\omega a_{AB}(\omega,z)]_{\omega=0}\).  The angular
operator and the retarded-time moment are therefore tied together: the
former chooses the electric or magnetic sphere pattern, while the latter
chooses the order in the soft expansion.  The identification with the
exponentiating soft factor then does one specific job.  It selects, among
all possible kernels of the same differential type, the one whose hard
Ward operator is the Kerr spin insertion in the selected exponentiating
projection.  This is the sense in which the differential form of the charge
and the amplitude-side Kerr exponent are not independent assumptions, but
two sides of the same Ward identity.

\subsection{Phase-space convention}

We use the shifted-news radiative phase space of Ref.~\cite{Freidel:2021ytz}.
The ordinary news is \(N_{AB}=\partial_u C_{AB}\), and
\begin{equation}
\News_{AB}=N_{AB}-\tau_{AB}
\label{eq:shifted-news-main}
\end{equation}
is the canonical momentum conjugate to \(C_{AB}\).  Here $\tau_{AB}$ is
the symmetric traceless Geroch tensor, defined by
\(D_A\tau^{AB}+(1/2)D^B R=0\), and \(R\) is the two-dimensional Ricci
scalar~\cite{Geroch77}.  The Ashtekar--Streubel
bracket is written in the convention~\cite{Ashtekar:1981bq}
\begin{equation}
\{F,G\}_{AS}=\lambda_{AS}
\int_{\Iplus}\dd u\,\dd^2z\sqrt\gamma
\left[
\frac{\delta F}{\delta C_{AB}}
\frac{\delta G}{\delta\News^{AB}}
-
\frac{\delta F}{\delta\News^{AB}}
\frac{\delta G}{\delta C_{AB}}
\right],
\label{eq:AS-bracket-main}
\end{equation}
Here \(\lambda_{AS}\) is the inverse symplectic normalization of the
Ashtekar--Streubel bracket.  Its numerical value depends on whether one
absorbs factors such as \(16\pi G\) into the definition of \(\News_{AB}\)
or into the charge.  We keep it explicit in intermediate functional
derivative checks and then choose the overall normalization of the charges
so that
\(\{C_{AB},\Q_s^{\rm soft}(t)\}_{AS}=u^s\K^{(s,0)}_{AB}[t]+O(C)\).
No algebraic statement below depends on the numerical value of
\(\lambda_{AS}\).

\subsection{Soft-charge kernels and their scope}

For each non-negative integer \(s\) we introduce a symmetric sphere tensor
parameter \(t^{A_1\cdots A_s}\).  For \(s=0\) this is a scalar
supertranslation parameter, for \(s=1\) it is a vector field on \(S^2\),
and for \(s\ge2\) it is a higher-rank parameter.  The soft charge in the
Kerr-selected sector is written in the form
\begin{equation}
\Q_s^{\rm soft}(t)
=
\int_{\Iplus}\dd u\,\dd^2z\sqrt\gamma\,
 u^s\K^{(s)}_{AB}[t;C]\News^{AB}.
\label{eq:soft-charge-main}
\end{equation}
Here \(\K^{(s)}_{AB}[t;C]\) is a symmetric trace-free tensor on the sphere,
constructed from \(t\), \(C_{AB}\), the sphere metric and covariant
derivatives.  The factor \(u^s\) extracts the \(s\)th retarded-time moment
of the soft graviton mode.  The kernel has a field expansion
\begin{equation}
\K^{(s)}_{AB}[t;C]
=
\K^{(s,0)}_{AB}[t]+\K^{(s,1)}_{AB}[t;C]+O(C^2).
\label{eq:kernel-expansion-main}
\end{equation}
The field-independent kernel is built from the maximally longitudinal
scalar
\begin{equation}
\chi_t^{(s)}(z):=
\begin{cases}
D_{A_1}\!\cdots D_{A_s}t^{A_1\cdots A_s}, & s\ \hbox{even},\\[2pt]
\epsilon^{BA_1}D_BD_{A_2}\!\cdots D_{A_s}t^{A_1\cdots A_s}, & s\ \hbox{odd},
\end{cases}
\label{eq:soft-kernel-differential-main}
\end{equation}
Its parity, however, is not a free choice.  Introduce an inverse-length
bookkeeper \(\lambda\), set to the physical soft frequency \(\omega\) after
the level expansion.  In the classical spin limit the helicity-\(\eta\)
Kerr three-point source and its level coefficients are defined by
\begin{equation}
S_{\eta,{\rm exp}}(\lambda)
:=S^{(0)}_\eta e^{\eta\lambda a\cdot q}
=\sum_{s\ge0}\lambda^sS^{(s)}_{\eta,{\rm exp}},
\qquad
S^{(s)}_{\eta,{\rm exp}}
=S^{(0)}_\eta\frac{\eta^s(a\cdot q)^s}{s!}.
\label{eq:helicity-levels-main}
\end{equation}
Here \(a^\mu=S^\mu/m\) is the ring-radius vector.  In particular,
\(S^{(s)}_{+,{\rm exp}}=S^{(0)}_+(a\cdot q)^s/s!\) and
\(S^{(s)}_{-,{\rm exp}}=S^{(0)}_-(-1)^s(a\cdot q)^s/s!\); the complete
sources are the corresponding sums over \(s\), not individual
coefficients.  Splitting the exponential into its parts
even and odd under \(\eta\to-\eta\),
\begin{equation}
e^{\eta\lambda a\cdot q}
=
\cosh(\lambda a\cdot q)+\eta\,\sinh(\lambda a\cdot q),
\label{eq:parity-alternation-main}
\end{equation}
the even part multiplies the two helicity components of
\(S^{(0)}_{AB}\) equally and yields a real tensor, while the odd part
multiplies them with opposite signs and yields \(i\) times a real tensor.
Since the \(n\)th term of the tower carries \((a\cdot q)^n\), the levels
alternate between these two behaviours; equivalently, in the helicity
variables of \eqref{eq:exponentiating-replacement-main},
\(\overline{S^{(s)}_{+,{\rm exp}}}=(-1)^sS^{(s)}_{-,{\rm exp}}\).  To make
the real scenario explicit, let \(m_A\) be a complex dyad normalized by
\(\gamma_{AB}=2m_{(A}\bar m_{B)}\), and define
\(\mathcal T^{(s)}_{AB}:=S^{(s)}_{+,{\rm exp}}m_Am_B+
S^{(s)}_{-,{\rm exp}}\bar m_A\bar m_B\).  Then
\(\overline{\mathcal T^{(s)}_{AB}}=(-1)^s\mathcal T^{(s)}_{AB}\), so the
real tensor used in the frame equation is
\begin{equation}
\mathscr S^{(s)}_{AB,{\rm exp}}:=
\begin{cases}
\mathcal T^{(s)}_{AB},&s\ {\rm even},\\
-i\,\mathcal T^{(s)}_{AB},&s\ {\rm odd}.
\end{cases}
\label{eq:real-parity-source-main}
\end{equation}
The sign in the odd line is fixed by the orientation and self-duality
conventions in \eqref{eq:W-eta-imaginary-main}.  This
matches the Kerr no-hair relation \eqref{eq:kerr-multipoles-main}, whose
mass moments \(M_\ell\) are supported on even \(\ell\) and whose current
moments \(S_\ell\) on odd \(\ell\), the latter sourcing odd-parity metric
data.  Accordingly the frame equation at level \(s\) fixes the electric
projection of the source for even \(s\) and its magnetic projection for
odd \(s\), and the kernels alternate,
\begin{equation}
\K^{(s,0)}_{AB}[t]
=
\begin{cases}
\Dop_{AB}\,\chi_t^{(s)}, & s\ \hbox{even},\\[4pt]
\Ddual_{AB}\,\chi_t^{(s)}, & s\ \hbox{odd},
\end{cases}
\qquad
\Ddual_{AB}:=\epsilon_{C(A}\Dop_{B)}{}^{C},
\label{eq:soft-kernel-alternating-main}
\end{equation}
and the differential form of the linear soft charge is
\begin{equation}
\Q_{s,{\rm lin}}^{\rm soft}(t)
=
\int_{\Iplus}\dd u\,\dd^2z\sqrt\gamma\,
u^s\,
\K^{(s,0)}_{AB}[t]
\News^{AB}.
\label{eq:soft-charge-differential-main}
\end{equation}
For \(s=0\), \(\chi_t^{(0)}=T\) and one recovers the usual
supertranslation kernel \(\Dop_{AB}T\).  For \(s=1\), the definition gives
directly \(\chi_t^{(1)}=\epsilon^{AB}D_AY_B\); this magnetic case is treated in
Section \ref{sec:leading-subleading-main}.  For \(s\ge2\) the same formula
gives the maximally longitudinal component of the parity fixed by
\eqref{eq:parity-alternation-main}.  A globally complete
higher-spin transformation may require trace descendants and curvature
terms; those lower-order terms are important for closure, but they do not
change the leading soft source that is matched below.

\begin{remark}[When the projection is exhaustive]
\label{rem:projection-exhaustive}
It is important to state how much of the source the projection captures.
Writing the potential as a function of the two invariants
\eqref{eq:aligned-invariants-main} and using
\eqref{eq:aligned-kernel-main},
\begin{equation}
\Dop_{AB}G(x,y)
=
G_{xx}\big(D_AxD_Bx\big)^{\rm STF}
+2G_{xy}\big(D_{(A}xD_{B)}y\big)^{\rm STF}
+G_{yy}\big(D_AyD_By\big)^{\rm STF},
\label{eq:two-variable-hessian-main}
\end{equation}
so matching a source proportional to \(\cosh(y)\,x^{-1}(D_AxD_Bx)^{\rm
STF}\) requires \(G_{xy}=G_{yy}=0\), hence \(G=f(x)+cy\), hence \(G_{xx}\)
independent of \(y\) --- which contradicts \(G_{xx}=\cosh(y)/x\) unless
\(y\) is itself a function of \(x\).  By
\eqref{eq:aligned-affine-main} this happens precisely in the aligned
configuration, and one checks directly that the transverse spin components
are the obstruction: \(\partial_\phi(a\cdot q)=0\) forces \(a^1=a^2=0\)
when \(p^\mu=(E,0,0,P)\).  Consequently, for aligned spin the even and odd
parts of the source are \emph{purely} electric and magnetic and the frame
equations lose nothing; for a generic spin orientation each level retains
a residual piece of the opposite parity which the exponentiating tower
does not fix.  All statements below about completeness of the tower refer
to the aligned case; the generic case remains a projection, in the same
sense as the other universality caveats of this section.
\end{remark}

\begin{remark}
Equation \eqref{eq:parity-alternation-main} resolves what would otherwise
be an inconsistency.  If one insists on the electric kernel at every
level, the odd-\(s\) frame equation has no real solution: the electric
projection of the odd-\(s\) source vanishes, and a formal solution comes
out purely imaginary.  The aligned-spin analysis of Section
\ref{sec:aligned-generating-main} exhibits exactly this behaviour, and it
is the diagnostic that fixes the alternation.  It also shows that the
alternation is not an obstruction to solving the tower: the two parities
are governed by the two halves of one hyperbolic generating function.
\end{remark}

The first term in \eqref{eq:kernel-expansion-main} produces the
inhomogeneous shift
\begin{equation}
\nu^{(0)}_{t\,AB}=u^s\K^{(s,0)}_{AB}[t].
\end{equation}
The term linear in \(C\) is the beginning of the homogeneous asymptotic
transformation.  More explicitly, the corresponding action on the shear
has the form
\begin{equation}
\delta_t C_{AB}
=
u^s\K^{(s,0)}_{AB}[t]
+
\mathbb L_t^{(s)}C_{AB}
+O(C^2),
\label{eq:first-Lt-definition-main}
\end{equation}
where \(\mathbb L_t^{(s)}\) denotes the homogeneous, \(C\)-linear part of
the spin-\(s\) asymptotic transformation.  It is a differential operator
on the shear whose highest angular-derivative part is controlled by
\(t^{A_1\cdots A_s}\).  This is the operator whose principal symbol will
be analyzed in Section \ref{sec:algebra-main}.

The universal identification is
\begin{equation}
\K^{(s,0)}_{AB}[t]=2G\,\mathscr S^{(s)}_{AB,{\rm exp}},
\label{eq:electric-identification-main}
\end{equation}
where \(\mathscr S^{(s)}_{AB,{\rm exp}}\) denotes the real projection
\eqref{eq:real-parity-source-main} of the exponentiating part of the
\(s\)th soft factor.  The factor \(2G\) is fixed by the \(s=0\) VV equation
\eqref{eq:vv-frame-equation-main}.  Equation
\eqref{eq:electric-identification-main} is
not meant to capture non-universal soft data.  It is the equation that
determines the generator.  Equivalently, the charge construction asks us to
solve
\begin{equation}
\K^{(s,0)}_{AB}[t]=2G\,\mathscr S^{(s)}_{AB,{\rm exp}}
\label{eq:inverse-kernel-problem-main}
\end{equation}
for the scalar potential \(\chi_t^{(s)}\) at the parity fixed by
\eqref{eq:parity-alternation-main}, and then to reconstruct the
corresponding part of \(t^{A_1\cdots A_s}\).  At \(s=0\) this is exactly
the VV equation for \(T_{\rm VV}\).  At \(s=1\) it determines
\(\epsilon^{AB}D_AY_B=D^2\Psi\), leaving the electric potential in
\(Y^A=D^A\Phi+\epsilon^{AB}D_B\Psi\) unfixed.  At higher rank it fixes
the maximally longitudinal component of one parity of the tensor
parameter.  This is why the soft-factor matching is not a decorative
identification: it is the practical rule for computing the asymptotic
generators selected by the Kerr soft sector.

The word ``universal'' has a precise role here.  The kernels we use are
generated by scalar-like, maximally longitudinal derivatives on the
sphere, of the parity dictated by \eqref{eq:parity-alternation-main}.
They are the natural higher-spin analogues of the trace-free Hessian
\(\Dop_{AB}T\) in the VV equation and of its parity dual.  A complete soft
theorem also contains lower-rank descendants and field-dependent radiative
corrections.  Those pieces are essential for the full gravitational phase
space, but they are not selected by the Kerr three-point exponential
alone.  The construction below should therefore be read as a clean
Kerr-selected subsector of the universal soft contribution, not as a
claim of completeness.

There are three reasons for using \(\mathscr S^{(s)}_{AB,{\rm exp}}\) as the source
in \eqref{eq:inverse-kernel-problem-main}.  First, a soft charge is a
Hamiltonian displacement of the shear, so its field-independent kernel is
literally the large-gauge frame shift.  Second, the same source appears on
the hard side of the Ward identity; choosing it fixes the hard and soft
charges to be two representations of the same soft theorem.  Third, the
exponentiating source is the classical limit of the spinning three-point
operator that generates the Kerr multipole tower.  Non-universal soft
terms may certainly modify the full asymptotic transformation, but they do
not define a theory-independent Kerr dressing.

\subsection{Universal soft contribution and the Kerr projection}

The previous paragraph hides an important point.  Beyond the leading and
subleading soft theorems, there is no statement that the complete soft
factor is universal.  What is universal is a distinguished operatorial
piece.  In the notation of the celestial and soft-symmetry literature, the
universal contribution for \(s\ge2\) may be represented as
\cite{Hamada:2018vrw,Li:2018gnc,Himwich:2023njb}
\begin{equation}
S^{(s)}_{AB,{\rm univ}}
=
-\frac{1}{s!}\,
\frac{(q^\rho J_{\mu\rho})(q^\sigma J_{\nu\sigma})}{p\cdot q}
\left(q\cdot\frac{\partial}{\partial p}\right)^{s-2}
\varepsilon_{AB}^{\mu\nu}(q),
\qquad s\ge2.
\label{eq:himwich-pate-universal-main}
\end{equation}
Here \(p^\mu\) is the momentum of the hard external leg, \(q^\mu(z)\) is
the null soft direction, \(J_{\mu\nu}\) is the Lorentz generator acting on
that hard leg, and
\begin{equation}
\varepsilon_{AB}^{\mu\nu}(q)
=D_{\{A}q^\mu D_{B\}}q^\nu
\end{equation}
is the sphere-index graviton polarization tensor.  Equation
\eqref{eq:himwich-pate-universal-main} should be read as an operator
acting on amplitudes.  The non-universal remainder is not discarded
because it is unimportant; it is omitted because it is not fixed by the
universal soft symmetry alone.

To compare with the spin-exponential form it is useful to choose helicity
polarizations.  We use
\(\eta_{\mu\nu}={\rm diag}(+,-,-,-)\) and
\(\epsilon^{0123}=+1\).  The label \(\eta=\pm1\) denotes the helicity of
the soft graviton.  We write \(\varepsilon^\mu_\eta(q)\) for a complex null
polarization vector satisfying
\begin{equation}
q\cdot\varepsilon_\eta=0,\qquad
\varepsilon_\eta\cdot\varepsilon_\eta=0,\qquad
\varepsilon_+\cdot\varepsilon_-=-1,
\end{equation}
and \(\varepsilon_{\eta}^{\mu\nu}=\varepsilon_\eta^\mu
\varepsilon_\eta^\nu\) for the factorized graviton polarization, with the
understanding that the trace-free sphere tensor is obtained by projecting
onto \(D_Aq^\mu D_Bq^\nu\).  In this basis the leading soft factor is
\begin{equation}
S_\eta^{(0)}=\frac{(p\cdot\varepsilon_\eta)^2}{p\cdot q}.
\label{eq:leading-helicity-soft-main}
\end{equation}
The part of the next terms that is generated by repeated insertions of
the angular momentum operator can be written as
\begin{align}
S_\eta^{(1)}&=-iS_\eta^{(0)}
\frac{q_\mu J^{\mu\nu}\varepsilon_{\eta\nu}}
{p\cdot\varepsilon_\eta},
\nonumber\\
S_\eta^{(2)}&=-S_\eta^{(0)}\frac{1}{2!}
\left(
\frac{q_\mu J^{\mu\nu}\varepsilon_{\eta\nu}}
{p\cdot\varepsilon_\eta}
\right)^2.
\label{eq:low-soft-helicity-main}
\end{align}
The exponentiating replacement used in this paper keeps precisely this
operatorial pattern,
\begin{equation}
\widehat S_{\eta,{\rm exp}}^{(s)}
=
S_\eta^{(0)}
\frac{(-i)^s}{s!}
\left(
\frac{q_\mu J^{\mu\nu}\varepsilon_{\eta\nu}}
{p\cdot\varepsilon_\eta}
\right)^s .
\label{eq:exponentiating-replacement-main}
\end{equation}
The hat distinguishes this operator acting on the hard leg from the
classical coefficient \(S_{\eta,{\rm exp}}^{(s)}\) defined in
\eqref{eq:helicity-levels-main}; replacing the spin operator by its
classical Kerr value gives the latter after the on-shell projection.
This is not the full soft
factor at order \(s\).  It is the part that becomes the Kerr multipole
generator after the hard spin operator is replaced by a classical spin
tensor.

The primary justification for the exponential dressing is not the
amplitude but the linearized solution.  The Kerr field at \(O(G)\) is the
Schwarzschild field complex-shifted by \(i a^\mu\), and by the Fourier
shift theorem a displacement \(\vec c=i\vec a\) multiplies the momentum
space field by \(e^{-i\vec k\cdot(i\vec a)}=e^{+\vec k\cdot\vec a}\).  The
classical source used throughout this paper is therefore
\begin{equation}
\mathscr S^{(s)}_{AB,{\rm exp}}
=\hbox{real parity projection of the coefficient of }\lambda^s
\hbox{ in }\;
S^{(0)}_\eta\,e^{\eta\lambda a\cdot q},
\qquad \lambda=\omega,
\label{eq:classical-source-main}
\end{equation}
valid at generic \(q^\mu\) on the celestial sphere.  The exponent
\(a\cdot k=\omega(a\cdot q)\) is dimensionless.  The retarded-time moment
extracts the coefficient of \(\omega^s\), while \(\lambda\) keeps track of
the dimensions in the generating function.  The level-\(s\) potential
therefore has dimension \(L^{s+1}\), and only
\(\omega^s\chi^{(s)}\) has the length of a Bondi-frame shift.  Accordingly
\(\Phi_\lambda\) and \(\Psi_\lambda\) below are generating functions: their
coefficients must be accompanied by the corresponding power of
\(\lambda\).  The self-dual/anti-self-dual split supplies the helicity
sign.  This is the
form matched by the frame equations of Section
\ref{sec:leading-subleading-main} and solved in Section
\ref{sec:aligned-generating-main}, and its multipole content is
\eqref{eq:complex-shift-legendre-main}.

The amplitude provides the on-shell confirmation.  In the Guevara--Ochirov--Vines
convention the amplitude-level object is an operator on massive spin
states \cite{ArkaniHamed:2017jhn,Guevara:2017csg,Guevara:2018wpp,Guevara:2019fsj}.  For a soft graviton \(k^\mu=\omega q^\mu\),
\begin{align}
\widehat{\mathcal A}_{3,\eta}^{(j)}
&=
\mathcal A_{3,\eta}^{(0)}
\exp\!\left(i\widehat{\mathcal W}_\eta\right),
\nonumber\\
\widehat{\mathcal W}_\eta
&=
\frac{k_\mu\varepsilon_{\eta\nu}J^{\mu\nu}}
{p\cdot\varepsilon_\eta}
=
\omega\,\widehat W_\eta(q).
\label{eq:gov-operator-main}
\end{align}
Here \(j\) labels the spin of the massive state and
\(\mathcal A_{3,\eta}^{(0)}\) is the scalar minimal-coupling amplitude.
The operator \(J^{\mu\nu}=L^{\mu\nu}+S^{\mu\nu}\) is the total Lorentz
generator acting on the massive spinor variables and hard kinematics:
\(L^{\mu\nu}\) is its orbital differential part and \(S^{\mu\nu}\) its
intrinsic-spin part.  The physical spin-\(j\) amplitude is obtained by sandwiching
\eqref{eq:gov-operator-main} between massive spin states.  After taking
the classical Kerr limit, the spin part of \(J^{\mu\nu}\) is replaced by a
classical spin tensor \(S_{\rm cl}^{\mu\nu}\).  Crossing to the
outgoing-soft convention used for the classical source gives
\begin{equation}
S_{\rm Kerr}^{\rm cl}(k)
=
S_\eta^{(0)}
\exp[-i\omega W_\eta(q)],
\qquad
W_\eta(q)=
\frac{q_\mu\varepsilon_{\eta\nu}S_{\rm cl}^{\mu\nu}}
{p\cdot\varepsilon_\eta}.
\label{eq:kerr-exponent-main}
\end{equation}
Thus the \(+i\) exponential belongs to the GOV operator acting on massive
spinors, while the \(-i\) exponential is the crossed classical Kerr/source
convention used for the soft dressing below; the two signs are not
silently identified.  Expanding the classical source gives
\begin{equation}
S_{\rm Kerr}
=
\sum_{n\ge0}S^{(0)}\frac{[-i\omega W_\eta(q)]^n}{n!}.
\label{eq:kerr-expansion-main}
\end{equation}
In the classical spin limit this is the Kerr multipole generating function,
\begin{equation}
M_\ell+iS_\ell=M(ia)^\ell,
\label{eq:kerr-multipoles-main}
\end{equation}
in the standard Geroch--Hansen/Thorne multipole language
\cite{Geroch1970,Hansen1974,Thorne:1980ru}.  Here
\(a^\mu=S^\mu/m\), \(S^\mu\) being the physical spin vector; in the
axisymmetric scalar notation \(a\) is its signed magnitude.  Hence the charge tower
\eqref{eq:soft-charge-main} is naturally paired with the Kerr multipole
tower.  The first term \(S^{(0)}\) gives the boosted Schwarzschild or mass
field.  After the powers of \(\omega\) are stripped, the term
\(S^{(0)}(-iW_\eta)\) gives the spin or frame-dragging data.  The quadratic
term \(S^{(0)}(-iW_\eta)^2/2!\) gives the electric
quadrupole contribution.  The general coefficient
\begin{equation}
S^{(0)}\frac{(-iW_\eta)^s}{s!}
\label{eq:kerr-soft-coefficient-main}
\end{equation}
is the soft imprint of the Kerr multipole relation
\eqref{eq:kerr-multipoles-main}.
This is the motivation for isolating the universal exponentiating
sector.

\subsection{Leading Faddeev--Kulish Matching}

\subsubsection{Soft charge as a zero-frequency operator}
\label{sec:soft-zero-frequency-operator-main}

The leading soft charge can be written in zero-frequency form as
\begin{equation}
\Q_0^{\rm soft}(T)
=
\int \dd\mu(k)\,\delta(\omega)\,
\widetilde{\Dop}^{\mu\nu}(T)
\left[
\varepsilon_{\mu\nu}^{\eta}(q)a_\eta(\omega q)
-
\varepsilon_{\mu\nu}^{\eta *}(q)a^\dagger_\eta(\omega q)
\right],
\label{eq:appendix-leading-soft-operator}
\end{equation}
where $k^\mu=\omega q^\mu$, $\dd\mu(k)$ is the Lorentz-invariant measure on
the null cone, \(a_\eta\) and \(a^\dagger_\eta\) respectively annihilate
and create a graviton of helicity \(\eta\), and
$\widetilde{\Dop}^{\mu\nu}$ denotes the projection of
$\Dop_{AB}$ onto the polarization tensor basis.  Normalization constants
depend on conventions and are immaterial for the statement.

The leading Faddeev--Kulish dressing of a one-particle state of momentum
$p$ has exponent
\begin{equation}
R^{(0)}(p)
\sim
\sum_\eta\int\dd\mu(k)\,\phi(\omega)
\frac{p^\mu p^\nu}{p\cdot k}
\left[
\varepsilon_{\mu\nu}^{\eta}a^\dagger_\eta(k)
-
\varepsilon_{\mu\nu}^{\eta *}a_\eta(k)
\right],
\label{eq:appendix-FK-leading}
\end{equation}
where $\phi(\omega)$ has support near $\omega=0$ and $\phi(0)=1$.  After
the standard zero-frequency identification, the exponent has the same
operator form as \eqref{eq:appendix-leading-soft-operator} precisely when
\begin{equation}
\Dop_{AB}T
\propto
S^{(0)}_{AB}.
\end{equation}
Thus the leading dressing selects $T_{\rm VV}$, in agreement with the
amplitude derivation of Ref.~\cite{Elkhidir:2024ward}.  This is the
amplitude version of the intrinsic/canonical frame shift.

It is useful to spell out the elementary coherent-state computation,
because it is the local mechanism behind the frame dictionary.  Write a
generic soft dressing as
\begin{equation}
R[f]
=
\sum_\eta\int\dd\mu(k)
\left[
f_\eta(k)a^\dagger_\eta(k)
-f_\eta^*(k)a_\eta(k)
\right],
\label{eq:FK-displacement-general}
\end{equation}
with the usual graviton commutator normalization absorbed into
\(\dd\mu(k)\).  Since \(R[f]\) is linear in creation and annihilation
operators, the Baker--Campbell--Hausdorff series terminates:
\begin{equation}
e^{-R[f]}a_\eta(k)e^{R[f]}
=a_\eta(k)+f_\eta(k),
\qquad
e^{-R[f]}a^\dagger_\eta(k)e^{R[f]}
=a^\dagger_\eta(k)+f_\eta^*(k).
\label{eq:FK-displaces-modes}
\end{equation}
For the linearized metric expansion this gives
\begin{align}
e^{-R[f]}h_{\mu\nu}(x)e^{R[f]}
&=h_{\mu\nu}(x)+h_{\mu\nu}^{f}(x),
\nonumber\\
h_{\mu\nu}^{f}(x)
&=
\sum_\eta\int\dd\mu(k)
\left[
\varepsilon^\eta_{\mu\nu}f_\eta(k)e^{-ik\cdot x}
 +{\rm c.c.}
\right].
\label{eq:FK-classical-profile}
\end{align}
Thus every choice of soft profile \(f_\eta\) defines a classical
radiative/Coulombic tail at null infinity.  For the leading gravitational
dressing,
\begin{equation}
f_\eta^{(0)}(p;k)
\propto
\phi(\omega)\,
\frac{p^\mu p^\nu\varepsilon^\eta_{\mu\nu}(k)}
{p\cdot k},
\label{eq:FK-leading-profile}
\end{equation}
the large-\(r\), fixed-\(u\) limit of
\eqref{eq:FK-classical-profile} has the shear component
\begin{equation}
C^f_{AB}(z)
=
-4G\,S^{(0)}_{AB}(p;q)
=
-2\Dop_{AB}T_{\rm VV},
\label{eq:FK-leading-shear-profile}
\end{equation}
up to ordinary translations and the common infrared prescription.  This
is the direct operator derivation of the statement that the leading FK
dressing prepares precisely the VV shear.

On the other hand, the spinning black-hole dressing is obtained by multiplying the leading
profile by the exponentiating spin factor,
\begin{equation}
f_\eta^{\rm Kerr}(p,a;k)
=
f_\eta^{(0)}(p;k)\,\exp[-i\mathcal W_\eta(p,a;k)].
\label{eq:FK-Kerr-profile}
\end{equation}
Substituting \eqref{eq:FK-Kerr-profile} into
\eqref{eq:FK-classical-profile} gives a classical profile whose expansion
is
\begin{equation}
h_{\mu\nu}^{\rm Kerr,lin}
=
\sum_{n\ge0}\frac{(-i)^n}{n!}\,
h_{\mu\nu}^{(n)}[\mathcal W_\eta^n f^{(0)}].
\label{eq:FK-Kerr-profile-expansion}
\end{equation}
Here \(h_{\mu\nu}^{(n)}[\mathcal W_\eta^n f^{(0)}]\) denotes the
linearized metric profile obtained by inserting the indicated
order-\(n\) source in the Fourier integral
\eqref{eq:FK-classical-profile}.
The \(n=0\) term is the boosted Schwarzschild field.  The \(n=1\) term is
the current-dipole frame-dragging field.  Higher terms reproduce the Kerr
multipole pattern after the standard gauge reconstruction of the Coulombic
field.  At null infinity we use, at each level, the shear projection of the parity
selected by \eqref{eq:parity-alternation-main}: electric at even \(s\),
magnetic at odd \(s\).  Consequently the FK computation does not by itself
determine the opposite-parity completion at a given level, nor
finite-frequency non-universal terms; it
determines the universal soft cloud that appears in the
canonical/intrinsic dictionary.

The leading example teaches us the rule:
\begin{equation}
\hbox{soft dressing exponent}
\quad\Longleftrightarrow\quad
\hbox{soft charge}
\quad\Longleftrightarrow\quad
\hbox{asymptotic frame shift}.
\label{eq:appendix-leading-rule}
\end{equation}
At higher spin the same rule is applied to the universal exponentiating
part of the soft factor.  The charge parameter is no longer just a scalar
$T$ but a symmetric tensor, or equivalently a polynomial symbol on
$T^\ast S^2$.

\subsection{Mode-Space Form of the Soft Charges}

Let the radiative graviton field be expanded near $\Iplus$ in helicity
modes as
\begin{equation}
C_{AB}(u,z)
\sim
\sum_{\eta=\pm}
\int_0^\infty\dd\omega\,
\left[
\varepsilon^{\eta}_{AB}(q)a_\eta(\omega q)e^{-i\omega u}
+
\varepsilon^{\eta *}_{AB}(q)a^\dagger_\eta(\omega q)e^{i\omega u}
\right],
\end{equation}
with conventional normalization suppressed.  The shifted news is
\begin{equation}
\News_{AB}
=
\partial_u C_{AB}-\tau_{AB}.
\end{equation}
The soft part of the charge is sensitive to the zero-frequency behavior of
the oscillator modes.  Formally,
\begin{equation}
\int_{-\infty}^{+\infty}\dd u\,u^s e^{-i\omega u}
=
2\pi i^s\delta^{(s)}(\omega).
\end{equation}
Thus a charge with retarded-time moment $u^s$ projects onto the
$s$th derivative of the soft mode at $\omega=0$.

For $s=0$ we have the leading mode operator
\begin{equation}
\Q_0^{\rm soft}(T)
\sim
\sum_\eta\int\dd\mu(k)\,\delta(\omega)\,
\widetilde{\Dop}^{\mu\nu}(T)
\left[
\varepsilon_{\mu\nu}^{\eta}a_\eta(\omega q)
-
\varepsilon_{\mu\nu}^{\eta *}a^\dagger_\eta(\omega q)
\right].
\end{equation}
This is the form compared to the leading Faddeev--Kulish exponent.  The
comparison fixes $\Dop_{AB}T$ to be the leading soft tensor.  As we discussed above, the
Veneziano--Vilkovisky potential is the unique Lorentz-covariant solution
modulo translations.

At $s=1$ the charge contains one power of $u$, and hence a derivative of
the zero-frequency delta function:
\begin{equation}
\Q_1^{\rm soft}(Y)
\sim
\sum_\eta\int\dd\mu(k)\,\delta'(\omega)\,
\widetilde{\Ddual}^{\mu\nu}(\epsilon^{AB}D_AY_B)
\left[
\varepsilon_{\mu\nu}^{\eta}a_\eta(\omega q)
+
\varepsilon_{\mu\nu}^{\eta *}a^\dagger_\eta(\omega q)
\right].
\end{equation}
Here \(\widetilde{\Ddual}^{\mu\nu}\) is the polarization-space image of
the magnetic operator \(\Ddual_{AB}\), just as
\(\widetilde{\Dop}^{\mu\nu}\) is that of \(\Dop_{AB}\).
The sign and factor of $i$ depend on the convention relating the
retarded-time moment to $\delta'(\omega)$.

For general $s$, a rank-$s$ mode operator, the universal mode operator has the
zero-frequency mode
\begin{equation}
\Q_s^{\rm soft}(t)
\sim
\sum_\eta\int\dd\mu(k)\,\delta^{(s)}(\omega)\,
\widetilde{\K}^{\mu\nu}_{(s)}[t]
\left[
\varepsilon_{\mu\nu}^{\eta}a_\eta(\omega q)
+(-1)^{s+1}
\varepsilon_{\mu\nu}^{\eta *}a^\dagger_\eta(\omega q)
\right].
\end{equation}
The kernel $\widetilde{\K}_{(s)}$ is the polarization-space image of
$\K^{(s,0)}_{AB}[t]$.  The universal frame equation is
\begin{equation}
\widetilde{\K}^{\mu\nu}_{(s)}[t]\varepsilon^\eta_{\mu\nu}
=
2G\,S^{(0)}_\eta\frac{(-iW_\eta)^s}{s!}.
\end{equation}
The two helicity equations are recombined according to
\eqref{eq:real-parity-source-main}.  This is the all-orders version of the
leading Veneziano--Vilkovisky problem.

\section{Frame Equations}
\label{sec:leading-subleading-main}

We now solve \eqref{eq:electric-identification-main} for the
parity-selected components of the higher-spin generators
\(t^{A_1\ldots A_s}\).

\subsection{Supertranslations}

At \(s=0\), the kernel \(\K^{(0,0)}_{AB}[T]\) is the trace-free Hessian
\(\Dop_{AB}T\). As mentioned above, the leading member of the tower is not a
new object: it is precisely the Veneziano--Vilkovisky supertranslation.

The inhomogeneous part of a supertranslation acts on the shear as
\begin{equation}
\delta_T C_{AB}=-2\Dop_{AB}T.
\end{equation}
The homogeneous part contains the transport term
\begin{equation}
\delta_T^{\rm hom}C_{AB}=T\partial_u C_{AB}.
\end{equation}
At the past boundary of $\Iplus$, the canonical frame condition is
\begin{equation}
C_{AB}\big|_{\Iplus_-}=0.
\end{equation}
If an intrinsic frame has nonzero boundary shear
$C^{\rm int}_{AB}|_{\Iplus_-}$, the supertranslation relating the two
frames satisfies
\begin{equation}
-2\Dop_{AB}T=C^{\rm int}_{AB}\big|_{\Iplus_-}.
\end{equation}
For the linearized boosted Schwarzschild source this becomes the
Veneziano--Vilkovisky equation.

This is the prototype for every higher member of the construction.  The
canonical frame condition removes the boundary shear by convention; the
intrinsic frame keeps the shear naturally sourced by the hard scattering
data.  The soft charge does not merely create a formal zero-frequency
graviton.  Its inhomogeneous action supplies the large diffeomorphism that
relates these two descriptions.  At \(s=0\), the tensor
\(\Dop_{AB}T\) is therefore simultaneously a soft kernel, a shear
displacement and the electric displacement-memory pattern measured at null
infinity.

\subsection{Sphere diffeomorphisms}

For a smooth vector field \(Y^A\) on \(S^2\), set
\(\alpha_Y=\tfrac12D_AY^A\).  In the generalized-BMS phase space the
celestial metric is allowed to vary, and the standard action is
\cite{Campiglia:2014yka,Campiglia:2015kxa}
\begin{align}
\delta_Y\gamma_{AB}
&=\Lie_Y\gamma_{AB}-2\alpha_Y\gamma_{AB},
\nonumber\\
\delta_Y C_{AB}
&=\Lie_Y C_{AB}-\alpha_Y C_{AB}
+\alpha_Yu\partial_uC_{AB}
-2u(D_AD_B\alpha_Y)^{\rm TF}.
\label{eq:generalized-bms-action-main}
\end{align}
The first three terms in the second line form the homogeneous action.  They
are what produce the non-Abelian bracket with a supertranslation shift.
The last term is the inhomogeneous electric shear shift associated with
the divergence of \(Y^A\); simultaneously,
\(\delta_Y\gamma_{AB}\neq0\) for a generic smooth vector.  This expanded
phase-space statement is the precise sense in which a
\(\mathrm{Diff}(S^2)\) parameter is a generalized-BMS transformation.

At subleading order the Kerr exponentiating source is magnetic by
\eqref{eq:parity-alternation-main}.  Its projection of the generalized-BMS
subleading Ward charge therefore fixes the curl of a sphere vector field:
\begin{equation}
\Ddual_{AB}\big(\epsilon^{CD}D_CY_D\big)
=2G\,\mathscr S^{(1)}_{AB,{\rm exp}}.
\label{eq:subleading-divergence-main}
\end{equation}
A distinction is essential.  Equation
\eqref{eq:subleading-divergence-main} is a parity projection of the soft
Ward kernel; it is not obtained by replacing the last term of
\eqref{eq:generalized-bms-action-main} by a magnetic tensor while holding
\(\gamma_{AB}\) fixed.  Calling its solution a generalized-BMS vector
means that \(Y^A\) is the smooth Ward parameter whose full phase-space
completion is \eqref{eq:generalized-bms-action-main}; the charge equation
fixes only its coexact part.
A generic solution is not a global conformal Killing vector and is not a
holomorphic or meromorphic vector field in the usual extended-BMS sense.
It is a smooth generator of generalized BMS, namely an element of
\(\mathrm{Diff}(S^2)\) \cite{Campiglia:2014yka,Campiglia:2015kxa,Compere:2018ylh}. Indeed, if $Y^A$ is a global conformal Killing vector, then in complex coordinates
the holomorphic and anti-holomorphic parts are highly constrained.  The
operator $\Ddual_{AB}(\epsilon^{CD}D_CY_D)$ then has the analytic structure appropriate
to global Lorentz transformations.  The right-hand side, however, contains the hard
angular-momentum insertion
\begin{equation}
W_\eta=
\frac{q\cdot J\cdot\varepsilon_\eta}{p\cdot\varepsilon_\eta},
\end{equation}
which is a nontrivial function of both $z$ and $\bar z$ for generic massive
kinematics.  Therefore the solution generically lives in
$\mathrm{Diff}(S^2)$ rather than in the global Lorentz subalgebra.

In turn, the vector admits the
Helmholtz decomposition
\begin{equation}
Y^A=D^A\Phi+\epsilon^{AB}D_B\Psi.
\label{eq:helmholtz-main}
\end{equation}
Only
\begin{equation}
\epsilon^{AB}D_AY_B=D^2\Psi
\end{equation}
appears in \eqref{eq:subleading-divergence-main}.  Therefore the
exponentiating sector fixes \(\Psi\), the magnetic potential, but it
does not fix \(\Phi\), the electric potential.  This is the first place
where the parity alternation \eqref{eq:parity-alternation-main} becomes
operational, and it is physically the expected answer: the \(s=1\) Kerr
datum is the current dipole, that is frame dragging, which is odd-parity
in the standard multipole classification.  The minimal choice is therefore
to set
\begin{equation}
D_A Y^A=0
\end{equation}
away from singular points, so that $Y^A$ is locally a divergence-free, or
Hamiltonian, vector field.

To repeat: for the Kerr problem \(Y^A\) should be treated as a smooth
generalized-BMS vector field, not as a holomorphic conformal generator.
A holomorphic vector field would be far too
restrictive: the source produced by a generic massive spinning hard leg is
a smooth function of the real direction \(q^\mu(z,\bar z)\), not a
meromorphic current insertion in a local complex patch.  The chiral
celestial presentation can be recovered later by projection, but the
global Kerr-frame problem naturally lives in the real
\(\mathrm{Diff}(S^2)\) phase space.  The exponentiating subleading source
is a smooth real function on the
sphere built from \(p^\mu\), \(q^\mu(z)\) and the spin tensor.  It fixes
only \(\epsilon^{AB}D_AY_B\), so it fixes only the magnetic potential in
the Hodge decomposition of \(Y^A\).  In other words, the subleading soft
sector does not select an extended-BMS meromorphic vector; it selects the
curl of a smooth generalized-BMS Ward parameter.  The electric part of the
vector is additional data, while its homogeneous weight is
\(\alpha_Y=\tfrac12D\cdot Y\) once the full generalized-BMS completion
\eqref{eq:generalized-bms-action-main} is chosen.

This is the first place where the Compere--Fiorucci--Ruzziconi
super-Lorentz phase-space viewpoint \cite{Compere:2018ylh} becomes directly
useful.  The subleading soft frame shift is naturally a
$\mathrm{Diff}(S^2)$ transformation, not merely a global conformal
transformation.  The corresponding dressing viewpoint is also close in
spirit to the subleading soft dressings studied in Ref.~\cite{Choi:2019rlz}.

\subsubsection{A solvable special case}

There is a useful special kinematic case in which the equation can be
solved explicitly.  We record it in the electric variable, because the
form of the answer is what diagnoses the parity of the source.  Let
\begin{equation}
x=p\cdot q,
\qquad
y=q\cdot J\cdot\partial_zq.
\end{equation}
Assume that the kinematics impose a linear relation
\begin{equation}
y=c\,\partial_zx.
\label{eq:appendix-special-constraint}
\end{equation}
This occurs, for example, when the spatial angular momentum and boost
vectors are aligned with the spatial momentum, with \(c\) a complex
kinematic coefficient determined by the
spin and boost parameters.  If one provisionally uses the electric kernel and lets $D\cdot Y$ depend
only on $x$, the subleading equation reduces to
\begin{equation}
F''(x)=\frac{ic}{x},
\end{equation}
with particular solution, modulo the kernel of $\Dop_{AB}$,
\begin{equation}
D\cdot Y=ic\,x\log x-ic\,x.
\label{eq:appendix-special-divergence}
\end{equation}
The explicit factor of \(i\) is the point.  A real supertranslation-type
generator cannot be imaginary, and the obstruction is precisely
\eqref{eq:parity-alternation-main}: the \(s=1\) source has no electric
part.  Replacing \(\Dop_{AB}\to\Ddual_{AB}\) and \(D\cdot Y\to
\epsilon^{AB}D_AY_B\) removes the factor of \(i\) and returns a real
magnetic potential with the same logarithmic profile.
One solution is
\begin{equation}
Y^z
=
-\frac{ic}{2}\,
\frac{\mathcal G(x)}{p\cdot\partial_zq},
\qquad
Y^{\bar z}
=
+\frac{ic}{2}\,
\frac{\mathcal G(x)}{p\cdot\partial_{\bar z}q},
\end{equation}
with
\begin{equation}
\mathcal G(x)=\frac{x^2}{2}\log x-\frac34x^2,
\end{equation}
the relative sign being the one appropriate to a divergence-free rather
than a gradient vector.
This example is not used as a general proof.  Its purpose is to show that
the subleading frame equation has nontrivial $\mathrm{Diff}(S^2)$
solutions, that spin data enter naturally, and that the parity of the
solution is fixed by \eqref{eq:parity-alternation-main} rather than
chosen.

\subsubsection{Higher-spin frame data}

For a rank-$s$ parameter the parity-selected inhomogeneous shift is
\begin{equation}
\nu^{(0)}_{t\,AB}=u^s\K^{(s,0)}_{AB}[t].
\end{equation}
The canonical higher-spin frame can be defined as one in which the relevant
parity-selected soft moments vanish at the boundary.  The intrinsic higher-spin
frame is then obtained by solving
\begin{equation}
\K^{(s,0)}_{AB}[t]=2G\,\mathscr S^{(s)}_{AB,{\rm exp}}.
\end{equation}
This is not a new gauge condition imposed in the bulk.  It is an
asymptotic definition motivated by the leading Veneziano--Vilkovisky
construction and by soft dressing.

The phrase ``higher-spin frame'' is meant in this restricted asymptotic
sense.  We are not introducing propagating higher-spin fields in the bulk.
The rank \(s\) label records the number of angular derivatives, or
equivalently the degree of the polynomial symbol on \(T^\ast S^2\), needed
to organize the \(s\)th soft moment.  The leading STF component of the
selected parity is fixed by the soft kernel; trace descendants and the
opposite-parity components enter only when this leading action is
completed to a global transformation.
The corresponding scalar, vector and tensor harmonic decompositions are
collected in Appendix \ref{app:spherical-harmonic-basis}, while the
electric/magnetic subleading equation is checked explicitly in Appendix
\ref{sec:component-mode-boundary-checks}.

Therefore, for \(s\ge2\) the Kerr exponentiating source fixes the maximally
longitudinal component of the rank-\(s\) parameter, at the parity fixed by
\eqref{eq:parity-alternation-main}: electric for even \(s\), magnetic for
odd \(s\).  The opposite-parity components, trace descendants, and
lower-rank completions require additional input.  The natural algebraic home for the full bookkeeping is
the polynomial algebra on \(T^\ast S^2\), discussed only after the charges
and Ward identity have been specified.

This is the part that produces the Kerr multipole generating function.
Non-universal corrections can modify the full soft expansion but are not
part of the universal Kerr dressing claim.
The separation is important.  The complete gravitational soft expansion
contains theory-dependent and state-dependent pieces beyond the universal
terms.  Those pieces may contribute to radiative tails, magnetic parity
data and nonlinear memory.  The charge tower constructed here keeps the
part that exponentiates with the spinning three-point operator.  That is
why it is sufficient for the Kerr multipole dressing, but not advertised
as the full all-orders soft theorem.

\section{Hard Charges and Ward Identities}
\label{sec:hard-main}

The soft charge by itself is not a symmetry generator of the scattering
problem.  It creates or annihilates a zero-frequency graviton.  A symmetry
statement requires a hard operator whose action on finite-energy states is
the hard side of the soft theorem.  The result established in this section
is therefore deliberately precise: assuming the tree-level gravitational
soft theorem and the standard null-infinity normalization, the soft charge
\(\Q_s^{\rm soft}(t)\) and the hard operator defined below obey the Ward
identity in the universal exponentiating sector.  We do not prove here an
independent all-loop quantum symmetry theorem, nor do we construct the
exact nonlinear hard flux form for every possible completion of
the radiative phase space.

The total charge is
\begin{equation}
\Q_s^{\rm total}(t)=\Q_s^{\rm soft}(t)+\Q_s^{\rm hard}(t),
\label{eq:total-charge-main}
\end{equation}
where the hard piece acts on finite-energy hard data.  The Ward identity is
\begin{equation}
\langle{\rm out}|[\Q_s^{\rm total}(t),S]|{\rm in}\rangle=0.
\label{eq:ward-main}
\end{equation}
Here \(S\) is the scattering matrix.
Equivalently,
\begin{equation}
\langle{\rm out}|\Q_s^{\rm soft}(t)S|{\rm in}\rangle
=
-\langle{\rm out}|\Q_s^{\rm hard}(t)S|{\rm in}\rangle.
\label{eq:soft-hard-ward-main}
\end{equation}
This equation is the charge form of the soft theorem.  The differential
expression for \(\Q_s^{\rm soft}(t)\) defines a family of soft smearings.
The condition
\(\K^{(s,0)}_{AB}[t]=2G\,\mathscr S^{(s)}_{AB,{\rm exp}}\) selects the
Kerr/exponentiating
member of that family and therefore fixes the hard Ward operator to be the
same exponentiating soft operator acting on hard states.

There is an important conceptual reason to emphasize the hard piece.  A
soft charge alone changes the radiative vacuum; it does not by itself
express conservation in a scattering process.  The Ward identity works
because the same large transformation also acts on the hard particles and
on the finite-frequency gravitational flux.  In the Kerr sector this hard
action carries the same spin exponential that appears in the three-point
amplitude.  The equality between soft insertion and hard differential
operator is therefore the bridge between the null-infinity charge and the
amplitude-side Kerr multipole generator.

\subsection{Soft theorem and smeared hard operator}

Let \(\mathcal M_n\) be the \(n\)-point hard amplitude and
\(\mathcal M_{n+1}^{(\eta)}(\omega q)\) the corresponding amplitude with
an additional outgoing soft graviton of helicity \(\eta=\pm\).  The \(s\)th soft theorem
defines leg-wise hard differential operators
\(\mathcal D^{(s)}_{q,\eta,i}\) by
\begin{equation}
\left.
\partial_\omega^s\big[\omega\mathcal M_{n+1}^{(\eta)}(\omega q)\big]
\right|_{\omega=0}
=
\sum_i\sigma_i\,\mathcal D^{(s)}_{q,\eta,i}\mathcal M_n ,
\label{eq:hard-soft-theorem-main}
\end{equation}
where \(\sigma_i=+1\) for outgoing hard legs and \(\sigma_i=-1\) for
incoming hard legs.  The charge smearing is
\begin{equation}
\mathcal D^{(s)}_{t,i}\mathcal M_n
:=
\sum_{\eta=\pm}\int\dd^2z\,\sqrt\gamma\,
\K^{(s,0)}_{AB}[t](z)\,
\varepsilon_\eta^{AB}(q(z))\,
\mathcal D^{(s)}_{q,\eta,i}\mathcal M_n .
\label{eq:smeared-hard-operator-main}
\end{equation}
This is the hard operator with exactly the same angular kernel as the soft
charge.  With the normalization of \(\Q_s^{\rm soft}\) used in
Section~\ref{sec:soft-kerr-main}, the soft insertion gives
\begin{equation}
\langle{\rm out}|\Q_s^{\rm soft}(t)S|{\rm in}\rangle
=
\sum_i\sigma_i\,
\mathcal D^{(s)}_{t,i}\mathcal M_n .
\label{eq:soft-insertion-derived-main}
\end{equation}
The hard charge is represented on external states by
\begin{equation}
\Q_s^{\rm hard}(t)|{\rm hard}\rangle
=
-\sum_i\sigma_i\,\mathcal D^{(s)}_{t,i}|{\rm hard}\rangle .
\label{eq:hard-state-main}
\end{equation}
Equations \eqref{eq:soft-insertion-derived-main} and
\eqref{eq:hard-state-main} imply the Ward identity
\eqref{eq:soft-hard-ward-main}.  This is a representation statement rather
than an independent theorem: once the soft theorem and the smearing kernel
are specified, every object in the Ward identity has been defined.

In the Kerr exponentiating sector the diagonal spin-dependent part is
\begin{equation}
\mathcal D^{(s)}_{q,\eta,i}\Big|_{\rm Kerr,diag}
=
S_i^{(0)}(q,\eta)\frac{(-iW_{\eta,i})^s}{s!},
\label{eq:hard-kerr-main}
\end{equation}
where \(W_{\eta,i}\) is the classical Kerr/source quantity defined in
\eqref{eq:kerr-exponent-main}.  Before the classical limit is taken, this
insertion is the operator \((i\widehat{\mathcal W}_{\eta,i})^s/s!\) acting
on the massive spin state as in \eqref{eq:gov-operator-main}; the orbital
part of \(J_i\) gives the usual differential action on hard kinematics.

\subsection{Flux form}

A convenient null-infinity form is
\begin{align}
\Q_s^{\rm hard}(t)
&=
\int_{\Iplus}\dd u\,\dd^2z\,\sqrt\gamma\,
\Big[
u^s t^{A_1\cdots A_s}
D_{A_1}\cdots D_{A_s}T^{\rm hard}_{uu}
\nonumber\\
&\hspace{2.4cm}
{}+
s\,u^{s-1}t^{A_1\cdots A_s}
D_{A_1}\cdots D_{A_{s-1}}T^{\rm hard}_{uA_s}
\Big]
\;+\;\Q^{\rm hard}_{s,i^\pm}(t).
\label{eq:hard-charge-main}
\end{align}
Here \(T^{\rm hard}_{uu}\) and \(T^{\rm hard}_{uA}\) are the independent
hard stress-tensor/flux components through \(\Iplus\), while
\(\Q^{\rm hard}_{s,i^\pm}\) denotes massive-particle contributions
entering through future/past timelike infinity \(i^\pm\).  The expression is a
Ward-operator form: integrations by parts, conformal-weight
terms and improvements depend on the Bondi-frame convention.  These
ambiguities do not affect the Ward identity, provided the total charge is
used consistently.

Let us spell out how \eqref{eq:hard-charge-main} becomes the hard soft
operator.  For a massless hard field, the mode expansion at
\(\Iplus\) has the form
\begin{equation}
\Phi(u,z)
=
\int_0^\infty\frac{\dd E}{2\pi}
\left[
a(E,z)e^{-iEu}+a^\dagger(E,z)e^{iEu}
\right],
\label{eq:hard-mode-expansion-main}
\end{equation}
so the flux \(T_{uu}^{\rm hard}\sim:\partial_u\Phi\partial_u\Phi:\)
contains bilinears \(a^\dagger(E,z)a(E,z)\) after normal ordering.  The
basic distributional identity
\begin{equation}
\int_{-\infty}^{\infty}\dd u\,u^s e^{iu(E-E')}
=
2\pi i^s\delta^{(s)}(E-E')
\label{eq:u-moment-delta-main}
\end{equation}
shows what the factor \(u^s\) does: after integrating by parts in energy,
it turns the flux charge into an \(s\)th-order differential operator on
the hard momentum variables.  The sphere derivatives in
\eqref{eq:hard-charge-main} act on the angular delta functions that
localize the hard particle at \(z=z_i\).  Thus, on an external hard
state, the radiative flux form has the structure
\begin{equation}
\Q_s^{\rm hard}(t)|p_1,\ldots,p_n\rangle
=
-\sum_{i=1}^n\sigma_i\,\mathcal S^{(s)}_{t,i}
|p_1,\ldots,p_n\rangle,
\label{eq:hard-action-derived-main}
\end{equation}
with \(\mathcal S^{(s)}_{t,i}=\mathcal D^{(s)}_{t,i}\) after the
normalization of the flux charge is matched to the soft theorem.  For
massive external legs the same operator is supplied by the \(i^\pm\) part
of the charge; this is the timelike-infinity completion of the same Ward
identity.

For \(s=0\), the radiative part of the hard charge agrees with the
Faddeev--Kulish analysis of Ref.~\cite{Elkhidir:2024ward}.  In their
notation,
\begin{equation}
Q^{\rm rad}_{h,0}(T)
=
-\frac{1}{32\pi G}
\int\dd u\,\dd^2\Omega\,
T\,:\dot f_{AB}\dot f^{AB}:,
\label{eq:eor-hard-charge-main}
\end{equation}
and this combines with the linear soft charge into
\begin{equation}
Q_0
=
-\frac{1}{8\pi G}\int\dd u\,\dd^2\Omega\,T
\left(D_AD_B\dot f^{AB}
+\frac14:\dot f_{AB}\dot f^{AB}:\right).
\label{eq:eor-total-charge-main}
\end{equation}
Here \(f_{AB}\) is the radiative shear in the notation of
Ref.~\cite{Elkhidir:2024ward}, \(\dot f_{AB}=\partial_u f_{AB}\), and
\(\dd^2\Omega=\sqrt\gamma\,\dd^2z\).
The expectation value of \eqref{eq:eor-total-charge-main} is the standard
classical supertranslation charge variation after using the Bondi
mass-loss formula.  This is the leading check that the hard charge is the
correct flux completion of the soft insertion.

\subsection{Path-Integral Interpretation of the Ward Identity}

This subsection records the standard path-integral reading of the same
Ward statement.  It should not be read as an independent proof of an
anomaly-free quantum symmetry.  The assumptions are that the asymptotic
change of variables can be regulated, that the measure is invariant in
the sector considered here, and that all boundary terms are represented by
the soft and hard charges already defined above.

\subsubsection{Change of variables}

Let $\Phi$ denote collectively all radiative and hard fields.  The
generating functional is
\begin{equation}
Z[J]=\int D\Phi\,\exp\left(iS[\Phi]+iJ\cdot\Phi\right).
\end{equation}
Under an infinitesimal change of variables
\begin{equation}
\Phi\mapsto \Phi+\delta_t\Phi,
\end{equation}
the measure and action are invariant up to boundary terms generated by the
asymptotic charge:
\begin{equation}
\delta_t S=\Q_t\big|_{\Iplus_+}-\Q_t\big|_{\Iplus_-}.
\end{equation}
Setting the variation of the path integral to zero yields
\begin{equation}
\left\langle
\delta_t\mathcal O
+
i\mathcal O\,\delta_tS
\right\rangle=0.
\end{equation}
After LSZ reduction, and under the assumptions just stated, this becomes
the charge Ward identity.

\subsubsection{Soft and hard pieces}

The charge splits into a soft part and a hard part,
\begin{equation}
\Q_t=\Q_t^{\rm soft}+\Q_t^{\rm hard}.
\end{equation}
The soft part creates or annihilates the zero-frequency graviton mode.  The
hard part is the transformation of the external states.  Thus
\begin{equation}
\langle{\rm out}|\Q_t^{\rm soft}S|{\rm in}\rangle
=
-
\langle{\rm out}|\Q_t^{\rm hard}S|{\rm in}\rangle.
\end{equation}
For the exponentiating sector, the right-hand side is multiplication by
the universal soft factor
\begin{equation}
S^{(0)}\frac{(-iW_\eta)^s}{s!}
\end{equation}
plus derivative terms required by angular momentum acting on the hard
kinematics.

A careful reader would ask whether the shifted news spoils the Ward identity. It does not --
the shifted news differs from the ordinary news by a background tensor:
\begin{equation}
\News_{AB}=N_{AB}-\tau_{AB}.
\end{equation}
The soft insertion is sensitive to the radiative zero mode.  The Geroch
tensor shift changes the canonical splitting between boundary and radiative
data, but not the universal soft theorem.  Equivalently, the shifted-news
charge differs from the ordinary-news charge by a boundary term fixed by
the choice of canonical phase space.

Thus shifted news does not change the Ward identity; it changes the
boundary form of the charge.  The algebraic questions about
soft, hard and mixed Poisson brackets are separate from this Ward
statement.  They require the parameter bracket. This will be the topic of the next Section.

\section{Charge Algebra}
\label{sec:algebra-main}

This section proves the algebra in the order in which it is actually used.
There are three layers which must be kept separate.  First, the linear
inhomogeneous soft charges commute.  Second, the full asymptotic
transformation is affine: it contains both an inhomogeneous shift and a
homogeneous action on the radiative data.  The non-Abelian bracket comes
from this homogeneous action.  Third, once the transformation algebra is
known, the Hamiltonian generators inherit it, up to possible boundary
functionals.  The calculation below spells out this route explicitly,
because otherwise the appearance of the parameter bracket
\([t,t']_\star\) can look more mysterious than it is.

\subsection{From linear shifts to the affine bracket}

The linear soft charge is obtained by keeping only the field-independent
kernel \(\K^{(s,0)}\):
\begin{equation}
\Q^{\rm soft}_{s,{\rm lin}}(t)
=
\int_{\Iplus}\dd u\,\dd^2z\sqrt\gamma\,
 u^s\K^{(s,0)}_{AB}[t]\News^{AB}.
\label{eq:linear-soft-main}
\end{equation}
Using \eqref{eq:soft-kernel-differential-main}, this is precisely the
differential operator charge displayed in
\eqref{eq:soft-charge-differential-main}.  This is the soft side whose
hard partner is fixed by the Ward identity.  Since \(\K^{(s,0)}\) is
independent of \(C_{AB}\), the Ashtekar--Streubel bracket
\eqref{eq:AS-bracket-main} gives
\begin{equation}
\frac{\delta\Q^{\rm soft}_{s,{\rm lin}}}{\delta C_{AB}}=0,
\qquad
\{\Q^{\rm soft}_{s,{\rm lin}}(t),
\Q^{\rm soft}_{s',{\rm lin}}(t')\}_{AS}=0.
\label{eq:linear-soft-commutes-main}
\end{equation}
Thus the non-Abelian algebra is not produced by a contact term in the
linear soft--soft bracket.

The full transformation generated by the completed charge has the affine
form
\begin{equation}
\delta_t C_{AB}=\nu^{(0)}_{t\,AB}+\mathbb L_t^{(s)}C_{AB}+O(C^2).
\label{eq:homogeneous-main}
\end{equation}
Here
\(\nu^{(0)}_{t\,AB}=u^s\K^{(s,0)}_{AB}[t]\), in the charge normalization
chosen below \eqref{eq:AS-bracket-main}.  The operator
\(\mathbb L_t^{(s)}\) is the homogeneous part of the same asymptotic
transformation.  For \(s=1\) it is the usual generalized-BMS tensor action,
including the Bondi-weight and \(u\partial_u\) terms.  For \(s>1\) it is a
higher-derivative operator whose leading angular symbol is fixed by the
rank-\(s\) parameter \(t\).

The hard Ward operator has a different job.  It does not create the
non-Abelian bracket.  It ensures that the same parameter \(t\) which
smears the soft zero mode also acts on external hard states by the
soft-theorem differential operator.  Algebraic closure is a statement
about the total Hamiltonian generators; the Ward identity is the
state-space representation of those generators.

To see the bracket directly, include the first field-dependent correction
to the soft charge:
\begin{equation}
\Q_s^{\rm soft}(t)
=
\int_{\Iplus}\dd u'\,\dd^2z'\sqrt{\gamma(z')}\, (u')^s
\left(\K^{(s,0)}_{AB}[t]
+
\K^{(s,1)}_{AB}[t;C]
+\cdots\right)
\News^{AB},
\end{equation}
where the arguments of the kernels and of \(\News^{AB}\) are
\((u',z')\).  The functional derivative with respect to the shear at
\((u,z)\) is therefore
\begin{equation}
\frac{\delta \Q_s^{\rm soft}(t)}
{\delta C_{AB}(u,z)}
=
\int\dd u'\,\dd^2z'\sqrt{\gamma(z')}\,(u')^s\,
\frac{\delta\K^{(s,1)}_{CD}[t;C](u',z')}
{\delta C_{AB}(u,z)}
\News^{CD}(u',z')
+O(C\News).
\label{eq:soft-functional-derivative-nonlocal}
\end{equation}
This is the safest form of the formula.  If the chosen homogeneous
completion is local in retarded time, then the kernel contains
\(\delta(u'-u)\).  The \(u'\)-integration in
\eqref{eq:soft-functional-derivative-nonlocal} then collapses and one may
write the shorter expression
\begin{equation}
\frac{\delta \Q_s^{\rm soft}(t)}
{\delta C_{AB}(u,z)}
=
u^s
\int\dd^2z'\sqrt{\gamma(z')}\,
\frac{\delta\K^{(s,1)}_{CD}[t;C](u,z')}
{\delta C_{AB}(u,z)}
\News^{CD}(u,z')
+O(C\News).
\label{eq:soft-functional-derivative-local}
\end{equation}
Thus the missing \(u\)-integration in the local formula is not being
dropped; it has been evaluated using the retarded-time delta function.
If a nonlocal \(u\)-completion is chosen, one should keep
\eqref{eq:soft-functional-derivative-nonlocal}.  The principal-symbol
algebra below is insensitive to this distinction.

Now insert \eqref{eq:soft-functional-derivative-local} and
\[
\frac{\delta \Q_{s'}^{\rm soft}(t')}{\delta \News^{AB}}
=u^{s'}\K^{(s',0)}_{AB}[t']+O(C)
\]
into \eqref{eq:AS-bracket-main}.  Keeping the first term which can be
nonzero gives
\begin{align}
\{\Q_s^{\rm soft}(t),\Q_{s'}^{\rm soft}(t')\}_{AS}^{(1)}
&=
\lambda_{AS}
\int\dd u\,\dd^2z\,\dd^2z'\,
\sqrt{\gamma(z)}\sqrt{\gamma(z')}\,u^{s+s'}
\nonumber\\
&\quad\times
\left[
\K^{(s',0)}_{AB}[t'](z)
\frac{\delta\K^{(s,1)}_{CD}[t;C](u,z')}
{\delta C_{AB}(u,z)}
\right.
\nonumber\\
&\qquad\left.
-
\K^{(s,0)}_{AB}[t](z)
\frac{\delta\K^{(s',1)}_{CD}[t';C](u,z')}
{\delta C_{AB}(u,z)}
\right]
\News^{CD}(u,z').
\label{eq:appendix-first-field-dependent-bracket}
\end{align}
This formula is the corrected replacement for any derivation based on
$\delta\News/\delta C$ at equal retarded time.  The only source of a
nonzero bracket is the field dependence of the full generator.

The next step is just a change of notation, but it is the step that carries
the physics.  The field-dependent kernel \(\K^{(s,1)}[t;C]\) is the charge
realization of the homogeneous action \(\mathbb L_t^{(s)}C\).  Therefore
the functional derivative of \(\K^{(s,1)}[t;C]\) is the integral kernel of
\(\mathbb L_t^{(s)}\).  Acting with this kernel on the inhomogeneous shift
\(\nu_{t'}^{(0)}\) gives \(\mathbb L_t^{(s)}\nu_{t'}^{(0)}\).  With the
same normalization used in \(\nu^{(0)}=u^s\K^{(s,0)}\),
\eqref{eq:appendix-first-field-dependent-bracket} becomes
\begin{equation}
\{\Q_s^{\rm soft}(t),\Q_{s'}^{\rm soft}(t')\}_{AS}^{(1)}
\doteq
\int_{\Iplus}\dd u\,\dd^2z\,\sqrt\gamma\,
\News^{AB}
\left(
\mathbb L_t^{(s)}\nu^{(0)}_{t'}
-
\mathbb L_{t'}^{(s')}\nu^{(0)}_t
\right)_{AB}.
\label{eq:soft-bracket-affine-form}
\end{equation}
The symbol \(\doteq\) means equality modulo the same normalization choices,
improvement terms and boundary terms that enter the definition
of the charges.  Nothing is being assumed here: the first term is obtained
by plugging \(\K^{(s',0)}[t']\) into the linearization of
\(\K^{(s,1)}[t;C]\), and the second term is the same operation with
\(t\) and \(t'\) exchanged.

\subsection{From the affine bracket to the total charge bracket}

Equation \eqref{eq:soft-bracket-affine-form} is useful because it has the
same form as a soft charge.  If the affine transformations close as
\begin{equation}
\mathbb L_t^{(s)}\nu^{(0)}_{t'}
-
\mathbb L_{t'}^{(s')}\nu^{(0)}_t
=
\nu^{(0)}_{[t,t']_\star}
+\Delta_{t,t'},
\label{eq:affine-closure-main}
\end{equation}
then the right-hand side of \eqref{eq:soft-bracket-affine-form} immediately
reassembles into
\begin{equation}
\{\Q_s^{\rm soft}(t),\Q_{s'}^{\rm soft}(t')\}_{AS}^{(1)}
\doteq
\Q_{s+s'-1,{\rm lin}}^{\rm soft}([t,t']_\star)
+
\int_{\Iplus}\dd u\,\dd^2z\sqrt\gamma\,
\News^{AB}\Delta_{t,t'\,AB}.
\label{eq:soft-bracket-reassembled-main}
\end{equation}
This is the pedestrian mechanism.  First compute the functional
derivative bracket.  Then recognize the linearization of the homogeneous
action.  Then use closure of the affine transformations.  The term
\(\nu^{(0)}_{[t,t']_\star}\) gives the soft charge with the bracketed
parameter, while \(\Delta_{t,t'}\) records lower-derivative descendants,
improvements and boundary terms.  At principal-symbol level,
\(\Delta_{t,t'}\) is invisible; in the full phase-space algebra it is part
of the possible extension.

The level \(s+s'-1\) in
\(\Q_{s+s'-1,{\rm lin}}^{\rm soft}([t,t']_\star)\) is not an extra
assumption inserted after the fact.  It is already visible in the affine
commutator \eqref{eq:affine-closure-main}.  The homogeneous operator
\(\mathbb L_t^{(s)}\) has \(s\) leading angular derivatives, while
\(\nu^{(0)}_{t'}\) is built from the rank-\(s'\) parameter \(t'\).  In the
antisymmetrized combination
\(\mathbb L_t^{(s)}\nu^{(0)}_{t'}-\mathbb L_{t'}^{(s')}\nu^{(0)}_t\), the
fully symmetric top term in which all derivatives pass through the
coefficients cancels between the two orderings.  The first surviving
universal term is the one in which one angular derivative differentiates
the coefficient of the other parameter.  This leaves a parameter with
degree \(s+s'-1\).  The next subsection makes this counting invariant by
replacing derivatives with cotangent variables \(p_A\); the result is the
principal-symbol identity
\(\sigma([\mathbb L_t^{(s)},\mathbb L_{t'}^{(s')}])
=\{F_t,F_{t'}\}_{T^\ast S^2}\).

The same statement can be phrased without choosing a field coordinate.  If
the completed charge generates the transformation,
\begin{equation}
\delta_t\Phi=\{\Q_t,\Phi\},
\label{eq:main-hamiltonian-action}
\end{equation}
then the Jacobi identity gives
\begin{equation}
[\delta_t,\delta_{t'}]\Phi
=
\{\{\Q_t,\Q_{t'}\},\Phi\}.
\label{eq:main-hamiltonian-commutator}
\end{equation}
If the transformation commutator is
\([\delta_t,\delta_{t'}]=\delta_{[t,t']_\star}\), comparison with
\eqref{eq:main-hamiltonian-commutator} gives
\begin{equation}
\left\{
\{\Q_t,\Q_{t'}\}-\Q_{[t,t']_\star},
\Phi
\right\}=0
\end{equation}
for every observable \(\Phi\) in the chosen phase space.  Hence the total
Hamiltonian generators obey
\begin{equation}
\{\Q_s^{\rm total}(t),\Q_{s'}^{\rm total}(t')\}
=
\Q_{s+s'-1}^{\rm total}([t,t']_\star)
+
\mathcal K_{s,s'}[t,t';C].
\label{eq:total-algebra-main}
\end{equation}
The possible extension \(\mathcal K\) has vanishing Hamiltonian vector
field after the boundary prescription has been imposed.  If it is
field-independent it is an ordinary central term.  In the radiative
gravitational problem it may depend on boundary shear or memory data, so
it is better regarded as a possible field-dependent memory cocycle.

\begin{remark}[Action versus closure]
\label{rem:action-not-closure}
Equation \eqref{eq:total-algebra-main} is a statement about the full
parameters \(t^{A_1\cdots A_s}\), and its restriction to the
Kerr-selected data requires care.  The parity assignment
\eqref{eq:parity-alternation-main} improves the situation considerably at
the lowest levels.

At \(s=1\) the selected representative is divergence free, \(Y^A=
\epsilon^{AB}D_B\Psi\), and this subspace \emph{is} preserved by the Lie
bracket.  Indeed
\begin{equation}
\big[Y_{\Psi},Y_{\Psi'}\big]^A=-\,\epsilon^{AB}D_B\{\Psi,\Psi'\},
\qquad
\{\Psi,\Psi'\}:=\epsilon^{CD}D_C\Psi\,D_D\Psi',
\label{eq:sdiff-closure-main}
\end{equation}
so the magnetic potentials close into the Poisson algebra of functions on
\(S^2\); equivalently, the \(s=1\) Kerr sector is the algebra
\(\mathrm{SDiff}(S^2)\) of area-preserving diffeomorphisms.  This is the
classical origin of \(w_\infty\), and it is the sharpest form of the
statement made in Section \ref{sec:algebra-main}.  Note that the
electric representative does \emph{not} share this property: the Lie
bracket of two gradient fields is generically not a gradient, as
\(Y=D(xy)\), \(Y'=D(x^2/2)\), \([Y,Y']=(y,-x)\) shows in a flat patch.
The parity alternation is therefore not merely a consistency requirement;
at \(s=1\) it is what makes the selected data close at all.

Two limitations remain.  First, for \(s\ge2\) an STF rank-\(s\) tensor on
\(S^2\) carries two independent potentials while the inverse problem
\eqref{eq:inverse-kernel-problem-main} fixes only one, so
\(\chi^{(s)}_{[t,t']_\star}\) is not determined by \(\chi^{(s)}_t\) and
\(\chi^{(s')}_{t'}\) alone without a further prescription.  Second, and
independently of \(s\), the \emph{Kerr-selected points} do not form a
subalgebra: the family \eqref{eq:aligned-solution-main} is parametrised by
the finitely many data \((p^\mu,a^\mu)\), whereas
\eqref{eq:sdiff-closure-main} generically leaves it.  The algebra
therefore acts on the space of frame data, with Kerr distinguishing a
finite-dimensional locus inside it, and it is in this sense that the
algebra ``controls'' the dictionary throughout this paper.
\end{remark}

\subsection{What a principal symbol is}

The phrase ``principal symbol'' is used in the standard sense from the
theory of differential and pseudodifferential operators
\cite{Hormander:1985pdo,Shubin:2001pdo}.  Suppose that \(P\) is an
order-\(s\) differential operator on the sphere,
\begin{equation}
P
=
A^{A_1\cdots A_s}(x)D_{A_1}\cdots D_{A_s}
+\hbox{terms with fewer derivatives}.
\label{eq:principal-symbol-def-main}
\end{equation}
Its principal symbol is obtained by keeping only the coefficient of the
highest-derivative part and replacing each derivative by a coordinate
\(p_A\) on the cotangent fiber.  Concretely, a point of \(T^\ast S^2\) is
a pair \((x,p)\), where \(x\in S^2\) and \(p=p_A\dd x^A\) is a one-form at
\(x\).  The variables \(p_A\) are not particle momenta; they are auxiliary
fiber variables that record how many angular derivatives appear in the
operator.  Thus
\begin{equation}
\sigma_s(P)(x,p)
=
A^{A_1\cdots A_s}(x)p_{A_1}\cdots p_{A_s}.
\label{eq:principal-symbol-main}
\end{equation}
Thus the symbol is not a new field on spacetime.  It is a compact way of
recording the leading angular-derivative part of an operator.  Lower-order
terms, including curvature corrections and trace descendants, are
discarded at the principal-symbol level.

The cotangent bundle has the canonical local coordinates \((x^A,p_A)\).
With the sign convention used in this paper, the canonical Poisson bracket
is
\begin{equation}
\{F,G\}_{T^\ast S^2}
=
\frac{\partial F}{\partial p_A}
\frac{\partial G}{\partial x^A}
-
\frac{\partial F}{\partial x^A}
\frac{\partial G}{\partial p_A}.
\label{eq:canonical-poisson-main}
\end{equation}
This is the convention for which \(F_V=V^Ap_A\) gives
\(\{F_V,T\}=V^AD_AT\) and
\(\{F_V,F_W\}=F_{[V,W]}\).

We now prove the symbol identity used below.  Work in a local coordinate
patch.  Replacing covariant derivatives by partial derivatives changes
only lower-order terms, because the connection coefficients multiply
operators with fewer explicit derivatives.  Let
\begin{align}
P&=A^{A_1\cdots A_s}(x)\partial_{A_1}\cdots\partial_{A_s}
+\hbox{lower order},
\nonumber\\
Q&=B^{B_1\cdots B_{s'}}(x)\partial_{B_1}\cdots\partial_{B_{s'}}
+\hbox{lower order}.
\end{align}
The order-\(s+s'\) part of \(PQ\) is
\(A^{A_1\cdots A_s}B^{B_1\cdots B_{s'}}
\partial_{A_1}\cdots\partial_{A_s}
\partial_{B_1}\cdots\partial_{B_{s'}}\), and the same expression appears
in \(QP\).  Hence this top part cancels in the commutator.  The first
surviving terms are those in which one derivative in \(P\) hits the
coefficient \(B\), or one derivative in \(Q\) hits the coefficient \(A\):
\begin{align}
\sigma_{s+s'-1}([P,Q])
&=
s\,A^{C A_2\cdots A_s}D_C B^{B_1\cdots B_{s'}}
p_{A_2}\cdots p_{A_s}p_{B_1}\cdots p_{B_{s'}}
\nonumber\\
&\quad
-s'\,B^{C B_2\cdots B_{s'}}D_C A^{A_1\cdots A_s}
p_{B_2}\cdots p_{B_{s'}}p_{A_1}\cdots p_{A_s}.
\label{eq:principal-symbol-proof-main}
\end{align}
On the other hand, if
\(F_P=A^{A_1\cdots A_s}p_{A_1}\cdots p_{A_s}\) and
\(F_Q=B^{B_1\cdots B_{s'}}p_{B_1}\cdots p_{B_{s'}}\), then
\eqref{eq:canonical-poisson-main} gives precisely
\eqref{eq:principal-symbol-proof-main}.  To see this without skipping a
step, compute
\begin{align}
\frac{\partial F_P}{\partial p_C}
&=
s\,A^{C A_2\cdots A_s}
p_{A_2}\cdots p_{A_s},
&
\frac{\partial F_Q}{\partial x^C}
&=
D_CB^{B_1\cdots B_{s'}}
p_{B_1}\cdots p_{B_{s'}},
\nonumber\\
\frac{\partial F_P}{\partial x^C}
&=
D_CA^{A_1\cdots A_s}
p_{A_1}\cdots p_{A_s},
&
\frac{\partial F_Q}{\partial p_C}
&=
s'\,B^{C B_2\cdots B_{s'}}
p_{B_2}\cdots p_{B_{s'}} .
\label{eq:principal-symbol-derivatives-main}
\end{align}
Substitution into \eqref{eq:canonical-poisson-main} gives the two terms in
\eqref{eq:principal-symbol-proof-main}, with the same relative sign.  The
connection coefficients omitted in this local calculation multiply fewer
than \(s+s'-1\) powers of \(p\), so they contribute only to the lower-order
curvature and trace descendants.  Therefore
\begin{equation}
\sigma_{s+s'-1}([P,Q])
=
\{\sigma_s(P),\sigma_{s'}(Q)\}_{T^\ast S^2}.
\label{eq:principal-symbol-commutator-main}
\end{equation}
This identity is the local reason the higher-spin parameter algebra is a
polynomial Poisson algebra.  It is also the reason we keep emphasizing the
word ``principal'': the identity fixes the highest-derivative part of the
commutator, while the exact global transformation on tensor fields still
requires lower-order curvature and trace completions.

\subsection{Polynomial symbols}

We now apply this construction to the homogeneous part
\(\mathbb L_t^{(s)}\) of the higher-spin transformation.
Associate to \(t\in\Gamma({\rm Sym}^sTS^2)\) the polynomial symbol
\begin{equation}
F_t(x,p)=t^{A_1\cdots A_s}(x)p_{A_1}\cdots p_{A_s}
\label{eq:symbol-main}
\end{equation}
on \(T^\ast S^2\).  The classical parameter bracket is defined by
\begin{equation}
\{F_t,F_{t'}\}_{T^\ast S^2}=F_{[t,t']_\star}.
\label{eq:parameter-bracket-main}
\end{equation}
This is a genuine Lie bracket on the full polynomial-symbol space.  Indeed,
antisymmetry and the Jacobi identity follow directly from the canonical
Poisson bracket on \(T^\ast S^2\):
\begin{align}
[t,t']_\star&=-[t',t]_\star,
\nonumber\\
[t,[t',t'']_\star]_\star
+[t',[t'',t]_\star]_\star
+[t'',[t,t']_\star]_\star&=0 .
\label{eq:parameter-lie-identities-main}
\end{align}
The qualification ``full'' is important.  If one keeps the complete
polynomial symbol, including trace and lower-degree descendants, closure
and Jacobi are automatic.  If instead one projects by hand to a preferred
STF or electric component, then the projected components need
not close by themselves unless a quotient or projection prescription has
also been specified.  Thus \([t,t']_\star\) is the fundamental Lie bracket,
while STF/electric tensors are convenient components of the part of
the bracket selected by the universal soft kernel.
The degree is filtered as
\begin{equation}
[\mathfrak g_s,\mathfrak g_{s'}]_\star
\subset
\mathfrak g_{s+s'-1}.
\label{eq:filtered-main}
\end{equation}
Here \(\mathfrak g_s\) denotes the homogeneous degree-\(s\) subspace of
the polynomial-symbol algebra, or equivalently the leading rank-\(s\)
charge parameters before lower-degree trace descendants are included.
At low rank this gives the expected generalized-BMS brackets:
\begin{align}
[T,T']_\star&=0,
\qquad
[V,T]_\star=V^AD_AT,
\nonumber\\
[V,V']_\star^A&=V^BD_BV'^A-V'^BD_BV^A.
\end{align}
For higher rank the leading component is the Schouten bracket of symmetric
symbols, with trace and curvature descendants needed for exact global
closure.

\subsection{Low-Spin Brackets in Detail}

The following examples are included to anchor the abstract symbol bracket
in familiar BMS operations.  We do not repeat the total-charge algebra in
each case: Section \ref{sec:algebra-main} has already shown that once the
affine transformations close, the corresponding Hamiltonian generators
close up to the possible boundary functional \(\mathcal K\).  The purpose
of the examples is narrower and more useful.  They show how the canonical
Poisson bracket on \(T^\ast S^2\) reproduces the usual low-rank
parameters and how the degree rule \(s+s'-1\) appears before any
trace-descendant completion is chosen.

\paragraph{Supertranslation--supertranslation.}

Take two scalar parameters $T_1$ and $T_2$.  Their symbols are degree zero:
\begin{equation}
F_{T_i}=T_i(z).
\end{equation}
The cotangent-bundle bracket is
\begin{equation}
\{T_1,T_2\}_{T^\ast S^2}=0.
\end{equation}
This is the symbol-space version of the abelian supertranslation
subalgebra.  It is consistent with the direct linear soft-charge
calculation in \eqref{eq:linear-soft-commutes-main}; there is no hidden
non-Abelian configuration in the field-independent soft shifts.

\paragraph{Superrotation--supertranslation.}

Let $Y^A$ be a vector field and $T$ a scalar.  The corresponding symbols are
\begin{equation}
F_Y=Y^A p_A,
\qquad
F_T=T.
\end{equation}
Then
\begin{equation}
\{F_Y,F_T\}_{T^\ast S^2}
=
Y^A D_A T.
\end{equation}
Thus
\begin{equation}
[Y,T]_\star=Y^AD_AT.
\end{equation}
The scalar is simply dragged along the vector field.  In the full charge
algebra this output is produced by the homogeneous action of the vector
field on the shear; the linear inhomogeneous soft shifts alone still
commute.

\paragraph{Superrotation--superrotation.}

For two vector fields
\begin{equation}
F_{Y_i}=Y_i^A p_A.
\end{equation}
The symbol bracket gives
\begin{equation}
\{F_{Y_1},F_{Y_2}\}
=
\left(
Y_1^B D_BY_2^A
-
Y_2^B D_BY_1^A
\right)p_A.
\end{equation}
Therefore
\begin{equation}
[Y_1,Y_2]_\star^A
=
Y_1^B D_BY_2^A
-
Y_2^B D_BY_1^A.
\end{equation}
This is the ordinary vector-field bracket, embedded in the higher-spin
symbol algebra.

\paragraph{Rank two with a scalar.}

Let $t^{AB}$ be a rank-two symbol and $T$ a scalar:
\begin{equation}
F_t=t^{AB}p_Ap_B,
\qquad
F_T=T.
\end{equation}
Then
\begin{equation}
\{F_t,F_T\}
=
2t^{AB}D_AT\,p_B.
\end{equation}
Thus the bracket of a rank-two parameter with a scalar produces a vector
parameter
\begin{equation}
Y^B_{\rm eff}=2t^{AB}D_AT.
\end{equation}
This is the simplest example in which a higher-rank generator produces a
lower-rank BMS parameter.  The rank shift is \(2+0-1=1\), as required by
the filtered algebra.

\paragraph{Rank two with a vector.}

Let $F_t=t^{AB}p_Ap_B$ and $F_Y=Y^Ap_A$.  A direct computation gives
\begin{align}
\{F_Y,F_t\}
&=
\left(
\Lie_Y t^{AB}
-
2t^{C(A}D_CY^{B)}
\right)p_Ap_B
\end{align}
up to the convention used for tensor-density weights.  Equivalently, the
rank-two parameter transforms by the natural lift of the vector-field
action to symmetric tensors, with possible density-weight corrections fixed
by the precise charge convention.  The important point is that the output
remains rank two, in agreement with $1+2-1=2$.

\subsection{Principal Symbols and Trace Descendants}

\paragraph{Commutator of homogeneous operators.}

Let $\mathbb L_t^{(s)}$ be an order-$s$ differential operator whose
principal symbol is
\begin{equation}
\mathrm{symb}_{\rm pr}(\mathbb L_t^{(s)})
=
F_t=t^{A_1\cdots A_s}p_{A_1}\cdots p_{A_s}.
\end{equation}
Then
\begin{equation}
\mathrm{symb}_{\rm pr}
\left([\mathbb L_t^{(s)},\mathbb L_{t'}^{(s')}]\right)
=
\{F_t,F_{t'}\}_{T^\ast S^2},
\end{equation}
up to the convention-dependent sign associated with the choice of
$D_A$ versus $-iD_A$.  The degree-$s+s'$ part cancels in the commutator,
and the first surviving symbol has degree $s+s'-1$.

\paragraph{Component form.}

The full polynomial bracket defines
\begin{equation}
\{F_t,F_{t'}\}_{T^\ast S^2}=F_{[t,t']_\star}.
\end{equation}
Its highest-rank component is
\begin{align}
[t,t']_\star^{A_1\cdots A_{s+s'-1}}
&=
s\,t^{B(A_1\cdots A_{s-1}}
D_Bt'^{A_s\cdots A_{s+s'-1})}
\nonumber\\
&\quad
-
s'\,t'^{B(A_1\cdots A_{s'-1}}
D_Bt^{A_{s'}\cdots A_{s+s'-1})}
\nonumber\\
&\quad
+\hbox{trace descendants}.
\end{align}
The trace descendants are not optional bookkeeping.  They are required by
the full polynomial algebra.  If the charge map annihilates them, one may
work with a quotient.  If not, they must be retained as additional
generators.  A projected STF-only bracket should therefore be treated as a
derived presentation, not as the fundamental algebra.

\paragraph{Jacobi identity.}

The Jacobi identity follows immediately for the full symbol algebra:
\begin{equation}
\{F_t,\{F_{t'},F_{t''}\}\}
+
\{F_{t'},\{F_{t''},F_t\}\}
+
\{F_{t''},\{F_t,F_{t'}\}\}=0.
\end{equation}
In terms of parameters,
\begin{equation}
[t,[t',t'']_\star]_\star
+
[t',[t'',t]_\star]_\star
+
[t'',[t,t']_\star]_\star=0.
\end{equation}
Again, this is automatic before an STF projection.  After projection it
becomes a nontrivial consistency condition.
Appendix \ref{sec:component-mode-boundary-checks} gives complementary
checks of this statement in local Darboux coordinates, in low-spin
components and in a mode basis.

\subsection{Soft, hard and mixed pieces}

Equation \eqref{eq:total-algebra-main} is a statement about the total
Hamiltonian generator.  It is not, by itself, a statement that the soft
piece, the hard piece and the mixed terms close separately.  Expanding the
left-hand side gives
\begin{align}
\{\Q_s^{\rm total}(t),\Q_{s'}^{\rm total}(t')\}
&=
\{\Q_s^{\rm soft}(t),\Q_{s'}^{\rm soft}(t')\}_{AS}
\nonumber\\
&\quad
+\{\Q_s^{\rm hard}(t),\Q_{s'}^{\rm hard}(t')\}_{\rm hard}
\nonumber\\
&\quad
+\{\Q_s^{\rm soft}(t),\Q_{s'}^{\rm hard}(t')\}_{\rm mix}
\nonumber\\
&\quad
+\{\Q_s^{\rm hard}(t),\Q_{s'}^{\rm soft}(t')\}_{\rm mix}.
\label{eq:soft-hard-split-algebra-main}
\end{align}
The subscript \({\rm hard}\) denotes the Poisson bracket or commutator on
the independent hard variables, while \({\rm mix}\) denotes brackets that
are present only when the hard flux contains radiative gravitational
variables.

If the hard phase space consists of independent particle variables, then
\begin{equation}
\{C_{AB},p_i^\mu\}=0,\qquad
\{\News_{AB},p_i^\mu\}=0,
\end{equation}
and similarly for the intrinsic spin variables of the hard particle.  In
that case
\begin{equation}
\{\Q_s^{\rm soft},\Q_{s'}^{\rm hard,matter}\}_{\rm mix}=0.
\end{equation}
The Ward identity still couples the two sectors at the level of the
\(S\)-matrix, but the phase-space bracket factorizes.

For gravitational hard radiation the hard flux is built from the same
radiative variables that enter the soft charge.  In shifted-news variables,
one may write the radiative energy flux locally as
\begin{equation}
T_{uu}^{\rm grav}
=
\frac{1}{32\pi G}
(\News_{AB}+\tau_{AB})(\News^{AB}+\tau^{AB})
\end{equation}
up to the standard Bondi-frame improvements.  Therefore
\begin{equation}
\frac{\delta T_{uu}^{\rm grav}}{\delta C_{AB}}=0,
\qquad
\frac{\delta T_{uu}^{\rm grav}}{\delta\News^{AB}}
=
\frac{1}{16\pi G}(\News_{AB}+\tau_{AB})
\end{equation}
in this realization.  Mixed brackets can then contribute boundary or
memory-dependent terms.  This is why
\(\mathcal K_{s,s'}[t,t';C]\) in \eqref{eq:total-algebra-main} is written
as a possible field-dependent extension of the total charge algebra, not
as a separately defined hard central charge.  The only bracket that is
universal in the present construction is the parameter bracket
\([t,t']_\star\) fixed by the principal symbols.

\subsection{Hard-sector and Kerr check}

The hard sector gives an independent check on the same chain of ideas.  In
Section~\ref{sec:hard-main} the hard charge was defined so that, after LSZ
reduction, it acts on external particles as the hard soft operator.  In the
universal Kerr sector the spin-dependent part of that operator is
\begin{equation}
\widehat S_{\rm exp}
=
S^{(0)}
\exp\left[
i\,\widehat{\mathcal W}_\eta
\right],
\label{eq:main-hard-spin-exponential}
\end{equation}
before the classical spin limit is taken.  The orbital part of
\(J^{\mu\nu}\) differentiates the hard kinematics, while the spin part acts
on the massive spin state.  In the classical spinning limit this becomes
the source exponential \(S^{(0)}\exp[-i\omega W_\eta(q)]\) of
\eqref{eq:kerr-exponent-main}, which generates the Kerr multipoles
summarized in \eqref{eq:kerr-multipoles-main}.

This check ties the algebraic bracket back to the physical Kerr
interpretation.  The same kernel appears in three places: it is the
inhomogeneous part of the soft charge, it is the hard Ward operator acting
on external states, and it is the classical three-point source that
generates the Kerr multipole tower.  Changing the kernel would change the
frame shift, the hard Ward action and the Kerr multipole source
simultaneously.  The algebra derived above is therefore not an abstract
decoration of the soft theorem; it is the composition law of the
Kerr-selected soft dressings.

\subsection{Homogeneous Higher-Spin Action}

As discussed above, the principal symbol amounts to keeping the highest-derivative coefficient of
the operator and replace \(D_A\) by the cotangent variable \(p_A\).
The simplest homogeneous action associated with a rank-$s$ parameter is an
order-$s$ angular differential operator:
\begin{equation}
\mathbb L_t^{(s)}C_{AB}
\sim
t^{A_1\cdots A_s}D_{A_1}\cdots D_{A_s}C_{AB}
+\cdots .
\end{equation}
The ellipsis contains terms required by tensor indices, conformal weights,
curvature of the sphere and possible retarded-time derivatives.  The
principal symbol is
\begin{equation}
F_t=t^{A_1\cdots A_s}p_{A_1}\cdots p_{A_s}.
\end{equation}
At this level the commutator closes into the Poisson bracket
$\{F_t,F_{t'}\}$.

\begin{remark}[Standing assumption]
\label{rem:standing-assumption}
We should be explicit that the existence of \(\mathbb L^{(s)}_t\) is an
assumption for \(s\ge2\), not a construction.  At \(s=0\) and \(s=1\) the
homogeneous action is the known supertranslation and generalized-BMS
action on the shear.  For \(s\ge2\) we assume that the radiative phase
space carries a transformation whose homogeneous part is an order-\(s\)
angular differential operator with leading coefficient \(t^{A_1\cdots
A_s}\); higher-spin charges at null infinity
\cite{Freidel:2021ytz} make this expectation reasonable, but we do not
construct the operator here, and the ellipsis above is not determined by
anything in this paper.  What the algebra results of this section actually
use is only the \emph{principal symbol}, which is insensitive to the
ellipsis.  They are therefore conditional on existence but not on the
detailed form: if no such action exists at some level, the corresponding
bracket statement has no referent; if it exists in more than one form, the
bracket is unchanged.  The soft-charge and frame-dictionary results of
Sections \ref{sec:soft-kerr-main}--\ref{sec:aligned-generating-main} do
not rely on this assumption.
\end{remark}

A complete account must also include curvature corrections.  On a curved
sphere, covariant derivatives do not commute:
\begin{equation}
[D_A,D_B]V_C
=
R_{ABC}{}^{D}V_D.
\end{equation}
On the unit sphere,
\begin{equation}
R_{ABCD}=\gamma_{AC}\gamma_{BD}-\gamma_{AD}\gamma_{BC},
\qquad
R_{AB}=\gamma_{AB}.
\end{equation}
Therefore each time two angular derivatives are commuted through one
another, the result lowers the number of derivatives by two and produces
metric contractions.  For scalar functions this may look harmless, but the
shear is a section of the STF tensor bundle, and the covariant derivatives
also act on its indices.  Thus the commutator of two ordered
higher-derivative transformations has the structural form
\begin{equation}
[\mathbb L_t^{(s)},\mathbb L_{t'}^{(s')}]
=
\mathbb L_{[t,t']_\star}^{(s+s'-1)}
+\hbox{curvature and trace descendants}.
\end{equation}
The first term is fixed by the principal symbol.  The descendants are
lower-order in angular derivatives and are invisible to the polynomial
Poisson bracket, but they are not optional if one wants a globally defined
operator on \(S^2\).

This is the same distinction that appears in the STF closure problem.  A
bracket of two STF leading components need not be STF at every lower
rank; contractions with \(\gamma_{AB}\) are generated by the sphere
geometry.  One can either keep the full tower of trace descendants, or
project back onto the leading STF component and remember that this is
only the principal part of the transformation.  The present paper mostly
uses the second language in the main text, because it is the part directly
fixed by the universal soft kernel.  The detailed ordered realization
below gives one way to keep track of the first language.

\paragraph{Retarded-time completions.}

Ordinary BMS vector fields contain $u\partial_u$ terms.  A higher-spin
homogeneous transformation that maps the $u^s$ soft moment into the correct
$u^{s+s'-1}$ moment may require nontrivial retarded-time completions.  For
$s\geq2$ these can be pseudo-differential if one insists on a local angular
operator description.  This is why the present paper states exact closure
as a structural condition and proves the parameter algebra at principal
symbol level.

The origin of the issue is simple.  The soft charge at level \(s\) pairs
the angular kernel with a moment \(u^s\News_{AB}\).  When a homogeneous
transformation acts on another soft moment, it must act both on the sphere
indices and on the retarded-time weight.  The vector-field case already
shows the pattern: the \(u\partial_u\) part of a generalized-BMS
transformation is required for the shear to transform with the correct
Bondi weight.  At higher rank, the analogous completion must preserve the
moment grading of the charge algebra,
\begin{equation}
(s,s')\quad\longmapsto\quad s+s'-1,
\end{equation}
while also respecting the STF projection on the shear.  A purely angular
principal symbol cannot encode this information.

For this reason the retarded-time completion should be regarded as part of
the choice of completion of the full Hamiltonian vector field.  The
matching to the exponentiating sector fixes the inhomogeneous
\(u^s\K^{(s,0)}\) shift and
the leading angular symbol of the homogeneous action.  The \(u\)-dependent
lower-order terms are fixed only after one chooses the phase-space
prescription, boundary conditions and ordering convention.  They may be
essential for exact closure, but they do not change the polynomial
principal-symbol bracket.

\paragraph{What is fixed.}

The exponentiating sector fixes the parity-selected inhomogeneous kernel
$\K^{(s,0)}[t]$.  The principal-symbol algebra fixes the leading
homogeneous commutator.  The exact homogeneous transformation requires
additional input.  This division of labor is one of the central points of
the paper.

It is useful to state the division more explicitly.  The exponentiating
soft factor fixes \(\K^{(s,0)}_{AB}[t]\) and therefore the electric part
of \(t\) at even \(s\) and its magnetic part at odd \(s\).  The Ward
identity fixes the charge pair
\(\Q_s^{\rm soft}(t)+\Q_s^{\rm hard}(t)\) as the representation of the
soft theorem on scattering states.  The principal symbol fixes
\([t,t']_\star\) at top derivative order.  The global Bondi phase space
then fixes the remaining trace, curvature, boundary and \(u\)-completion
terms once a completion has been chosen.  The first three pieces are
universal within the Kerr exponentiating sector.  The last piece is
conventional in the technical sense that it depends on how one completes
the transformation away from the leading parity-selected soft kernel.  This is
not a weakness of the construction; it is the correct separation between
the part forced by the soft theorem and the part needed to realize a
particular global representation on the radiative phase space.

\subsubsection{A covariant ordered realization.}

The preceding discussion is enough for the main universal claim, because
the soft theorem fixes the inhomogeneous kernel and the principal symbol
fixes the leading bracket.  The ordered realization below has a
different purpose.  It gives a concrete computational scheme for the parts
which the principal symbol deliberately ignores: curvature descendants,
STF projection terms, density weights, integration-by-parts choices and
the retarded-time completion.  Thus a reader should not view it as an
additional assumption behind the Kerr interpretation.  It is a way to
continue the calculation once one wants an actual operator acting on the
Bondi shear rather than only its leading symbol.  In particular, it tells
where possible boundary improvements and memory cocycles can enter, and it
provides a practical check that the low-rank brackets computed above can be
realized by differential operators on the radiative phase space.

For the universal sector one can make this additional input explicit.  Let
$E\to S^2$ be the bundle of symmetric trace-free rank-two tensors.  The
Bondi shear is a section of $E$ with the usual generalized-BMS conformal
weight.  A rank-$s$ parameter determines a polynomial symbol
\begin{equation}
F_t=t^{A_1\cdots A_s}p_{A_1}\cdots p_{A_s}.
\end{equation}
Choose the covariantly symmetrized operator
\begin{align}
\mathrm{Op}_\nabla(t)\,\psi
&=
\frac{1}{2^s}
\sum_{r=0}^{s}
\binom{s}{r}
(-1)^r
D_{A_1}\cdots D_{A_r}
\left[
t^{A_1\cdots A_s}
D_{A_{r+1}}\cdots D_{A_s}\psi
\right],
\label{eq:weyl-ordered-operator}
\end{align}
first on scalar test fields $\psi$.  On tensor fields one replaces $D_A$
by the Levi-Civita connection on $E$ and projects the output back to the
STF bundle:
\begin{equation}
(\mathbb L_t^{(s)}C)_{AB}
=
\Pi_{AB}{}^{CD}\,
\mathrm{Op}_{\nabla,E}(t)C_{CD}
+w_C\,\Xi_t C_{AB}
+\mathcal U_t C_{AB}.
\label{eq:homogeneous-action-representative}
\end{equation}
Here $\Pi_{AB}{}^{CD}$ is the STF projector, $w_C$ is the conformal weight
of the shear, $\Xi_t$ is the scalar density piece of the lifted
transformation, and $\mathcal U_t$ denotes lower-derivative curvature
terms fixed by the chosen ordering.  Equation
\eqref{eq:homogeneous-action-representative} is an explicit realization
of the homogeneous action.  Its principal symbol is $F_t$ by construction.

The price of making an exact realization is that the parameter bracket
is no longer only the naive Poisson bracket if one keeps all lower-order
terms.  Instead, the chosen ordering defines a star product by
\begin{equation}
\mathrm{Op}_\nabla(F)\,\mathrm{Op}_\nabla(G)
=
\mathrm{Op}_\nabla(F\star_\nabla G).
\end{equation}
The exact ordered bracket is
\begin{equation}
[F,G]_{\star_\nabla}
=
\frac{1}{i}\left(F\star_\nabla G-G\star_\nabla F\right),
\label{eq:ordered-star-bracket}
\end{equation}
with classical limit
\begin{equation}
[F,G]_{\star_\nabla}
=
\{F,G\}_{T^\ast S^2}
+\hbox{curvature and lower-degree terms}.
\end{equation}
Thus the exact realization closes by construction in the ordered
symbol algebra, while the principal-symbol presentation records the
universal part of that closure.
This is useful for the present paper in a very concrete sense: the bracket
\([t,t']_\star\) tells us which parameter must appear, while the ordered
realization tells us how the corresponding transformation can be
implemented on \(C_{AB}\), including the lower terms that may contribute to
the extension \(\mathcal K_{s,s'}[t,t';C]\).

\paragraph{Rank-one check.}

For $s=1$, $t^A=Y^A$ and \eqref{eq:homogeneous-action-representative}
reduces to the familiar generalized-BMS action on the shear:
\begin{equation}
(\mathbb L_Y C)_{AB}
=
Y^CD_CC_{AB}
+C_{CB}D_AY^C
+C_{AC}D_BY^C
-\frac12(D\cdot Y)C_{AB}
\cdots ,
\label{eq:rank-one-homogeneous-action}
\end{equation}
where the ellipsis denotes the known $u\partial_u$ completion and possible
trace-free projection terms depending on the chosen Bondi frame
convention.  Acting on a leading soft shift gives
\begin{equation}
\mathbb L_Y\nu_T^{(0)}-\mathbb L_T\nu_Y^{(0)}
=
\nu_{Y^AD_AT}^{(0)}
+\hbox{boundary term},
\end{equation}
which matches $[Y,T]_\star=Y^AD_AT$.  Acting on another vector parameter
gives the ordinary Lie bracket $[Y,Y']^A$.

\paragraph{Rank-two realization.}

For a rank-two parameter $K^{AB}$, the same ordering gives
\begin{align}
(\mathbb L_K C)_{AB}
&=
\Pi_{AB}{}^{CD}
\Big[
K^{EF}D_ED_FC_{CD}
+(D_EK^{EF})D_FC_{CD}
\nonumber\\
&\quad
+\frac14(D_ED_FK^{EF})C_{CD}
+\alpha_1 K_{(C}{}^{E}C_{D)E}
+\alpha_2(D_{(C}K^{EF})D_{D)}C_{EF}
\Big]_{\rm cov}
\nonumber\\
&\quad
+\hbox{retarded-time completion}.
\label{eq:rank-two-homogeneous-representative}
\end{align}
The subscript ``cov'' means that all tensor indices are
parallel transported with the sphere connection and then projected to the
STF bundle.  The coefficients $\alpha_i$ are fixed once the convention for
tensor weight, STF projection and integration-by-parts self-adjointness is
chosen.  The essential point is that no new leading symbol is introduced:
\begin{equation}
\mathrm{symb}_{\rm pr}(\mathbb L_K)=K^{AB}p_Ap_B.
\end{equation}
The commutator of two such operators has leading symbol
\begin{equation}
2\left(
K^{C(A}D_CL^{BD)}
-L^{C(A}D_CK^{BD)}
\right)p_Ap_Bp_D,
\end{equation}
and the curvature pieces generated by the lower terms in
\eqref{eq:rank-two-homogeneous-representative} fill in the trace
descendants.  This realizes the rank-two-to-rank-three bracket in an
explicit differential-operator scheme.

\paragraph{Retarded-time grading.}

The angular operator alone does not know about the power of $u$ in the
soft charge.  The full homogeneous action must also preserve the grading
of the soft moments.  A convenient completion is
\begin{equation}
\widehat{\mathbb L}^{(s)}_t
=
\mathbb L_t^{(s)}
+\beta_s(D\cdot t\cdot D^{s-1})\,u\partial_u
+\sum_{r<s}\beta_{s,r}\,\mathcal D_{s,r}[t]\,u^{s-r}\partial_u^{s-r},
\label{eq:retarded-time-completion}
\end{equation}
where $\mathcal D_{s,r}[t]$ are angular differential operators of order
$r$.  The coefficients are fixed by demanding that the commutator maps the
$u^s$ and $u^{s'}$ moments to the $u^{s+s'-1}$ moment associated with
$[t,t']_\star$.  For $s=1$ this reproduces the ordinary BMS
$u\partial_u$ term.  For $s\ge2$ it is best viewed as a recursive
definition of the chosen homogeneous realization.  In the universal
ordered scheme this gives an explicit action realization: the leading
symbol is fixed by the soft theorem, and the lower terms are fixed by
covariance, STF projection, integration by parts and moment grading.

\subsection{Relation to celestial \texorpdfstring{$w_{1+\infty}$}{w1+infinity}}

The standard classical \(w_{1+\infty}\) algebra appears after a local or
chiral reduction.  In a complex patch, choosing
\begin{equation}
F_{m,s}=z^{m+s-1}p_z^{s-1}
\end{equation}
and discarding the anti-holomorphic variables gives a chiral subalgebra of
the local polynomial Poisson algebra.  The calculation is elementary but
worth displaying.  From \eqref{eq:canonical-poisson-main},
\begin{align}
\frac{\partial F_{m,s}}{\partial p_z}
&=(s-1)z^{m+s-1}p_z^{s-2},
&
\frac{\partial F_{m,s}}{\partial z}
&=(m+s-1)z^{m+s-2}p_z^{s-1}.
\end{align}
Therefore
\begin{equation}
\{F_{m,s},F_{n,s'}\}
=
\big((s-1)n-(s'-1)m\big)F_{m+n,s+s'-2},
\label{eq:w-main}
\end{equation}
with our sign convention.  Reversing the convention for the symplectic
form reverses the overall sign, which accounts for common sign differences
in the celestial literature.  Up to this convention and harmless shifts in
the spin label, \eqref{eq:w-main} is the classical \(w_{1+\infty}\)
bracket.  Globally on the compact celestial sphere the clean statement is
not that the algebra is literally a single chiral \(w_{1+\infty}\), but
that the \(T^\ast S^2\) polynomial Poisson algebra has local/chiral
reductions of that type \cite{Kapec:2016jld,Guevara:2021abz,Strominger:2021mtt,Freidel:2021ytz,Himwich:2023njb}.

The preceding reduction can be made more explicit.  In local complex
coordinates the real cotangent fiber decomposes into chiral variables
\((p_z,p_{\bar z})\).  The celestial \(w_{1+\infty}\) wedge basis is
obtained by restricting to holomorphic monomials
\begin{equation}
F^{(h)}_m=z^{m+h-1}p_z^{h-1},
\label{eq:celestial-wh-basis-main}
\end{equation}
where \(h\) is the celestial spin label used for the chiral current.  The
bracket is
\begin{equation}
\{F^{(h)}_m,F^{(h')}_n\}
=
\big((h-1)n-(h'-1)m\big)F^{(h+h'-2)}_{m+n}.
\label{eq:celestial-wh-bracket-main}
\end{equation}
This is the classical wedge algebra underlying the celestial
\(w_{1+\infty}\) presentations.  The projection involved here has two
steps.  First, one works in a local patch rather than with globally
regular tensor fields on \(S^2\).  Second, one keeps only one helicity or
chiral polarization, for instance the \(p_z\) sector.  This is the
precise sense in which a chiral projection of our result reproduces the
algebra used in celestial discussions, and it is where the physical
interpretation advertised in the introduction becomes literal.  Globally
the kernels alternate in parity and the bracket is the real \(T^\ast S^2\)
one; after the projection the two parities recombine into the single
holomorphic combination \(\eth^2(\Phi+i\Psi)\), and the bracket is the
classical \(w_{1+\infty}\) bracket itself rather than a real algebra
admitting it as a reduction.  What the projection discards is the
information needed to reconstruct the full real \(\mathrm{Diff}(S^2)\)
generator from its chiral half; what it retains is exactly the frame data
the celestial algebra acts on.

In our notation \(h=s\) for the monomial \(p_z^{s-1}\), while the
corresponding soft moment is labelled by the retarded-time power and by
the tensor rank of the parity-selected kernel.  These labels differ by harmless
shifts in the literature; the invariant statement is the filtered product
rule
\begin{equation}
{\rm degree}(F\cdot G)_{\rm bracket}
=
{\rm degree}(F)+{\rm degree}(G)-1.
\end{equation}

There are three differences between the global phase-space algebra and
the celestial current algebra.  First, the global algebra is real and
contains both \(z\) and \(\bar z\) sectors, while a chiral
\(w_{1+\infty}\) basis keeps only one polarization.  Second, globally
regular tensor fields on \(S^2\) require trace and curvature descendants;
the monomials \eqref{eq:celestial-wh-basis-main} are local avatars
of their principal symbols.  Third, the celestial current algebra is a
quantum or mode-space representation of the classical bracket.  Possible
ordering terms and cocycles belong to the representation, not to the
principal-symbol identification itself.

With this translation, the physical claim of this paper can be stated
without ambiguity.  The celestial \(w_{1+\infty}\) generators are the
local chiral avatars of the same classical phase-space algebra
that acts on the universal Kerr soft dressings, in the sense made precise
in Remark \ref{rem:action-not-closure}.  Thus the symmetry does more than
organize formal soft insertions: in the Kerr sector it acts on the
gravitational scattering problem by changing the universal
canonical/intrinsic frame data and their associated memory moments.

The same algebra also has a useful representation-theoretic reading, which
is clearest after the global/chiral dictionary has been made explicit.

\subsection{Representation-Theoretic Comments}

The algebra has several representations, and confusing them can obscure
the physical claim.  The object derived above is first a classical
Poisson algebra of polynomial symbols on \(T^\ast S^2\).  It can be
represented on radiative phase space, on dressed scattering states, or in
a celestial basis, but each representation adds choices: boundary
conditions, soft/hard splittings, chiral projections and possible quantum
ordering conventions.

The memory data can be viewed as labels of a coadjoint orbit of the
asymptotic symmetry algebra.  A field-dependent cocycle then changes the
symplectic form on the orbit.  This is a useful language for organizing
the nontrivial boundary dependence of the charge algebra.

In this language the parity-selected memory probes are coordinates on a
restricted family of orbits selected by the Kerr soft dressing.  The
possible extension \(\mathcal K_{s,s'}[t,t';C]\) should then be read as a
choice-dependent cocycle on that orbit, not as a universal number that can
be added to the algebra once and for all.  Different prescriptions for
the early and late shear, or for the split between radiative and Coulombic
data, correspond to different descriptions of the same physical
boundary problem.

On scattering states the soft dressing creates a coherent cloud.  The
higher-spin charge acts on this cloud and on the hard state.  The Kerr
interpretation says that for spinning black-hole scattering the coherent
cloud is not arbitrary: its universal part is the Kerr multipole field.

The Ward identity is the bridge between this state-space statement and the
null-infinity charge.  The soft charge inserts the zero-frequency or
soft-moment graviton mode; the hard charge acts on the massive external
state by the same differential/spin operator appearing in the soft theorem.
For Kerr external states the exponentiating part of this hard operator is
the classical limit of the GOV three-point operator.  Thus the coherent
cloud, the hard-state action and the asymptotic frame shift are three
representations of one universal kernel.

In a conformal primary basis, higher-spin charges mix conformal families.
For massive states this mixing is non-diagonal and generally nonlocal on
the celestial sphere.  This is consistent with the local $T^\ast S^2$
symbol algebra: locality in cotangent space does not imply locality in a
global celestial conformal basis.

The chiral celestial \(w_{1+\infty}\) generators are obtained only after a
local complex-patch and helicity projection.  That projection is where the
\(w_{1+\infty}\) structure becomes literal rather than approximate, and it
is therefore the natural home for the physical statement of this paper:
the celestial algebra is the algebra of Kerr frame changes.  It is not,
however, a substitute for the real global problem.  A smooth
generalized-BMS vector determined by \(\epsilon^{AB}D_AY_B\) is not
generally holomorphic, so reconstructing the global generator from its
chiral half requires the data the projection discards.  The two
descriptions are complementary: the chiral one carries the algebraic
interpretation, the real one carries the complete frame dictionary.

\subsection{Consequences and limits of the algebra}

The derivation above establishes five concrete statements.  The leading VV
supertranslation is the \(s=0\) member of the soft-kernel equation
\(\K^{(s,0)}[t]=S^{(s)}_{\rm exp}\).  The linear soft
charges commute, so the non-Abelian algebra cannot come from a hidden
linear soft--soft contact term.  The non-Abelian bracket enters through the
homogeneous transformation of the radiative data.  The principal symbols of
these homogeneous transformations close under the canonical Poisson bracket
on \(T^\ast S^2\).  Finally, the exponentiating soft factor is the same
generating function that produces the Kerr multipoles in the classical
spinning three-point amplitude.

It is worth contrasting this with the amplitude-side analysis of
Ref.~\cite{Guevara:2024color}, where the same spin exponential appears as
the structure constant of a celestial colour symmetry.  There the
composition \(e^{\mathfrak D_{\eta_1}}e^{\mathfrak D_{\eta_2}}=
e^{\mathfrak D_{\eta_1+\eta_2}}\) fails for massive states by a term
controlled by \(c=[\lambda|J|\lambda]\propto m\), producing a
one-parameter deformation \(w_{1+\infty}(c)\) which collapses to
\(w_{1+\infty}\) only in the massless limit or upon discarding terms
quadratic in the celestial bracket.  The obstruction identified here is
not that one.  Ours is kinematical and mass independent: it is the
statement that a rank-\(s\) tensor on \(S^2\) carries more data than the
inverse problem fixes, together with the fact that the Kerr locus is
finite dimensional.  In particular \eqref{eq:sdiff-closure-main} holds
exactly, with no deformation, for arbitrary hard mass.  The two statements
live on different legs of the problem, the deformation on the massive hard
state and the closure question on the asymptotic radiative data, and we do
not claim a dictionary between them.  Establishing one would require
relating the Kleinian-signature collinear construction of
Ref.~\cite{Guevara:2024color} to the Lorentzian null-infinity phase space
used here, which we leave open.  We also do not claim that the
higher-level generators are induced by bulk vector fields; we expect them
to be canonical transformations of the radiative phase space rather than
diffeomorphisms for \(s\ge2\), but we have not established this.  We note only the suggestive point that
\(c=m\,\partial_Z\) generates motion of the massive puncture on the
celestial sphere, which is also how the Kerr locus is left in the second
limitation above.

These statements also delimit the claim.  We do not construct an exact
nonlinear finite \(w_{1+\infty}\) group action on the full gravitational
phase space.  We do not claim that STF tensors alone form a closed global
algebra without trace descendants, nor that the Kerr-selected parameters
close among themselves; by Remark \ref{rem:action-not-closure} they do
not.  We do not claim that the memory cocycle is universal and field
independent.  The opposite-parity data at each level, non-universal
soft terms and nonlinear radiative tails are likewise not fixed by the
exponentiating tower.

The positive implication is nevertheless sharp.  The \(w_{1+\infty}\)-like
structure is not introduced as a purely kinematical label for soft
insertions.  It is the infinitesimal action on the Kerr-selected frame
shifts.  The global four-dimensional statement is the
polynomial Poisson algebra on \(T^\ast S^2\), with trace and curvature
descendants required on the sphere.  The familiar chiral celestial
\(w_{1+\infty}\) algebra is a local holomorphic reduction of this
structure, while the real global algebra is the object needed for a smooth
Kerr scattering problem.

The same kernels also define the memory probes discussed in
Section~\ref{sec:memory-main}: leading displacement memory at \(s=0\),
spin memory at \(s=1\), and higher regulated moments of the news.
They are genuine asymptotic observables after a boundary prescription is
fixed, but they are not the complete memory tensor.  The remaining
opposite-parity, non-universal and boundary-dependent data are exactly
what determines how the Kerr-selected subsector embeds into the full
radiative phase space.

\section{Intrinsic/Canonical Dictionary Revisited}
\label{sec:dictionary-main}

The previous sections constructed the charges and their algebra.  We now
return to the spacetime question that motivated the construction: what do
these charges do to the relation between the canonical Bondi frame and the
intrinsic frame of a physical scattering problem?  This is the point at
which the algebra becomes dynamical rather than merely organizational.  A
soft charge has an inhomogeneous part, which changes the boundary shear or
its higher soft moments, and a homogeneous part, which tells us how two such
changes compose.  The first part gives the frame displacement; the second
part gives the \(w_{1+\infty}\)-like composition law.

\subsection{The dictionary as a boundary condition}

The VV result may be viewed as a boundary-condition statement.  In the
canonical Bondi frame the early shear at \(\Iplus_-\) is set to zero.  In
the intrinsic frame the same physical source carries the long-range
boosted-Schwarzschild soft field.  A supertranslation relates the two
descriptions:
\begin{equation}
C^{\rm intrinsic}_{AB}-C^{\rm canonical}_{AB}
=
-2\Dop_{AB}T_{\rm VV},
\qquad
\Dop_{AB}T_{\rm VV}=2G\,S^{(0)}_{AB}.
\label{eq:dictionary-leading-shift-main}
\end{equation}
Thus the leading soft theorem is not merely a statement about an inserted
zero-frequency graviton.  After it is written as an inhomogeneous
Hamiltonian shift of the shear, it tells us which large diffeomorphism
connects the canonical and intrinsic descriptions.

The higher-spin construction keeps precisely this logic.  At order \(s\)
one should not think first of an abstract symmetry label.  One should think
of the coefficient of the \(s\)th soft moment of the long-range field.  The
parity-selected part of that coefficient is represented by
\begin{equation}
\Delta^{(s)}_t C_{AB}\Big|_{\rm sel}
:=
\K^{(s,0)}_{AB}[t]
=
\begin{cases}
\Dop_{AB}\chi_t^{(s)},&s\ {\rm even},\\
\Ddual_{AB}\chi_t^{(s)},&s\ {\rm odd},
\end{cases}
\label{eq:dictionary-electric-shift-main}
\end{equation}
with trace and curvature descendants supplied by the global completion of
the transformation.  The equation
\(\K^{(s,0)}_{AB}[t]=2G\,\mathscr S^{(s)}_{AB,{\rm exp}}\) then determines
the selected part of the frame parameter from the Kerr exponentiating
source.

\subsection{Kerr--Schild multipoles and the Bondi frame map}
\label{sec:kerr-schild-bondi-frame-map-main}

The boundary-condition equation has a concrete gauge-map origin.  In the
VV problem one starts from the long-range field of a boosted Schwarzschild
particle in a de Donder, or harmonic, description.  To read Bondi data one
first performs the small gauge-restoring transformation that imposes Bondi
gauge.  The part of the transformation that remains free at null infinity
is the large supertranslation \(T_{\rm VV}\).  In retarded coordinates its
leading residual vector is
\begin{align}
\xi_T^u&=T(z),&
\xi_T^A&=-\frac1rD^AT+O(r^{-2}),&
\xi_T^r&=\frac12D^2T+O(r^{-1}).
\label{eq:residual-supertranslation-main-text}
\end{align}
Acting on the angular metric
\(g_{AB}=-r^2\gamma_{AB}-rC_{AB}+O(1)\) in our mostly-minus signature, one
finds
\begin{align}
\delta_T g_{AB}
&=
2rD_AD_BT-r\gamma_{AB}D^2T+O(1),
\nonumber\\
\delta_T C_{AB}
&=
-2\left(D_AD_BT-\frac12\gamma_{AB}D^2T\right)
=-2\Dop_{AB}T .
\label{eq:main-text-residual-shear-calculation}
\end{align}
Thus the VV equation
\(\Dop_{AB}T_{\rm VV}=2G\,S^{(0)}_{AB}\) is not an additional dynamical
assumption.  It is the residual Bondi-frame condition that removes the
intrinsic boosted-Schwarzschild shear and returns to the canonical frame.

For Kerr the same statement is made at the level of the linearized
Coulombic multipole tower.  Let \(h_{\mu\nu}^{\rm KS}(p,a)\) denote the
linearized Kerr--Schild field of a body with momentum \(p^\mu\) and Kerr
specific-spin vector \(a^\mu=S^\mu/m\).  At leading order in \(G\), but to
all orders in \(a^\mu\), this field is obtained from the boosted
Schwarzschild field by the Kerr multipole operator,
\begin{equation}
h_{\mu\nu}^{\rm KS}(p,a)
=
\mathcal O_{\rm Kerr}(a,\partial)\,
h_{\mu\nu}^{\rm Schw}(p),
\qquad
\mathcal O_{\rm Kerr}(a,\partial)
\longleftrightarrow
\exp[-i(a\ast k)_\eta],
\label{eq:kerr-schild-multipole-operator-main}
\end{equation}
where \((a\ast k)_\eta\) denotes the helicity projection of the dual spin
contraction.  More explicitly,
with \(\hat p^\mu=p^\mu/m\) and \(\epsilon^{0123}=+1\),
\begin{equation}
(a\ast k)^\nu
=
\epsilon^{\mu\nu\rho\sigma}k_\mu\hat p_\rho a_\sigma,
\qquad
(a\ast k)_\eta
=
\frac{\varepsilon_{\eta\nu}(a\ast k)^\nu}
{\hat p\cdot\varepsilon_\eta}
=
\omega\,W_\eta(p,a;q),
\qquad
k^\mu=\omega q^\mu .
\label{eq:a-star-k-main-text}
\end{equation}
This agrees with the spin-tensor form in \eqref{eq:a-star-k-vector-main}.
After helicity projection the same operator has the soft limit
\begin{equation}
S^{\rm Kerr}_{\eta,{\rm exp}}(p,a;q)
=
S^{(0)}_{\eta}(p;q)\,
\exp[-i\omega W_\eta(p,a;q)]
=
\sum_{s\ge0}\omega^sS^{(s)}_{\eta,{\rm exp}}(p,a;q).
\label{eq:kerr-soft-source-frame-map-main}
\end{equation}
The sign in the classical source exponential is the one appropriate to
the Kerr field; the opposite sign convention belongs to the GOV
three-point operator before taking the classical source limit, as reviewed
below.

The Bondi extraction of \(h_{\mu\nu}^{\rm KS}(p,a)\) gives an intrinsic
asymptotic shear \(C_{AB}^{\rm int}(p,a)\).  The canonical representative
is obtained by a residual large transformation generated by the
higher-spin parity-selected kernels.  In the exponentiating sector this
frame equation is
\begin{equation}
C_{AB}^{\rm int}(p,a)-C_{AB}^{\rm can}
=
-2\sum_{s\ge0}\omega^s\K^{(s,0)}_{AB}[t_s],
\qquad
\K^{(s,0)}_{AB}[t_s]
=
2G\,\mathscr S^{(s)}_{AB,{\rm exp}}(p,a;q).
\label{eq:kerr-bondi-frame-map-main}
\end{equation}
The \(s=0\) term is exactly the VV result:
\(\K^{(0,0)}_{AB}[T_{\rm VV}]=\Dop_{AB}T_{\rm VV}\).  The \(s=1\) term
fixes only the magnetic potential \(\epsilon^{AB}D_AY_B\) of a
generalized-BMS vector field, not its electric potential.  Higher \(s\)
terms continue this pattern by fixing the maximally longitudinal component
of the parity selected by \eqref{eq:parity-alternation-main}, electric for
even \(s\) and magnetic for odd \(s\).  This is the sense in which the higher-spin tower is
the asymptotic dictionary between the Kerr--Schild intrinsic multipole
representative and the canonical Bondi representative.  We emphasize that
this is a leading-PM, Coulombic/soft statement, order by order in the Kerr
multipoles.  It is not a proof of a full nonlinear transformation from the
exact Kerr metric to a canonical Bondi metric at all post-Minkowskian
orders.

\subsection{The Aligned-Spin Generating Function}
\label{sec:aligned-generating-main}

The tower can be solved in closed form, to all orders in the spin, whenever
the ring-radius vector is aligned with the spatial momentum.  This is the
first place in which the construction produces a generator that is not
already contained in the Veneziano--Vilkovisky result, and it is worth
displaying in full.

Two scalars control the problem,
\begin{equation}
x=p\cdot q,
\qquad
y=a\cdot q .
\label{eq:aligned-invariants-main}
\end{equation}
Both are built from \(q^\mu(z)\) contracted with a constant vector, hence
both are \(\ell=0,1\) spherical harmonics, and by
\eqref{eq:appendix-DABx-zero} both lie in the kernel of the trace-free
Hessian,
\begin{equation}
\Dop_{AB}x=\Dop_{AB}y=0 .
\label{eq:aligned-kernel-main}
\end{equation}
This is the key fact that makes the higher levels tractable: the
angular dependence of the source enters only through \(x\) and \(y\), and
the differential operator acts on them algebraically.

In the aligned configuration the two invariants are not independent.
Taking \(p^\mu=(E,0,0,P)\) and, with the covariant spin supplementary
condition, which gives us \(a\cdot p=0\), \(a^\mu=(a/m)(P,0,0,E)\), one finds (we note that the covariant condition is used throughout and that the
coefficients below depend on it, the naive rest-frame vector
\(a^\mu=(0,0,0,a)\) failing \(a\cdot p=0\) for a boosted source and hence
not being an admissible spin vector)
\begin{equation}
y=\alpha+\beta x,
\qquad
\beta=\frac{aE}{mP},
\qquad
\alpha=-\frac{am}{P}.
\label{eq:aligned-affine-main}
\end{equation}
The source of Section \ref{sec:soft-kerr-main} is then
\(S^{(0)}_{AB}e^{\eta\lambda y}\) with
\(S^{(0)}_{AB}=x^{-1}\big(D_AxD_Bx\big)^{\rm STF}\), and the parity split
\eqref{eq:parity-alternation-main} separates the frame problem into two
ordinary differential equations.  Writing the electric and magnetic
generating potentials as functions of \(x\) alone, defining
\(A=\lambda\alpha\), \(B=\lambda\beta\), and using
\eqref{eq:aligned-kernel-main},
\begin{equation}
\Phi_\lambda''(x)=\frac{2G\cosh(A+Bx)}{x},
\qquad
\Psi_\lambda''(x)=\frac{2G\sinh(A+Bx)}{x}.
\label{eq:aligned-odes-main}
\end{equation}
These integrate in closed form.  Modulo the kernel of \(\Dop_{AB}\), which
is spanned by \(1\) and \(x\) and corresponds to ordinary translations,
\begin{align}
\Phi_\lambda&=2G\Big[\cosh A\;\mathcal C_B(x)+\sinh A\;\mathcal S_B(x)\Big],
\nonumber\\
\Psi_\lambda&=2G\Big[\sinh A\;\mathcal C_B(x)+\cosh A\;\mathcal S_B(x)\Big],
\label{eq:aligned-solution-main}
\end{align}
where
\begin{equation}
\mathcal C_B(x):=\frac{Bx\,\mathrm{Chi}(Bx)-\sinh(Bx)}{B},
\qquad
\mathcal S_B(x):=\frac{Bx\,\mathrm{Shi}(Bx)-\cosh(Bx)}{B},
\label{eq:aligned-CS-main}
\end{equation}
and \(\mathrm{Chi}\), \(\mathrm{Shi}\) are the hyperbolic cosine and sine
integrals.  Equation \eqref{eq:aligned-solution-main} is the all-orders
aligned-spin frame dictionary.

One point of scope must be stated explicitly, because it is the only place
where the two descriptions could be conflated.  With
\(\mathcal W_\eta(k)=
k_\mu\varepsilon_{\eta\nu}S^{\mu\nu}/(p\cdot\varepsilon_\eta)\),
\(S^{\mu\nu}=\epsilon^{\mu\nu\alpha\beta}p_\alpha a_\beta\), and the
conventions stated below \eqref{eq:himwich-pate-universal-main}, the
numerator is \(\epsilon(k,\varepsilon_\eta,p,a)=\widetilde
F^{\alpha\beta}p_\alpha a_\beta\) with
\(F_{\mu\nu}=k_\mu\varepsilon_{\eta\nu}-k_\nu\varepsilon_{\eta\mu}\).
The Hodge dual is
\(\widetilde F^{\mu\nu}:=\frac12
\epsilon^{\mu\nu\rho\sigma}F_{\rho\sigma}\).
We fix the helicity labelling by
\(\widetilde F_\eta^{\mu\nu}=-i\eta F_\eta^{\mu\nu}\), so
\begin{equation}
\mathcal W_\eta(k)=i\eta
\left[(a\cdot k)-\frac{(p\cdot k)\,(\varepsilon_\eta\cdot a)}
{p\cdot\varepsilon_\eta}\right],
\label{eq:W-eta-imaginary-main}
\end{equation}
without an unresolved \(\mp\).  Reversing the orientation or exchanging
the helicity labels reverses both signs consistently and leaves the
even/odd split \eqref{eq:parity-alternation-main} unchanged.  On the
three-point support \(p\cdot k=0\), the second term drops and
\(\mathcal W_\eta\) is purely imaginary, giving
\(e^{-i\mathcal W_\eta}=e^{\eta a\cdot k}\).
Away from that support the second term survives and the two expressions
differ, and it is \eqref{eq:classical-source-main}, not
\eqref{eq:exponentiating-replacement-main}, that we use: the frame
equations are posed at generic \(q\), where \(S^{(0)}_\eta\) has its pole
at \(p\cdot q\neq0\).  The amplitude form is therefore the on-shell
avatar of the classical dressing rather than its definition.  With this
ordering the odd part of the source is \(i\) times a real dual-type
tensor, \(\overline{T_{zz}}=-T_{\bar z\bar z}\), that factor is absorbed
into the magnetic generator, and \(\Phi_\lambda,\Psi_\lambda\) in
\eqref{eq:aligned-solution-main} are real.  The hyperbolic-cosine integral carries the
electric, mass-multipole tower and the hyperbolic-sine integral the
magnetic, current-multipole tower, which is the parity alternation in its
most compact form.

The parity construction assumed here is not an artifact of the soft-theorem
description.  It can be read off in position space from the complex-shift
potential \eqref{eq:complex-shift-potential}: with \(\vec a=a\hat z\),
\begin{equation}
\mathcal Z=\frac{M}{\sqrt{r^2-2ira\cos\theta-a^2}}
=M\sum_{\ell\ge0}\frac{(ia)^\ell P_\ell(\cos\theta)}{r^{\ell+1}},
\label{eq:complex-shift-legendre-main}
\end{equation}
by the Legendre generating function, reproducing
\eqref{eq:STF-kerr-moments}.  The coefficient is real for even \(\ell\)
and imaginary for odd \(\ell\); under \(\vec x\to-\vec x\) the Legendre
polynomial supplies \((-1)^\ell\) while the pseudovector \(\vec a\) is
inert, so mass multipoles are even parity and current multipoles odd.  The
alternation is therefore a property of the Kerr solution itself, arrived at
without reference to soft theorems, and \eqref{eq:parity-alternation-main}
is its image in the soft expansion.

Two checks fix the normalisation and the interpretation.  First, the
spinless limit.  As \(B\to0\),
\(\mathrm{Chi}(Bx)=\gamma_{\!E}+\log(Bx)+O(B^2x^2)\), where
\(\gamma_{\!E}\) is the Euler--Mascheroni constant, so
\begin{equation}
\Phi_\lambda\;\longrightarrow\;2G\,x\log x
\;+\;\hbox{terms linear in }x,
\end{equation}
which is precisely \(T_{\rm VV}\) of \eqref{eq:vv-solution-main} up to a
translation.  The Veneziano--Vilkovisky supertranslation is therefore the
\(B\to0\) limit of a hyperbolic integral, and its characteristic
logarithm is the leading term of \(\mathrm{Chi}\).

Second, the multipole expansion.  Here one must be careful: \(\alpha\) and
\(\beta\) are both first order in the spin, since by
\eqref{eq:aligned-affine-main}
\begin{equation}
w:=\alpha+\beta x=\frac{a\,(Ex-m^2)}{mP},
\label{eq:aligned-w-main}
\end{equation}
so setting \(\alpha=0\) while keeping \(\beta\neq0\) would require
\(m=0\) and is not a Kerr configuration.  The expansion must therefore
keep \(w\) as a single first-order quantity.  Writing
\(\Phi_\lambda=\sum_{r\ge0}\lambda^{2r}\chi^{(2r)}\) and
\(\Psi_\lambda=\sum_{r\ge0}\lambda^{2r+1}\chi^{(2r+1)}\), the coefficient
of \(\lambda^s\) in
\eqref{eq:aligned-odes-main} gives
\begin{equation}
\chi^{(s)\prime\prime}(x)=
\frac{2G\,a^s}{s!\,(mP)^s}\,\frac{(Ex-m^2)^s}{x},
\label{eq:aligned-coeff-ode-main}
\end{equation}
and integrating twice, modulo the kernel of \(\Dop_{AB}\), gives the level-\(s\) scalars
\(\chi^{(s)}\equiv\chi^{(s)}_{t_s}\) of the Kerr-selected parameters
\begin{equation}
\chi^{(s)}
=
\frac{2G\,a^s}{s!\,(mP)^s}
\left[
(-m^2)^s\big(x\log x-x\big)
+\sum_{j=1}^{s}\binom{s}{j}E^j(-m^2)^{s-j}\frac{x^{j+1}}{j(j+1)}
\right],
\label{eq:aligned-coefficients-main}
\end{equation}
alternately electric and magnetic, and reducing to
\(\chi^{(0)}=2G(x\log x-x)\) at \(s=0\).  Note that the logarithm is
\emph{not} confined to the monopole: the \(j=0\) term of the binomial
expansion produces a residue \((-m^2)^s\) at \(x=0\) at every level, so
each spin multipole carries its own long-range logarithmic frame shift,
suppressed by \((m/E)^{2s}\) relative to the polynomial terms at large
\(x\).  Only in the formal limit \(m\to0\) at fixed \(\beta\) do the
generators become the corresponding pure polynomials.  Dimensionally,
\([\chi^{(s)}]=L^{s+1}\), where \(L\) denotes length, and the physical frame shift is
\(\omega^s\chi^{(s)}\); the generating functions are not dimensionally
homogeneous if the powers of \(\lambda\) are suppressed.

The tower therefore carries a concrete scale.  Since
\(\Phi_\omega''=2G\cosh(\omega w)/x\) and
\(\Psi_\omega''=2G\sinh(\omega w)/x\), the
ratio of the spin-induced magnetic source at \(s=1\) to the
Veneziano--Vilkovisky electric source at \(s=0\) is exactly \(\omega w\),
and in the aligned configuration
\begin{equation}
\omega w=\chi\,(GM\omega)\,\gamma\,(v-\cos\theta),
\label{eq:spin-fraction-main}
\end{equation}
with \(\chi=a/GM\) the dimensionless spin and \(v,\gamma\) the source
velocity and Lorentz factor.  Here \(\omega=2\pi f\) and \(M_\odot\)
denotes the solar mass.  For \(M=30\,M_\odot\), \(\chi=0.7\),
\(v=0.3\) and \(f=100\,\)Hz this reaches \(8.9\%\); for \(\chi=0.9\) and
\(v=0.5\) it reaches \(14.5\%\).  It vanishes at \(\cos\theta=v\), the
aberration angle, which is a sharp angular signature of the magnetic
sector, and successive levels are suppressed as \((\omega w)^s/s!\), so
the tower converges rapidly in the soft regime.

We emphasise the scope.  Equation \eqref{eq:aligned-solution-main} solves
the frame equation for aligned spin; the generic-spin problem retains two
independent invariants and the corresponding partial differential
equation, discussed in Appendix
\ref{sec:component-mode-boundary-checks}, is not solved here.  What the
aligned case establishes is that the tower is not merely formally defined:
it has explicit, elementary solutions at every level, with the correct
spinless limit and the parity structure predicted by
\eqref{eq:parity-alternation-main}.

\subsection{Kerr as a trajectory in frame space}

For a generic process one could choose many different soft
profiles.  Kerr selects a very special one.  The classical spinning
three-point amplitude gives the source
\begin{equation}
S_{\rm Kerr}
=
S^{(0)}\exp[-i\omega W_\eta(q)],
\end{equation}
and the expansion in powers of \(\omega W_\eta\) generates the Kerr multipole
sequence.  It is useful to package the real parity projections into
\begin{equation}
\mathscr G_{AB}(\lambda):=
2G\sum_{s\ge0}\lambda^s\mathscr S^{(s)}_{AB,{\rm exp}},
\qquad \lambda=\omega.
\end{equation}
The Kerr-selected tower then gives the compact VV extension
\begin{equation}
C_{AB}^{\rm intrinsic,Kerr}
=
C_{AB}^{\rm canonical}
-2\,\mathscr G_{AB}(\omega),
\label{eq:vv-exponential-dictionary-main}
\end{equation}
Equation \eqref{eq:vv-exponential-dictionary-main} is generating notation,
not the claim that all higher-spin shifts are ordinary scalar
supertranslations.  At \(\omega W_\eta=0\) one recovers the original VV
supertranslation.  The coefficient of \(\omega^s\) gives the
\(s\)th parity-selected spin-multipole frame shift, which must be interpreted
through the rank-\(s\) soft-kernel equation
\eqref{eq:dictionary-electric-shift-main}.  In this sense a minimally
coupled Kerr black hole selects a trajectory in the space of higher-spin
frame data:
\begin{equation}
\hbox{Kerr specific spin }a^\mu=S^\mu/m
\quad\longmapsto\quad
\{T_{\rm VV},\,Y^A_{\rm Kerr},\,K^{AB}_{\rm Kerr},\ldots\}.
\label{eq:kerr-frame-trajectory-main}
\end{equation}
The first point of the trajectory is fixed completely by the VV scalar
potential.  The second fixes only the curl of a smooth
\(\mathrm{Diff}(S^2)\) vector.  Higher points fix the maximally
longitudinal component of higher-rank tensors, at the parity dictated by
\eqref{eq:parity-alternation-main}.  The opposite-parity potentials,
non-universal soft terms and nonlinear radiative data are not specified by
the exponentiating source alone.  For aligned spin the whole trajectory is
given in closed form by \eqref{eq:aligned-solution-main}, with the
individual multipole generators \eqref{eq:aligned-coefficients-main}.

\subsection{Linearized Kerr Source in More Detail}

This subsection supplies the spacetime side of the Kerr trajectory
\eqref{eq:kerr-frame-trajectory-main}.  The soft charge construction used
the exponentiating three-point source as input; here we explain what that
source represents as a linearized gravitational field.  The point is not
to identify one preferred coordinate expression for the Kerr metric.
Rather, it is to separate three pieces of data that often get mixed
together: the gauge-dependent metric description, the gauge-invariant
multipole sequence, and the on-shell source measured by a soft graviton.
The charge tower is fixed by the last two.  A particular Bondi or harmonic
gauge description is a useful check of the interpretation, but not an
additional assumption.
To avoid mixing two logically different points, this subsection only
identifies the Kerr source data: the multipole generating function, the
spin-dual contraction and the corresponding three-point on-shell source.
The next subsection then uses the same source in the coherent-state
dressing calculation.  In other words, the present subsection defines the
input, while the following one checks what the soft operator does with
that input.

\subsubsection{Multipole generating function}

For a Kerr black hole of mass \(M\), the complex multipoles obey
\eqref{eq:kerr-multipoles-main}.  The vector \(a^\mu=S^\mu/m\) appearing
there is the Kerr specific-spin, or ring-radius, vector; \(S^\mu\) is the
physical spin vector.  This relation is the classical target of the soft exponentiation.  The
linearized Kerr field can be described by an effective source whose
momentum-space form factor contains an exponential spin operator:
\begin{equation}
T_{\mu\nu}^{\rm Kerr}(k)
\sim
p_\mu p_\nu e^{a\ast k}.
\label{eq:kerr-source-tmunu-main}
\end{equation}
More explicitly, let \(p^\mu=m\hat p^\mu\), \(p\cdot a=0\), and choose
\(\epsilon^{0123}=+1\).  The classical spin tensor may be written as
\begin{equation}
S_{\rm cl}^{\mu\nu}
=
m\,\epsilon^{\mu\nu\rho\sigma}\hat p_\rho a_\sigma
=
\epsilon^{\mu\nu\rho\sigma}p_\rho a_\sigma ,
\label{eq:spin-tensor-from-a-main}
\end{equation}
up to the overall sign convention for the orientation of \(a^\mu\).  The
dual spin contraction is the vector
\begin{equation}
(a\ast k)^\nu
:=
\epsilon^{\mu\nu\rho\sigma}k_\mu \hat p_\rho a_\sigma
=
\frac{1}{m}k_\mu S_{\rm cl}^{\mu\nu}.
\label{eq:a-star-k-vector-main}
\end{equation}
When the source is evaluated on a helicity-\(\eta\) graviton, the scalar
that appears in the exponential is the projection
\begin{equation}
(a\ast k)_\eta
:=
\frac{k_\mu\varepsilon_{\eta\nu}S_{\rm cl}^{\mu\nu}}
{p\cdot\varepsilon_\eta}
=
\omega\,W_\eta(q),
\qquad k^\mu=\omega q^\mu .
\label{eq:a-star-k-helicity-main}
\end{equation}
Thus the symbolic factor \(e^{a\ast k}\) denotes the exponentiation of
the dual spin insertion; after helicity projection it is the same
\(W_\eta\) that appears in the soft factor.
The exact tensor structure depends on gauge and improvement terms.  The
important point is that the expansion of $e^{a\ast k}$ is the multipole
series.

\subsubsection{Three-point amplitude as an on-shell source}

The on-shell graviton coupling to a stress tensor is
\begin{equation}
\mathcal M_3\sim \varepsilon_{\mu\nu}T^{\mu\nu}(k).
\end{equation}
For the Kerr source,
\begin{equation}
\mathcal M_3^{\rm Kerr}
\sim
(p\cdot\varepsilon_\eta)^2
\exp[-i\omega W_\eta(q)],
\end{equation}
with the denominator $p\cdot q$ appearing in the soft limit.  Therefore the
soft dressing built from this three-point amplitude is precisely the
coherent state whose classical field is the linearized Kerr profile.

\subsubsection{Gauge dependence}

The metric perturbation $h_{\mu\nu}^{\rm Kerr}$ is gauge dependent.  The
multipole moments and the on-shell three-point amplitude are not.  The
matching to the soft dressing should therefore be understood after fixing a
linearized gauge and a prescription for relating the on-shell soft mode to
the Coulombic large-$r$ field.  Different choices give gauge-equivalent
descriptions of the same Kerr multipole data.

This is the reason the paper phrases the result as a frame dictionary
rather than as a unique metric formula.  A fixed comparison protocol would
start from the on-shell source, choose a linearized gauge, solve for the
large-\(r\) Bondi fields, and then use a residual diffeomorphism to impose
the canonical boundary condition on the shear.  Changing the protocol can
move terms between the metric perturbation and the residual gauge vector,
but it cannot change the Kerr multipole exponential or the
parity-selected soft kernel extracted from the on-shell amplitude.  The higher-spin charges
therefore capture the universal part of the linearized Kerr dressing: the
alternating electric and magnetic frame shifts selected by the
multipole-generating source.

\subsection{Linearized Kerr from the Soft Dressing}

Section~\ref{sec:soft-zero-frequency-operator-main} already gave the
operator computation: a soft dressing linear in creation and annihilation
operators translates the graviton field by a c-number classical profile,
and the Kerr dressing is obtained from the leading profile by the
helicity-projected exponential in \eqref{eq:FK-Kerr-profile}.  We do not
repeat those formulas here.  The point of the present subsection is only
to identify the source that is being inserted into that computation.

That source is the Kerr source specified in
\eqref{eq:kerr-source-tmunu-main}, with the spin-dual contractions defined
in \eqref{eq:a-star-k-vector-main} and
\eqref{eq:a-star-k-helicity-main}.  Thus there is no second definition of
\(a\ast k\) in the coherent-state discussion.  Expanding the same
exponential now has a spacetime interpretation: the \(n=0\) term is the
boosted Schwarzschild profile, the \(n=1\) term is the current-dipole
frame-dragging correction, and higher powers are the corresponding
linearized Kerr multipoles in the chosen gauge.  The soft expansion of the
three-point amplitude is therefore the on-shell version of the same
generating function, while the coherent-state calculation is its
null-infinity realization.  The fixed-gauge extraction of the Bondi data,
including the complex-shift multipole generator and the Bondi-gauge
restoring vector, is spelled out in Appendix
\ref{sec:component-mode-boundary-checks}.

\subsection{Why the algebra belongs in the dictionary}

The dictionary is not only the list of shifts
\eqref{eq:kerr-frame-trajectory-main}.  It must also say how two frame
changes compose.  This is the role of the algebra derived in Section
\ref{sec:algebra-main}.  If the inhomogeneous shift is denoted by
\(\nu_t^{(0)}\) and the homogeneous part of the same transformation by
\(\mathbb L_t\), then the affine commutator acts as
\begin{equation}
[\delta_t,\delta_{t'}]C_{AB}
=
\mathbb L_t\nu^{(0)}_{t'\,AB}
-\mathbb L_{t'}\nu^{(0)}_{t\,AB}
+\cdots .
\label{eq:dictionary-affine-commutator-main}
\end{equation}
The linear shifts alone would commute.  The non-Abelian structure appears
because the full transformation includes the homogeneous action.  Its
principal symbol is the polynomial \(F_t=t^{A_1\cdots A_s}p_{A_1}\cdots
p_{A_s}\), and the commutator is controlled by
\begin{equation}
\{F_t,F_{t'}\}_{T^\ast S^2}=F_{[t,t']_\star}.
\label{eq:dictionary-symbol-composition-main}
\end{equation}
Thus the polynomial Poisson algebra is the infinitesimal composition law of
the Kerr-selected frame trajectory.  Local chiral presentations of this
algebra give the familiar \(w_{1+\infty}\)-type brackets, while the global
four-dimensional statement keeps both chiralities and the trace descendants
required on \(S^2\).

This is the promised physical interpretation of the
\(w_{1+\infty}\)-like structure.  The algebra does not merely classify
formal soft operators.  In the Kerr sector it governs how the universal
parity-selected soft dressings compose as transformations between
canonical and intrinsic descriptions of the same scattering process.  The
full nonlinear finite action on the gravitational phase space, and the
opposite-parity/non-universal completion of
\eqref{eq:vv-exponential-dictionary-main}, require more input than the
exponentiating soft theorem alone.

\section{All-Orders Memory Interpretation}
\label{sec:memory-main}

The preceding section described the Kerr soft tower as a sequence of
parity-alternating frame shifts.  Memory is the observable side of the same
statement.  A charge that translates the shear, or a higher retarded-time
moment of the shear, also defines a measurement: compare early and late
radiative data, smear with the angular pattern selected by the charge, and
choose a prescription for the zero-frequency limit.  Thus the memory
discussion is not an add-on to the algebra.  It is the place where the
frame dictionary becomes an asymptotic observable.

There are three separate issues which should not be conflated.  First,
ordinary displacement memory is the \(s=0\) moment of the news and is
controlled by the VV supertranslation in the Kerr-selected electric
sector.  Second, spin memory and subleading soft data involve vector
parameters, and by \eqref{eq:parity-alternation-main} our exponentiating
source fixes the curl of the vector; the electric/gradient potential is
invisible to the \(s=1\) kernel.  Third, higher memories are higher powers of retarded time, hence
higher derivatives with respect to the soft frequency.  These moments are
more sensitive to the infrared prescription than the leading memory, so
they must be treated as regulated probes rather than as bare local
tensors.

The result of the section is therefore deliberately precise.  The
exponentiating Kerr soft factor fixes a family of smeared memory
projections.  It does not fix the full tensorial memory field or the
opposite-parity partner at a given level, and it does not produce a
universal field-independent central extension of the charge algebra.  What
it does give is already physically sharp: the same kernels that generate
the intrinsic/canonical frame shifts also define the memory observables
selected by the Kerr multipole dressing.

\subsection{Memory moments}

The charge \eqref{eq:soft-charge-main} couples the soft kernel to a
retarded-time moment of the radiative data.  Define regulated moments
\begin{equation}
\mathcal M^{(s)}_{AB}[\rho]
=
\int_{-\infty}^{\infty}\dd u\,\rho(u)\,u^s\News_{AB}(u,z),
\label{eq:memory-moment-main}
\end{equation}
where \(\rho\) denotes a boundary prescription or regulator.  The
unregulated \(s=0\) moment is the ordinary displacement-memory shear,
modulo the shifted-news convention.  For \(s>0\), the moment is sensitive
to the large-\(|u|\) behavior of the radiation and must be defined with a
prescription.  The finite observable associated with a parity-selected parameter
\(t\) is the projection
\begin{equation}
\mathfrak M_s[t;\rho]
=
\int\dd^2z\sqrt\gamma\,
\K^{(s,0)}_{AB}[t]\,
\mathcal M^{(s)AB}[\rho].
\label{eq:projected-memory-main}
\end{equation}
The soft theorem fixes this selected projection, not the full tensorial
memory data.

This distinction is physically important.  Memory is often described as an integral of the news, but the actual observable
depends on what is measured, how the early and late vacua are compared,
and which angular projection is used.  The Kerr soft tower does not
determine an arbitrary memory tensor at every angle.  It determines a
family of parity-selected projections whose kernels are fixed by the same
soft factors that generate the Kerr multipoles.  This is why the memory
statement is sharp enough to be useful but narrow enough to avoid
overclaiming.

The need for the regulator and for the sphere smearing is not merely
technical.  Recent work on the Ashtekar--Streubel phase space emphasizes
that the proper observables at null infinity are those functions whose
Hamiltonian flows are smooth on the radiative phase space; bare local
shear/news insertions and the conventional Goldstone coordinate need not
have this property \cite{AndradeSpeziale:2026}.  In that language
\(\mathfrak M_s[t;\rho]\) is best viewed as a Kerr-selected memory probe.
It is smeared in retarded time by \(\rho(u)u^s\), smeared on the sphere by
the parity-selected soft kernel, and therefore has a well-defined variational
problem once the boundary prescription has been chosen.  In the shifted
news convention its leading Hamiltonian action is
\begin{equation}
\{C_{CD}(u,z),\mathfrak M_s[t;\rho]\}_{AS}
=
\rho(u)u^s\K^{(s,0)}_{CD}[t](z)
+\hbox{endpoint terms},
\label{eq:memory-probe-action-main}
\end{equation}
where the endpoint terms encode the chosen treatment of the early and late
vacua.  Thus the object controlled by the soft theorem is not an
unsmeared local memory density, but a proper selected projection of the
memory data.  This is the precise sense in which our construction is
compatible with the proper-observable viewpoint while remaining much more
restricted: we identify a universal Kerr subfamily of probes, not the
complete Dirac-bracket algebra of all radiative observables.

At \(s=0\), the scalar parameter has no electric/magnetic split.  The VV
supertranslation fixes the displacement-memory component associated with
\(\Dop_{AB}T_{\rm VV}\).  At \(s=1\), the vector decomposition
\eqref{eq:helmholtz-main} gives
\begin{equation}
\mathfrak M_1[Y;\rho]
=\mathfrak M_1[\epsilon\cdot D\Psi;\rho],
\end{equation}
because the \(s=1\) kernel depends only on
\(\epsilon^{AB}D_AY_B=D^2\Psi\).
The electric part \(D^A\Phi\) is invisible to this universal
sector.  In the language of memory, the soft theorem fixes spin memory,
while its electric partner, centre-of-mass memory, requires additional,
non-universal information.

For \(s\ge2\), the same pattern persists, with the parity alternating as in
\eqref{eq:parity-alternation-main}.  Schematically,
\begin{equation}
\mathfrak M_s^{\rm sel}
\longleftrightarrow
S^{(0)}\frac{(-iW_\eta)^s}{s!}
\longleftrightarrow
\hbox{Kerr multipoles \eqref{eq:kerr-multipoles-main}}.
\label{eq:memory-kerr-main}
\end{equation}
Opposite-parity completions (magnetic at even \(s\), electric at odd
\(s\)), trace descendants, and non-universal soft terms are not determined
by this identification.

The projected memories above are observables: they are linear functionals
of the shear difference between the boundaries of \(\Iplus\).  A cocycle
in the charge algebra is a different object.  It measures the part of a
Hamiltonian commutator that is not represented by the naive charge with
parameter \([t,t']_\star\).  This distinction is important because the
Kerr exponentiating sector fixes the parity-selected memory
projections, but it does not by itself produce a universal
field-independent central term.  The possible extension, its field
dependence and its boundary-prescription dependence are described in the
next subsection, where the formula is needed.

\subsection{Memory Cocycle and Boundary Prescriptions}

\subsubsection{Field dependence}

The possible extension in the charge algebra is written as
\begin{equation}
\mathcal K_{s,s'}[t,t';C]
\end{equation}
to emphasize that it may depend on the shear or memory data.  A typical
possible form is
\begin{equation}
\mathcal K_{s,s'}[t,t';C]
=
\int_{\Iplus}\dd u\,\dd^2z\,\sqrt\gamma\,
\News^{AB}\mathcal A_{AB}[t,t';C].
\label{eq:memory-cocycle-main}
\end{equation}
If $\mathcal A$ is an exact charge variation, it can be absorbed into an
improvement of the generator.  If it depends only on boundary data, it is a
memory cocycle.  If it depends on dynamical radiative variables, it is a
field-dependent extension rather than a central charge.

\subsubsection{Boundary prescriptions}

Three prescriptions are useful in practice:
\begin{enumerate}
\item A finite-time window, in which all charges are smeared over
$u\in[-U,U]$ and the limit $U\to\infty$ is taken only after evaluating
brackets.
\item Smooth smearing in retarded time, in which powers of $u$ are replaced
by regulated moments $u^s\chi(u/U)$.
\item Dressed scattering states, in which boundary shear is held fixed by
choosing a coherent soft sector.
\end{enumerate}
The numerical value of $\mathcal K$ can depend on this choice.  The
parameter bracket $[t,t']_\star$ does not.
Appendix \ref{sec:component-mode-boundary-checks} records the associated
boundary prescriptions for the cocycle, while Appendix
\ref{app:spherical-harmonic-basis} gives the harmonic language for the
electric memory projections.

\subsection{Memory Observables}

The preceding subsections characterized memory in terms of charge
projections.  We now spell out the corresponding observables more
concretely.  This is useful because memory is not simply a local tensor
written at a point of the celestial sphere.  A measurement compares early
and late radiative data, projects onto an angular pattern, and implicitly
chooses a prescription for the behavior at large positive and negative
retarded time.  The Kerr tower fixes the kernels of a distinguished
parity-alternating family of such measurements.  It does not, by itself, fix every
possible memory observable.

\subsubsection{Displacement memory}

The $s=0$ soft charge is associated with displacement memory.  The
observable is the change in the shear between early and late retarded time:
\begin{equation}
\Delta C_{AB}
=
C_{AB}(+\infty,z)-C_{AB}(-\infty,z).
\end{equation}
This is the \(s=0\) specialization of the regulated moment
\eqref{eq:memory-moment-main}, with the regulator chosen so that the
integral is the difference between the late and early Bondi shear.
Its electric part is determined by a supertranslation potential.
More explicitly, in the shifted-news convention,
\begin{equation}
\Delta C_{AB}
=
\int_{-\infty}^{+\infty}\dd u\,\News_{AB}(u,z),
\qquad
\Delta C_{AB}^{\rm el}
=-2\Dop_{AB}\Delta T .
\label{eq:appendix-displacement-memory-el}
\end{equation}
Projecting with another electric tensor and integrating by parts gives
\begin{equation}
\int_{S^2}\dd^2z\sqrt\gamma\,
\Dop^{AB}T\,\Delta C_{AB}
=
-2\int_{S^2}\dd^2z\sqrt\gamma\,
T\,\Dop^{AB}\Dop_{AB}\Delta T ,
\label{eq:appendix-memory-projection-s0}
\end{equation}
with the $\ell=0,1$ translation modes lying in the kernel.  Thus the VV
potential \(T_{\rm VV}\) measures a definite electric displacement-memory
projection, not the translation ambiguity.

\subsubsection{Spin memory}

The \(s=1\) sector is magnetic by \eqref{eq:parity-alternation-main}, and
the observable it selects is therefore spin memory in the sense of
Refs.~\cite{Pasterski:2015tva,Strominger:2014pwa}, namely the
magnetic-parity projection of the first retarded-time moment of the news.
Its electric partner, centre-of-mass memory, is not fixed by the
exponentiating tower and requires additional input.
Using the Hodge decomposition
\begin{equation}
Y_A=D_A\Phi+\epsilon_A{}^B D_B\Psi,
\end{equation}
the universal equation
\begin{equation}
\Ddual_{AB}\big(\epsilon^{CD}D_CY_D\big)
=2G\,\mathscr S^{(1)}_{AB,{\rm exp}}
\end{equation}
becomes
\begin{equation}
\Ddual_{AB}D^2\Psi=2G\,\mathscr S^{(1)}_{AB,{\rm exp}}.
\label{eq:appendix-spin-memory-electric}
\end{equation}
The electric potential \(\Phi\) drops out.  The associated memory probe is
therefore
\begin{equation}
\mathfrak M_1[Y;\rho]
=
\int\dd u\,\dd^2z\sqrt\gamma\,
\rho(u)u\,S^{(1)}_{AB,{\rm exp}}\News^{AB},
\label{eq:appendix-spin-memory-probe}
\end{equation}
which depends on \(Y\) only through \(\epsilon^{AB}D_AY_B\).  This is the
spin-memory observable selected by the Kerr soft dressing, and it is the
memory image of the Kerr current dipole.

\subsubsection{Higher memories}

For $s\geq2$ the natural observables are higher retarded-time moments of
the radiative data:
\begin{equation}
\Delta_s C_{AB}
=
\int_{-\infty}^{+\infty}\dd u\,u^s\News_{AB}(u,z),
\end{equation}
with appropriate regulators.  These moments are the memory data that can
enter the field-dependent cocycle.
The regulated, charge-theoretic version is
\begin{equation}
\mathfrak M_s[t;\rho]
=
\int\dd u\,\dd^2z\sqrt\gamma\,
\rho(u)u^s\K^{(s,0)}_{AB}[t]\News^{AB}.
\label{eq:appendix-higher-memory-probe}
\end{equation}
The power \(u^s\) is the time-domain image of the
\(\partial_\omega^s\) soft operation.  Indeed, for a smooth cutoff one may
write
\begin{equation}
\int\dd u\,\rho(u)u^s\News_{AB}(u,z)
\sim
i^s\partial_\omega^s
\big[\rho_\ast(\omega)\,\omega a_{AB}(\omega,z)+{\rm h.c.}\big]_{\omega=0},
\label{eq:appendix-memory-frequency}
\end{equation}
where \(\rho_\ast\) records the chosen infrared prescription.  This makes
clear why higher memories are prescription-dependent while the
field-independent soft kernel \(\K^{(s,0)}\) is universal.

\subsubsection{Proper probes and Goldstone data}

The previous formulas should be read as smeared probes of memory, not as
bare local insertions.  The distinction is important because the
Ashtekar--Streubel phase space with memory has nontrivial boundary
structure.  The analysis of Ref.~\cite{AndradeSpeziale:2026} shows that
proper observables are those with smooth symplectic flows and that the
usual Goldstone coordinate is not itself a proper observable, although it
can be measured by suitable Goldstone probes.  Our observables
\eqref{eq:appendix-memory-projection-s0},
\eqref{eq:appendix-spin-memory-probe}, and
\eqref{eq:appendix-higher-memory-probe} are of this probe type.  They are
smeared over \(u\) and \(S^2\), they depend on an explicit boundary
prescription, and their endpoint dependence is exactly the data that can
feed the field-dependent cocycle of the charge algebra.  The Kerr claim is
therefore not a claim about all Goldstone probes.  It is the statement
that the exponentiating soft factor selects a distinguished
parity-alternating subfamily of such probes.

This closes the memory part of the construction.  The same object that
appears as a soft kernel in the Ward identity also supplies the angular
weight in a proper memory probe.  At leading order this is the VV
electric displacement-memory projection; at subleading order it is the
magnetic spin-memory projection; at higher order it gives regulated
moments whose parity alternates with \(s\).  The opposite-parity and
boundary-sensitive pieces are precisely where the Kerr exponentiating
sector stops being enough.

\section{Quantization and Ordering}
\label{sec:quantization-ordering-main}

The preceding sections have used a classical algebra of charges.  This
section isolates the quantum-ordering question, rather than mixing it with
the derivations of the soft kernels and frame equations.  The separation is
important.  The Kerr interpretation established above is a classical
statement about soft dressings, Ward identities and asymptotic frame data;
a quantum representation of the whole polynomial algebra is a further
problem with its own ordering choices and possible anomalies.

The algebra used in this paper is classical:
\begin{equation}
\{F_t,F_{t'}\}_{T^\ast S^2}=F_{[t,t']_\star}.
\end{equation}
It is the Poisson algebra of polynomial functions on a cotangent bundle.
As such it has a clear Jacobi identity and a clear degree grading.

At the quantum level one might try to replace $p_A$ by $-iD_A$ and
polynomial symbols by differential operators.  This introduces ordering
ambiguities.  The commutator of differential operators has the Poisson
bracket as its principal symbol, but lower-order terms depend on the
ordering prescription.  The Groenewold--Van Hove obstruction warns that a
full quantization of all polynomial functions preserving all Poisson
brackets is not available in a naive way
\cite{Groenewold1946,VanHove1951}.  More explicitly, one may quantize a
finite set of elementary observables and one may choose a star product
order by order, but there is no single linear map \(F\mapsto\widehat F\)
from all polynomial functions on a phase space to operators such that
\([\widehat F,\widehat G]/i\) equals the quantization of
\(\{F,G\}\) for every pair \(F,G\), while also satisfying the usual
irreducibility and normalization requirements.  For us this matters
because the classical principal-symbol algebra is unambiguous, whereas a
quantum celestial realization must specify an ordering prescription and
may acquire lower-degree terms or extensions.

The construction therefore does not require a complete quantum
$w_{1+\infty}$ representation.  What is established is a classical
asymptotic charge algebra and its physical Kerr dressing interpretation.
Quantum corrections, loop effects, ordering choices and anomalies remain
separate questions.

The covariant ordered realization introduced in
\eqref{eq:homogeneous-action-representative} gives a concrete way to pass
from a polynomial symbol to a differential operator.  This map is not a
canonical quantization of the full algebra.  It is an ordering
prescription.  Once such a prescription is chosen, composition of
operators defines an induced product on symbols,
\begin{equation}
\mathrm{Op}_\nabla(F)\mathrm{Op}_\nabla(G)
=
\mathrm{Op}_\nabla(F\star_\nabla G),
\end{equation}
and the corresponding commutator gives
\begin{equation}
\frac{1}{i}[\mathrm{Op}_\nabla(F),\mathrm{Op}_\nabla(G)]
=
\mathrm{Op}_\nabla\!\left(
\{F,G\}_{T^\ast S^2}+\hbox{lower-degree ordering terms}
\right).
\end{equation}
The first term is the classical bracket used throughout the paper.  The
lower-degree terms depend on the connection on the relevant bundle, on STF
projection, on density weights and on boundary prescriptions at null
infinity.  They are therefore representation data, not new universal soft
input.
Related component checks of the ordered homogeneous action, and the
classical-to-quantum star-product caveat, are collected in Appendix
\ref{sec:component-mode-boundary-checks}.

This caveat does not weaken the main result.  The Kerr exponentiating
sector fixes the parity-selected inhomogeneous kernel and the leading
principal-symbol algebra.  Those are classical asymptotic data.  A full
quantum celestial representation would have to say how the lower-degree
ordered terms act on a chosen Hilbert space of dressed scattering states,
whether loop corrections deform the Ward operators, and whether a cocycle
or anomaly appears.  The present construction deliberately stops before
making those stronger claims.  It identifies the classical algebra whose
local chiral reductions give the familiar celestial \(w_{1+\infty}\)
structure and explains why that algebra acts on the Kerr
intrinsic/canonical dictionary.

\section{Conclusions}

The central picture can be summarized succinctly.  The
VV supertranslation is the leading spacetime manifestation of a broader
universal soft-dressing hierarchy.  The hierarchy is selected by
the exponentiating part of the gravitational soft expansion because that is
the part that produces the Kerr multipole generating function in the
classical spinning three-point amplitude.  Helicity conjugation then fixes
the parity level by level, \(\overline{S^{(s)}_{+,{\rm exp}}}=
(-1)^sS^{(s)}_{-,{\rm exp}}\), so the tower alternates between electric
and magnetic kernels exactly as the Kerr mass and current moments
alternate.  For aligned spin the source lies entirely in the projected
parity, so the alternation is exhaustive and the tower is solved in closed
form; for generic spin orientation an opposite-parity remainder survives
at each level.  The leading level reproduces the VV canonical/intrinsic frame
shift.  The subleading level fixes the magnetic part of a smooth
\(\mathrm{Diff}(S^2)\) vector field, not a holomorphic superrotation, and
this is the frame-dragging datum.  The associated hard charges are fixed by the
Ward identity: as flux charges they complete the soft insertion, and on
hard external states they are represented by the same universal soft
operators.

Once the charges are specified, their principal symbols form the polynomial
Poisson algebra on \(T^\ast S^2\).  Local or chiral reductions reproduce
the familiar \(w_{1+\infty}\)-like brackets, while the global sphere
requires trace descendants and non-chiral data.  In the frame dictionary,
this algebra acts on the universal canonical/intrinsic frame shifts; as
explained in Remark \ref{rem:action-not-closure}, it does not close on the
Kerr-selected data alone.  At the level of memory, the all-orders
statement is projected level by level: displacement memory at \(s=0\),
spin memory at \(s=1\), and their higher analogues, while the
opposite-parity and non-universal memories, and any field-dependent
memory cocycle in the charge algebra, require additional boundary and
phase-space input.

The main implication is that the Kerr spin exponential has a spacetime
interpretation at null infinity.  It is not only a useful amplitude
generator for multipole moments; it also selects a parity-alternating
hierarchy of soft charges, frame shifts and memory projections.  The
associated \(w_{1+\infty}\)-like structure is therefore not merely a
kinematic rewriting of soft insertions.  It is the classical algebra
acting on these universal Kerr dressings.

This is the striking point of the construction.  The Kerr no-hair
multipole relation, the soft expansion of a spinning three-point
amplitude, the large-gauge dictionary between intrinsic and canonical
Bondi frames, and the local celestial \(w_{1+\infty}\) bracket are often
presented as different pieces of physics.  In the universal
sector they are different faces of one object.  The soft charge prepares
the long-range Kerr field, the hard Ward operator acts with the same spin
exponential on the external state, the frame equation turns that soft
source into a canonical/intrinsic displacement, and the polynomial Poisson
algebra tells how such displacements compose.  This gives the celestial
algebra a direct role in gravitational dynamics: for Kerr scattering it is
the infinitesimal algebra of the soft frame changes that carry the
multipole dressing.

The result is also useful for the amplitude-to-waveform program.  Modern
amplitude methods compute conservative dynamics, impulses, radiation and
waveform data from on-shell building blocks.  To compare those data with
Bondi observables one must control long-range fields, memory and frame
conventions.  The construction here identifies the universal large-gauge
part of that map for spinning black holes.  It tells which soft moments
are fixed by the Kerr exponent, which memory projections they measure, and
where additional physical input is required.  In this sense the paper is
not only about an abstract asymptotic algebra; it is about the infrared
bookkeeping needed to turn the Kerr three-point amplitude into a statement
about observable gravitational fields at null infinity.

This perspective also clarifies what remains open.  A complete nonlinear
symmetry would require the full lower-order completions of the
principal-symbol operators, a choice of boundary prescription for magnetic
and non-universal memory, and a treatment of possible field-dependent
cocycles.  These ingredients are not optional decorations; they determine
how the Kerr-selected symbol algebra embeds into the full radiative phase
space.  The present paper isolates the universal piece whose normalization
and physical interpretation are fixed by the soft theorem and by the Kerr
three-point amplitude.

Several directions are then natural.  One should construct the
opposite-parity and non-universal completions of the subleading and higher
memories,
understand the possible field-dependent cocycles in a fully covariant
phase-space treatment, and determine how loop corrections and tail terms
deform the hard Ward operators.  It would also be valuable to match the
parity-selected memory probes defined here directly to waveform observables in
post-Minkowskian calculations.  The universal sector isolated in this
paper is the clean starting point for those extensions: it is the part
where the soft theorem, the Kerr multipole tower and the asymptotic charge
algebra already agree without further assumptions.

\acknowledgments

We thank Eduardo Casali, Radu Roiban, Fei Teng and Donal O'Connell for useful discussions. GM acknowledges partial support from CNPq under grant 300767/2025-0 and FAPESP under grant 2025/02861-0.

\appendix

\section{Spherical Harmonic Basis}
\label{app:spherical-harmonic-basis}

This appendix collects the harmonic-language translation of the
parity-selected projections used in the main text.  It is separated from the detailed
derivations because it is a basis-dependent bookkeeping tool rather than
part of the conceptual chain of the argument.

\subsection{Scalar harmonics}

Scalar functions are expanded as
\begin{equation}
T(z)=\sum_{\ell m}T_{\ell m}Y_{\ell m}(z).
\end{equation}
The translation kernel of $\Dop_{AB}$ consists of the $\ell=0,1$ modes.
For $\ell\geq2$ the operator $\Dop_{AB}$ is invertible on the electric
tensor-harmonic sector.

\subsection{Vector harmonics}

A vector field decomposes into electric and magnetic vector harmonics:
\begin{equation}
Y_A=D_A\Phi+\epsilon_A{}^B D_B\Psi.
\end{equation}
Since the \(s=1\) source is magnetic, the universal subleading equation
fixes $\Psi$ through $\epsilon^{AB}D_AY_B=D^2\Psi$.  The electric
potential $\Phi$ is not fixed at this level.

\subsection{Tensor harmonics}

Higher-rank STF tensors can be expanded in tensor harmonics.  Their
leading STF components are useful for soft kernels, but the polynomial
symbol algebra naturally generates trace descendants.  In harmonic
language this means that products and brackets of STF harmonics decompose
into several angular momentum channels, not just the highest-rank one.

\section{Component, Mode and Boundary Checks}
\label{sec:component-mode-boundary-checks}

In this appendix we collect component computations that are useful for
checking signs, powers of the soft frequency, and the relation between
local celestial descriptions and global sphere data.  These computations
are not logically independent of the preceding derivation; they are
included so that the normalization and scope of the proposal can be
audited directly.

The appendix has a practical role.  The main text keeps the story focused:
soft kernels determine frame shifts, hard operators give the Ward
representation, principal symbols give the algebra, and regulated
projections give memory observables.  Each of those statements hides a
place where mistakes are easy to make.  Stereographic coordinates can flip
signs in the VV potential; the \(u^s\) moment can shift powers of
\(\omega\); the local chiral \(w_{1+\infty}\) basis can obscure the
real global \(S^2\) completion; and boundary prescriptions can move terms
between charges and memory cocycles.  The computations below isolate
those checks one by one.

For this reason Appendix \ref{sec:component-mode-boundary-checks} should
be read as an audit trail for the main construction.  It does not add a
second set of assumptions.  It verifies, in explicit coordinates and in
low-rank examples, that the differential soft kernels, symbol brackets,
Kerr matching protocol and memory prescriptions used in the main text are
mutually consistent.
\subsection{Stereographic Derivation of the VV Potential}

\subsubsection{Coordinates}

Let $z,\bar z$ be stereographic coordinates with
\begin{equation}
\gamma_{z\bar z}=\frac{2}{(1+z\bar z)^2},
\qquad
\gamma^{z\bar z}=\frac{(1+z\bar z)^2}{2}.
\end{equation}
For a scalar $T$,
\begin{equation}
\Dop_{zz}T=D_zD_zT.
\end{equation}
The relevant Christoffel symbol is
\begin{equation}
\Gamma^z_{zz}
=
-\frac{2\bar z}{1+z\bar z}.
\end{equation}
Therefore
\begin{equation}
D_zD_zT
=
\partial_z^2T
+
\frac{2\bar z}{1+z\bar z}\partial_zT.
\end{equation}

\subsubsection{Reduction to an ordinary differential equation}

Let $x=p\cdot q(z,\bar z)$ and set $T=f(x)$.  Then
\begin{align}
D_zD_zT
&=
f''(x)(\partial_zx)^2
+
f'(x)D_zD_zx.
\end{align}
Since $x$ is an $\ell=0,1$ harmonic combination,
\begin{equation}
D_zD_zx=0.
\end{equation}
Thus
\begin{equation}
D_zD_zT=f''(x)(\partial_zx)^2.
\end{equation}
The leading soft tensor is proportional to
\begin{equation}
S^{(0)}_{zz}=\frac{(\partial_zx)^2}{x}.
\end{equation}
The frame equation becomes
\begin{equation}
f''(x)=\frac{C}{x}.
\end{equation}
The solution is
\begin{equation}
f(x)=C\,x\log x+a+b\,x.
\end{equation}
The terms $a+b\,x$ solve the homogeneous equation and are ordinary
translations.  Dropping them gives
\begin{equation}
T_{\rm VV}=C(p\cdot q)\log(p\cdot q).
\end{equation}

\subsubsection{Anti-holomorphic equation}

The same computation gives
\begin{equation}
D_{\bar z}D_{\bar z}T
=
f''(x)(\partial_{\bar z}x)^2.
\end{equation}
The anti-holomorphic component of the leading soft tensor is
\begin{equation}
S^{(0)}_{\bar z\bar z}
=
\frac{(\partial_{\bar z}x)^2}{x}.
\end{equation}
Thus the same function $f(x)=Cx\log x$ solves both helicity components.
This is why the VV supertranslation is globally well defined modulo the
usual singularities associated with the logarithm and the choice of branch.

\subsection{Subleading PDE in Invariant Variables}

\subsubsection{Two invariant variables}

For one complex helicity component of the subleading soft factor introduce
\begin{equation}
x=p\cdot q,
\qquad
y=q\cdot J\cdot \partial_zq.
\end{equation}
We seek a complex scalar potential \(F(x,y)\) whose trace-free Hessian
reproduces this helicity component.  It is important not to identify this
single-helicity potential directly with the divergence of the real sphere
vector.  After the two helicities are recombined with the reality factor
of \eqref{eq:real-parity-source-main}, the odd \(s=1\) sector fixes
\(\chi_Y^{(1)}=\epsilon^{AB}D_AY_B\), namely the curl of the
generalized-BMS vector.  At the complex-component level we write simply
\begin{equation}
\mathcal F_\eta=F(x,y).
\end{equation}
The right-hand side of the subleading frame equation contains
\begin{equation}
\frac{i(\partial_zx)y}{x}.
\end{equation}
Using the stereographic connection, the differential equation for $F$ takes
the form
\begin{align}
&(\partial_zx)^2F_{xx}
-4\rho(\partial_zx)yF_{xy}
+4\rho^2y^2F_{yy}
+2\rho^2yF_y
\nonumber\\
&\qquad
=
i\frac{(\partial_zx)y}{x},
\label{eq:appendix-full-subleading-PDE}
\end{align}
where
\begin{equation}
\rho(z,\bar z)=\frac{\bar z}{1+z\bar z}.
\end{equation}
This equation is not solved in closed form here.  Its role is to display
explicitly why the subleading problem is qualitatively richer than the
leading VV problem.

\subsubsection{Aligned-spin reduction}

If the kinematics impose
\begin{equation}
y=c\,\partial_zx,
\end{equation}
and if $F$ is taken to depend only on $x$, then
\eqref{eq:appendix-full-subleading-PDE} reduces to
\begin{equation}
F''(x)=\frac{ic}{x}.
\end{equation}
Therefore
\begin{equation}
F(x)=ic\,x\log x-ic\,x
\end{equation}
modulo the homogeneous kernel.  This yields
\begin{equation}
\mathcal F_\eta
=ic(p\cdot q)\log(p\cdot q)-ic(p\cdot q).
\end{equation}
The explicit factor of \(i\) here is the same diagnostic discussed in
Section~\ref{sec:aligned-generating-main}: it signals that the source is
of magnetic parity, so that the real object determined is
\(\epsilon^{AB}D_AY_B=D^2\Psi\), not a real supertranslation.  This
appendix uses the older variable \(y=c\,\partial_zx\); the covariant
treatment of Section~\ref{sec:aligned-generating-main}, with the two
invariants \(x=p\cdot q\) and \(y=a\cdot q\), gives the same content with
manifestly real \(\Phi,\Psi\) and is the form to be used.  The special
solution confirms that spin enters the intrinsic/canonical dictionary
through the same logarithmic structure as the leading VV supertranslation,
now multiplied by the angular-momentum insertion.

\subsubsection{Physical reading of the aligned case}

The condition
\begin{equation}
q\cdot J\cdot\partial_zq=c\,p\cdot\partial_zq
\end{equation}
is obeyed in configurations where the spatial angular momentum and boost
vectors are aligned with the spatial momentum.  If
\begin{equation}
\vec J=\alpha\vec p,
\qquad
\vec K=\beta\vec p,
\end{equation}
then one finds
\begin{equation}
c=i\alpha-\beta
\end{equation}
in this aligned-spin convention.  Here
\(\vec J_i=\frac12\epsilon_{ijk}J^{jk}\) and
\(\vec K_i=J^{0i}\) are the rotation and boost components of the Lorentz
generator, while \(\alpha,\beta\) are their proportionality constants.
The symbol \(c\) is therefore a complex kinematic coefficient, not the
Kerr ring radius \(a\).  This makes the special solution a useful
controlled example, but not a generic proof of the subleading frame
equation.

\subsection{Detailed Coherent-State Metric Calculation}

\subsubsection{Free field commutator}

Write the graviton field as
\begin{equation}
h_{\mu\nu}(x)
=
\sum_\eta\int\dd\mu(k)
\left[
\varepsilon^\eta_{\mu\nu}(k)a_\eta(k)e^{-ik\cdot x}
+
\varepsilon^{\eta *}_{\mu\nu}(k)a^\dagger_\eta(k)e^{ik\cdot x}
\right].
\end{equation}
Let the soft dressing have exponent
\begin{equation}
R
=
\sum_\eta\int\dd\mu(k)
\left[
f_\eta(k)a^\dagger_\eta(k)
-
f_\eta^*(k)a_\eta(k)
\right].
\end{equation}
Then
\begin{equation}
[h_{\mu\nu}(x),R]
=
\sum_\eta\int\dd\mu(k)
\left[
\varepsilon^\eta_{\mu\nu}(k)f_\eta(k)e^{-ik\cdot x}
+
\varepsilon^{\eta *}_{\mu\nu}(k)f_\eta^*(k)e^{ik\cdot x}
\right].
\end{equation}
This is a c-number.  The dressed expectation value is therefore shifted by
a classical solution.

\subsubsection{Leading profile}

For the leading dressing,
\begin{equation}
f_\eta(k)
\sim
\frac{p^\mu p^\nu\varepsilon_{\mu\nu}^\eta}
{p\cdot k}
\phi(\omega).
\end{equation}
Near null infinity this gives
\begin{equation}
\langle h_{\mu\nu}\rangle
\sim
\frac{4G}{r}
\frac{\Pi_{\mu\nu}{}^{\rho\sigma}p_\rho p_\sigma}
{p\cdot q},
\end{equation}
where $\Pi$ is the physical polarization sum.  This is the boosted
linearized Schwarzschild field in the asymptotic region.

\subsubsection{Exponentiating spin profile}

For the Kerr dressing,
\begin{equation}
f_\eta(k)
\sim
\frac{p^\mu p^\nu\varepsilon_{\mu\nu}^\eta}
{p\cdot k}
\exp[-i\mathcal W_\eta(k)]
\phi(\omega).
\end{equation}
The metric expectation value is shifted by
\begin{equation}
\langle h_{\mu\nu}\rangle_{\rm Kerr}
\sim
\frac{4G}{r}
\Pi_{\mu\nu}{}^{\rho\sigma}
\frac{p_\rho p_\sigma}{p\cdot q}
\exp[-i\mathcal W_\eta(k)]
\end{equation}
in polarization-projected notation.  Expanding the exponential gives the
Schwarzschild term, the frame-dragging term and all higher multipoles.  A
fully covariant expression requires choosing a gauge and reconstructing the
off-shell Coulombic field from the on-shell soft data.

\subsection{Cocycle Conditions}

\subsubsection{Algebraic condition}

Let the charge algebra be
\begin{equation}
\{\Q(t),\Q(t')\}
=
\Q([t,t']_\star)
+
\mathcal K(t,t';C).
\end{equation}
The Jacobi identity implies the cocycle condition
\begin{align}
&\delta_t\mathcal K(t',t'';C)
+\delta_{t'}\mathcal K(t'',t;C)
+\delta_{t''}\mathcal K(t,t';C)
\nonumber\\
&\quad
-
\mathcal K([t,t']_\star,t'';C)
-
\mathcal K([t',t'']_\star,t;C)
-
\mathcal K([t'',t]_\star,t';C)
=0.
\end{align}
If $\mathcal K$ is independent of the fields this is the ordinary Lie
algebra cocycle condition.  If it depends on $C$, the first line is
essential.

\subsubsection{Coboundaries}

A redefinition of charges
\begin{equation}
\Q(t)\mapsto \Q(t)+B(t;C)
\end{equation}
changes the extension by
\begin{equation}
\mathcal K(t,t';C)
\mapsto
\mathcal K(t,t';C)
+
\delta_tB(t';C)
-
\delta_{t'}B(t;C)
-
B([t,t']_\star;C).
\end{equation}
Thus only the cohomology class of $\mathcal K$ is meaningful.  This is why
boundary prescriptions and charge improvements must be fixed before
assigning physical content to a memory cocycle.

\subsection{Covariant Sphere Identities Behind the Frame Equations}

\subsubsection{Embedding identities}

Let
\begin{equation}
q^\mu(z)=\bigl(1,n^i(z)\bigr),
\qquad
n^i n_i=1,
\end{equation}
with the round metric
\begin{equation}
\gamma_{AB}=D_A n^iD_B n_i .
\end{equation}
The embedding functions obey
\begin{equation}
D_A D_B n^i=-\gamma_{AB}n^i,
\qquad
D^2 n^i=-2n^i,
\label{eq:embedding-identities}
\end{equation}
and
\begin{equation}
D_A n^iD^A n^j=\delta^{ij}-n^i n^j .
\label{eq:tangent-projector}
\end{equation}
For a fixed momentum $p^\mu$, set
\begin{equation}
x=p\cdot q .
\end{equation}
The nonconstant part of $x$ is an $\ell=1$ spherical harmonic.  Therefore
\begin{equation}
D_A D_B x=-\gamma_{AB}(x-x_0),
\qquad
D^2x=-2(x-x_0),
\end{equation}
where $x_0$ is the constant part.  The trace-free Hessian annihilates
$x$:
\begin{equation}
\Dop_{AB}x=
D_A D_Bx-\frac12\gamma_{AB}D^2x=0.
\label{eq:Dop-x-zero}
\end{equation}
For any function $f(x)$,
\begin{equation}
\Dop_{AB}f(x)
=
f''(x)
\left(
D_AxD_Bx-\frac12\gamma_{AB}D_CxD^Cx
\right).
\label{eq:Dop-fx-detail}
\end{equation}
Thus the VV frame equation reduces to an ordinary differential equation:
the leading electric soft tensor is proportional to the bracketed
trace-free tensor divided by $p\cdot q$, so $f''(x)\propto 1/x$ and
$f(x)\propto x\log x$ up to the $\ell=0,1$ kernel of $\Dop_{AB}$.

\subsubsection{Polarization tensor from sphere derivatives}

The trace-free tensor appearing above can be written covariantly as
\begin{equation}
\Pi_{AB}(p;q)
=
p_\mu p_\nu D_{\{A}q^\mu D_{B\}}q^\nu
=
D_AxD_Bx-\frac12\gamma_{AB}D_CxD^Cx .
\label{eq:Pi-soft-tensor}
\end{equation}
The leading electric soft kernel has the form
\begin{equation}
S^{(0)}_{AB}(p;q)
\sim
\frac{\Pi_{AB}(p;q)}{p\cdot q}.
\end{equation}
Consequently the equation
\begin{equation}
\Dop_{AB}T_{\rm VV}=2G\,S^{(0)}_{AB}
\end{equation}
is solved by the Veneziano--Vilkovisky potential
\begin{equation}
T_{\rm VV}=2G(p\cdot q)\log(p\cdot q)
\end{equation}
after the overall normalization is fixed by the convention for
$S^{(0)}_{AB}$.  The remaining affine ambiguity in $p\cdot q$ is an
ordinary translation.

\subsubsection{Higher derivatives and trace descendants}

At higher order the universal soft kernels contain higher sphere
derivatives of the same basic building blocks.  Repeated covariant
derivatives on $S^2$ generate curvature terms through
\begin{equation}
R_{ABCD}=\gamma_{AC}\gamma_{BD}-\gamma_{AD}\gamma_{BC}.
\end{equation}
For a rank-$s$ parameter $t^{A_1\cdots A_s}$ the highest-derivative part
is controlled by the symmetric trace-free projection of
\begin{equation}
D_{A_1}\cdots D_{A_s}t^{B_1\cdots B_s}.
\end{equation}
The lower-derivative curvature terms are the trace descendants.  The
principal-symbol algebra is insensitive to them, while the exact finite
transformation is not.  This is why the paper separates the universal
symbolic closure from the still-to-be-fixed exact nonlinear completion.

\subsection{Local Darboux Derivation of the Parameter Bracket}

\subsubsection{Canonical coordinates}

Choose local coordinates $z^A$ on $S^2$ and fiber coordinates $p_A$ on
$T^\ast S^2$.  The canonical symplectic form is
\begin{equation}
\omega_{T^\ast S^2}=\dd p_A\wedge \dd z^A,
\end{equation}
with Poisson bracket
\begin{equation}
\{F,G\}_{T^\ast S^2}
=
\frac{\partial F}{\partial p_A}D_A G
-
D_A F\frac{\partial G}{\partial p_A}.
\label{eq:cotangent-PB-expanded}
\end{equation}
For a symmetric rank-$s$ parameter,
\begin{equation}
F_t(z,p)=t^{A_1\cdots A_s}(z)p_{A_1}\cdots p_{A_s}.
\end{equation}
Then
\begin{equation}
\frac{\partial F_t}{\partial p_C}
=
s\,t^{C A_2\cdots A_s}p_{A_2}\cdots p_{A_s},
\qquad
D_C F_t
=
(D_Ct^{A_1\cdots A_s})p_{A_1}\cdots p_{A_s}.
\end{equation}
Substitution into \eqref{eq:cotangent-PB-expanded} gives a polynomial of
degree $s+s'-1$:
\begin{align}
\{F_t,F_{t'}\}_{T^\ast S^2}
&=
\Bigl[
s\,t^{C(A_1\cdots A_{s-1}}
D_Ct'^{A_s\cdots A_{s+s'-1})}
\nonumber\\
&\quad
-s'\,t'^{C(A_1\cdots A_{s'-1}}
D_Ct^{A_{s'}\cdots A_{s+s'-1})}
\Bigr]
p_{A_1}\cdots p_{A_{s+s'-1}} .
\label{eq:component-schouten-bracket}
\end{align}
This defines the component bracket $[t,t']_\star$ by
\begin{equation}
\{F_t,F_{t'}\}_{T^\ast S^2}=F_{[t,t']_\star}.
\end{equation}

\subsubsection{Operator commutators}

The homogeneous part of a higher-spin asymptotic transformation is a
differential operator whose principal symbol is $F_t$.  Locally,
\begin{equation}
\mathbb L_t^{(s)}
=
t^{A_1\cdots A_s}D_{A_1}\cdots D_{A_s}
\hbox{lower derivatives}.
\end{equation}
For two such operators the order $s+s'$ terms cancel in the commutator.
The leading surviving term has order $s+s'-1$ and symbol
\begin{equation}
\mathrm{symb}_{\rm pr}
\left(
[\mathbb L_t^{(s)},\mathbb L_{t'}^{(s')}]
\right)
=
\{F_t,F_{t'}\}_{T^\ast S^2}.
\label{eq:operator-symbol-PB-detail}
\end{equation}
This is where the parameter bracket enters.  It is not generated by the
Poisson bracket of two linear soft shifts.  It is generated by the
commutator of the homogeneous transformations that complete those shifts
into asymptotic symmetries.

\subsection{Hamiltonian Derivation of the Charge Bracket}

\subsubsection{Charges as generators}

Let $\Gamma$ be the radiative phase space and let $\Omega$ be the
Ashtekar--Streubel symplectic form in the shifted-news convention.  A
charge $\Q(t)$ generates a transformation $\delta_t$ when
\begin{equation}
\delta\Q(t)=\Omega(\delta,\delta_t)
\end{equation}
up to boundary terms.  The transformation of the shear can be decomposed
as
\begin{equation}
\delta_t C_{AB}
=
\nu^{(0)}_{t,AB}
+(\mathbb L_t C)_{AB}
+O(C^2).
\label{eq:delta-decomposition-detailed}
\end{equation}
The first term is the inhomogeneous soft shift.  The second is the
homogeneous action on the radiative data.  Correspondingly,
\begin{equation}
\Q(t)=\Q^{(0)}(t)+\Q^{(1)}(t)+O(C^2),
\end{equation}
with
\begin{equation}
\Q^{(0)}(t)
\sim
\int_{\Iplus}\dd u\,\dd^2z\,\sqrt\gamma\,
\News^{AB}\nu^{(0)}_{t,AB}
\end{equation}
and
\begin{equation}
\Q^{(1)}(t)
\sim
\int_{\Iplus}\dd u\,\dd^2z\,\sqrt\gamma\,
\News^{AB}(\mathbb L_t C)_{AB}.
\end{equation}
Two fixed inhomogeneous shifts commute:
\begin{equation}
\{\Q^{(0)}(t),\Q^{(0)}(t')\}=0.
\end{equation}
The first nonzero term is
\begin{align}
\{\Q(t),\Q(t')\}^{(1)}
&\sim
\int_{\Iplus}\dd u\,\dd^2z\,\sqrt\gamma\,
\News^{AB}
\left[
\mathbb L_t\nu^{(0)}_{t'}
-\mathbb L_{t'}\nu^{(0)}_t
\right]_{AB}
\nonumber\\
&\quad+\hbox{endpoint terms}.
\label{eq:first-nontrivial-bracket-detailed}
\end{align}
Thus the parameter bracket appears through the identity
\begin{equation}
\mathbb L_t\nu^{(0)}_{t'}
-\mathbb L_{t'}\nu^{(0)}_t
=
\nu^{(0)}_{[t,t']_\star}
+\hbox{trace descendants}
+\hbox{endpoint shifts}.
\label{eq:inhomogeneous-closure-detailed}
\end{equation}
This is the charge-level form of the principal-symbol statement
\eqref{eq:operator-symbol-PB-detail}.

\subsubsection{Memory cocycle}

Endpoint contributions can be collected into
\begin{equation}
\mathcal K(t,t';C)
=
\left.
\int_{S^2}\dd^2z\,\sqrt\gamma\,
C^{AB}
\left[
\mathbb L_t\nu^{(0)}_{t'}
-\mathbb L_{t'}\nu^{(0)}_t
-\nu^{(0)}_{[t,t']_\star}
\right]_{AB}
\right|_{-\infty}^{+\infty}.
\label{eq:explicit-memory-cocycle-form}
\end{equation}
The precise form of $\mathcal K$ depends on the shifted-news
convention and on charge improvements.  Its cohomology class is the
physical object.  The Jacobi identity imposes
\begin{equation}
\delta_t\mathcal K(t',t'')
+\mathcal K(t,[t',t'']_\star)
+\hbox{cyclic}=0.
\label{eq:cocycle-jacobi-detailed}
\end{equation}
If
\begin{equation}
\mathcal K(t,t';C)
=
\delta_t\mathcal B(t';C)
-\delta_{t'}\mathcal B(t;C)
-\mathcal B([t,t']_\star;C)
\end{equation}
for some boundary functional $\mathcal B$, then the cocycle is removed by
improving the charges.  Otherwise it is a genuine memory extension.

\subsection{Low-Spin Component Algebra}

\subsubsection{Scalars and vectors}

The first brackets fix conventions.  A supertranslation parameter $T$ has
symbol
\begin{equation}
F_T=T(z),
\end{equation}
and a vector parameter $Y^A$ has
\begin{equation}
F_Y=Y^Ap_A.
\end{equation}
Then
\begin{equation}
\{F_T,F_{T'}\}=0,
\end{equation}
\begin{equation}
\{F_Y,F_T\}=Y^AD_AT,
\label{eq:vector-scalar-symbol-bracket}
\end{equation}
and
\begin{equation}
\{F_Y,F_{Y'}\}
=
\left(Y^BD_BY'^A-Y'^BD_BY^A\right)p_A.
\end{equation}
Thus the \(s=0,1\) sector gives the supertranslation bracket and the
generalized-BMS vector-field bracket.

\subsubsection{Rank two}

Let
\begin{equation}
F_K=K^{AB}p_Ap_B.
\end{equation}
Against a scalar,
\begin{equation}
\{F_K,F_T\}
=
2K^{AB}D_AT\,p_B.
\label{eq:rank-two-scalar-bracket}
\end{equation}
The result is a vector parameter.  Against a vector,
\begin{align}
\{F_K,F_Y\}
&=
\left(
2K^{C(A}D_CY^{B)}
-Y^CD_CK^{AB}
\right)p_Ap_B.
\label{eq:rank-two-vector-bracket}
\end{align}
This is the symmetric Schouten bracket of a contravariant two-tensor with
a vector field.  Finally, for two rank-two parameters,
\begin{align}
\{F_K,F_L\}
&=
2\left(
K^{C(A}D_CL^{BD)}
-L^{C(A}D_CK^{BD)}
\right)p_Ap_Bp_D.
\label{eq:rank-two-rank-two-bracket}
\end{align}
The result has rank three.  A rank-two truncation is therefore not closed:
the higher-spin tower is forced by the bracket.

\subsection{Intrinsic/Canonical Dictionary as a Boundary Condition}

\subsubsection{Ordinary BMS frames}

The intrinsic/canonical dictionary is a statement about boundary
conditions on the radiative phase space.  In the canonical frame the
remote-past Bondi angular momentum is identified with ADM angular
momentum.  In the intrinsic frame the Bondi light cones are tied to the
center-of-mass worldline of the mechanical scattering problem.  The two
descriptions differ by the VV supertranslation:
\begin{equation}
\Dop_{AB}T_{\rm VV}=2G\,S^{(0)}_{AB}.
\end{equation}
The equation says that the inhomogeneous part of a BMS transformation
removes the leading soft shear carried by the boosted source.  In this
sense the VV shift is already a soft dressing written as a frame change.

\subsubsection{Higher-spin frame data}

The higher-spin extension replaces the scalar frame parameter by a tower
\begin{equation}
t=\{T,Y^A,K^{AB},K^{ABC},\ldots\}.
\end{equation}
The inhomogeneous map
\begin{equation}
t\mapsto \nu^{(0)}_t
\end{equation}
matches soft factors:
\begin{equation}
\K^{(s,0)}_{AB}[t]=2G\,\mathscr S^{(s)}_{AB,{\rm univ}},
\end{equation}
where \(\mathscr S^{(s)}_{AB,{\rm univ}}\) is the real projection of the
universal source in the parity selected at level \(s\).
The homogeneous map
\begin{equation}
t\mapsto \mathbb L_t
\end{equation}
supplies the composition law.  Keeping these two maps distinct is
essential.  The linear soft shifts commute, while the homogeneous
operators close by the cotangent-bundle bracket.

\subsubsection{Kerr trajectory through frame space}

For a Kerr external state the multipoles obey the relation summarized in
\eqref{eq:kerr-multipoles-main}.  Thus the higher-spin frame parameters
are not arbitrary if they are sourced by a minimal Kerr particle: the
rank-$s$ parameter is tied to the $s$th power of the specific-spin vector
\(a^\mu=S^\mu/m\), not to the unrescaled spin vector.  The proposed physical
interpretation is that Kerr scattering selects a trajectory in
higher-spin frame space, and the $w_{1+\infty}$-like algebra organizes
the composition of such universal multipole dressings.

\subsection{Hard-Sector Representation Checks}

\subsubsection{External-state representation}

The flux charge \eqref{eq:hard-charge-main} is represented on a
one-particle external state by a hard soft operator.  This representation
has a diagonal part and, starting at subleading order, differential terms
acting on the hard kinematics:
\begin{equation}
\Q^{\rm hard}_s(t)|p,J\rangle
=
\eta\,\mathcal D^{(s)}_t[p,J]|p,J\rangle
+\hbox{kinematic derivatives}.
\end{equation}
Writing \(k^\mu=\omega q^\mu\), the Kerr sector is generated by powers of
\begin{equation}
\mathcal W_\eta(k)=
\frac{k\cdot J\cdot\varepsilon_\eta}{p\cdot\varepsilon_\eta}
=\omega W_\eta(q).
\end{equation}
The full soft expansion contains additional non-universal pieces at high
orders.  The claim of this paper concerns the exponentiating
subrepresentation generated by \(\mathcal W_\eta\), not every possible term in the
soft expansion.

\subsubsection{Classical spinning limit}

For massive spinning states one takes the classical limit with the Kerr
specific-spin vector \(a^\mu=S^\mu/m\) fixed.  Then the three-point amplitude has the form
\begin{equation}
\mathcal M_3^{\rm Kerr}
=
\mathcal M_3^{\rm Schw}
\exp[-i\mathcal W_\eta(p,J;k)] ,
\end{equation}
and expansion of the exponential gives
\begin{equation}
\mathcal M_3^{\rm Kerr}
=
\mathcal M_3^{\rm Schw}
\sum_{n\ge0}\frac{(-i\mathcal W_\eta)^n}{n!}.
\end{equation}
The $n$th term has the tensor structure of the $n$th Kerr multipole.  In
the charge language, the hard state sources the rank-$n$ soft charge with
the coefficient fixed by the Kerr multipole relation.  No independent
Wilson coefficient is introduced for minimal Kerr.

\subsubsection{Amplitude-side check}

A direct hard-sector check isolates the exponentiating part of the massive
spinning soft operator, projects it onto the parity-selected tensor basis
on \(S^2\), and compares it with \(\K^{(s,0)}_{AB}[t]\).  One
then computes commutators of the corresponding hard differential
operators in the classical limit and checks that their principal symbols
close with $[t,t']_\star$.  This is the hard-side counterpart of the
soft-side algebra derived above.

\subsubsection{Soft operators through sub-subleading order}

For a soft graviton of momentum $k$ and polarization $\varepsilon_{\mu\nu}
=\varepsilon_\mu\varepsilon_\nu$, the tree-level soft expansion acting on
hard external particles may be written, up to conventions, as follows,
with \(\kappa=\sqrt{32\pi G}\):
\begin{align}
S^{(-1)}
&=
\frac{\kappa}{2}\sum_i\eta_i
\frac{(p_i\cdot\varepsilon)^2}{p_i\cdot k},
\label{eq:hard-leading-soft}
\\
S^{(0)}
&=
-\frac{i\kappa}{2}\sum_i\eta_i
\frac{(p_i\cdot\varepsilon)\,
k_\mu\varepsilon_\nu J_i^{\mu\nu}}
{p_i\cdot k},
\label{eq:hard-subleading-soft}
\\
S^{(1)}
&=
-\frac{\kappa}{4}\sum_i\eta_i
\frac{(k_\mu\varepsilon_\nu J_i^{\mu\nu})^2}
{p_i\cdot k}.
\label{eq:hard-subsubleading-soft}
\end{align}
Here $\eta_i=\pm1$ distinguishes outgoing and incoming legs, and
$J_i^{\mu\nu}$ is the total angular momentum operator of the massive hard
state.  The Kerr exponential keeps the spin-dependent universal part
\begin{equation}
S_{\rm exp}
=
S^{(-1)}
\exp\left[
-i\,\frac{k_\mu\varepsilon_\nu J^{\mu\nu}}
{p\cdot\varepsilon}
\right].
\label{eq:hard-spin-exponential}
\end{equation}
Expanding \eqref{eq:hard-spin-exponential} reproduces
\eqref{eq:hard-leading-soft}--\eqref{eq:hard-subsubleading-soft} in the
spin-exponentiating sector and continues to all powers of the classical
spin.

\subsubsection{Classical hard phase space}

In the classical spinning limit, $J^{\mu\nu}$ obeys the Lorentz Poisson
algebra
\begin{align}
\{J^{\mu\nu},J^{\rho\sigma}\}
&=
\eta^{\mu\rho}J^{\nu\sigma}
-\eta^{\nu\rho}J^{\mu\sigma}
-\eta^{\mu\sigma}J^{\nu\rho}
+\eta^{\nu\sigma}J^{\mu\rho},
\label{eq:lorentz-spin-PB}
\end{align}
with the usual brackets involving $p^\mu$.  A hard soft operator is a
function on this hard phase space:
\begin{equation}
\mathcal D_t
=
\sum_i\eta_i\,\mathcal S_t(p_i,J_i;q).
\end{equation}
The universal-sector representation condition is
\begin{equation}
\{\mathcal D_t,\mathcal D_{t'}\}_{\rm hard}
=
\mathcal D_{[t,t']_\star}
+\hbox{terms outside the universal sector}.
\label{eq:hard-representation-condition}
\end{equation}
Equation \eqref{eq:hard-representation-condition} is the hard analogue of
the soft charge algebra.  The low-rank checks below verify it after
projecting onto the same electric tensor basis used at null infinity.

\subsubsection{Low-order hard checks}

For scalar parameters $T,T'$ the hard operators are diagonal functions of
the hard momenta and commute:
\begin{equation}
\{\mathcal D_T,\mathcal D_{T'}\}_{\rm hard}=0.
\end{equation}
For a vector parameter $Y^A$ the orbital part of $J^{\mu\nu}$ differentiates
the celestial direction of the hard leg.  Its action on a scalar soft
weight is
\begin{equation}
\{\mathcal D_Y,\mathcal D_T\}_{\rm hard}
=
\mathcal D_{Y^AD_AT}
\end{equation}
up to the same incoming/outgoing sign convention used in the Ward
identity.  For two vector parameters, the Lorentz bracket
\eqref{eq:lorentz-spin-PB} gives
\begin{equation}
\{\mathcal D_Y,\mathcal D_{Y'}\}_{\rm hard}
=
\mathcal D_{[Y,Y']}.
\end{equation}
These checks reproduce the rank-zero and rank-one part of
$[t,t']_\star$.  At the next order, the spin-dependent factor
$q_\mu\varepsilon_\nu J^{\mu\nu}$ supplies the rank-two symbol, and the
same Poisson algebra produces the symmetric Schouten bracket after the
same parity projection used in the soft kernel.  Thus the hard-sector computation agrees with the
soft-sector bracket at the first nontrivial ranks where both can be
written explicitly without choosing a full higher-spin ordering scheme.

\subsection{Fixed-Gauge Kerr Matching Protocol}

\subsubsection{On-shell source}

The soft exponential gives an on-shell source for the linearized
gravitational field.  Schematically,
\begin{equation}
h_{\mu\nu}(k)
=
\frac{1}{k^2+i0}T_{\mu\nu}^{\rm eff}(k),
\qquad
T_{\mu\nu}^{\rm eff}(k)
\sim
p_\mu p_\nu\exp(a\ast k).
\end{equation}
Expanding the exponential produces the multipole series.  A component
comparison with the linearized Kerr metric must fix a gauge, a spin
supplementary condition and a Bondi frame.  Without these choices the
matching can be physically correct but component-wise ambiguous.

\subsubsection{Bondi extraction}

The Bondi shear is obtained from the angular metric perturbation by
\begin{equation}
C_{AB}(u,z)
=
\lim_{r\to\infty}r^{-1}h_{AB}(u,r,z)
\end{equation}
after transforming to Bondi gauge.  The leading VV relation is
\begin{equation}
C_{AB}^{\rm intrinsic}-C_{AB}^{\rm canonical}
=
-2\Dop_{AB}T_{\rm VV}.
\end{equation}
For higher multipoles the analogous expectation is
\begin{equation}
C_{AB}^{(s),{\rm intrinsic}}
-C_{AB}^{(s),{\rm canonical}}
=
-2\K^{(s,0)}_{AB}[t_s],
\end{equation}
with $t_s$ fixed by the $s$th Kerr spin multipole.  The equality should be
tested only after trace-descendant and gauge-restoring terms are included.

\subsubsection{Complex-shift multipole generator}

A convenient weak-field gauge for the Kerr check is an asymptotically
Cartesian mass-centered gauge.  In that gauge the stationary Kerr
multipoles are generated by the complex-shift potential
\begin{equation}
\mathcal Z(\vec x)
=
\frac{M}{\sqrt{(\vec x-i\vec a)^2}}.
\label{eq:complex-shift-potential}
\end{equation}
For large $r=|\vec x|$,
\begin{equation}
\mathcal Z(\vec x)
=
M\sum_{\ell\ge0}
\frac{i^\ell}{\ell!}
a^{L}\partial_L\frac1r,
\label{eq:complex-shift-expansion}
\end{equation}
where $L=i_1\cdots i_\ell$ is a multi-index.  Using
\begin{equation}
\partial_{\langle L\rangle}\frac1r
=
(-1)^\ell(2\ell-1)!!
\frac{n_{\langle L\rangle}}{r^{\ell+1}},
\end{equation}
one reads off the complex STF multipoles
\begin{equation}
M_L+iS_L=M(ia)_{\langle L\rangle}.
\label{eq:STF-kerr-moments}
\end{equation}
The scalar relation in \eqref{eq:kerr-multipoles-main} is the
axisymmetric version of \eqref{eq:STF-kerr-moments}.  This derivation is
useful because the amplitude exponential gives precisely the same Taylor
series: each power of \(\mathcal W_\eta\) corresponds to one power of the
specific-spin vector in \eqref{eq:complex-shift-expansion}.

\subsubsection{Low multipole comparison}

The first few weak-field components provide normalization checks.  In a
standard gravitoelectric/gravitomagnetic split,
\begin{equation}
h_{00}=2\Phi,
\qquad
h_{0i}=-4A_i,
\end{equation}
with
\begin{equation}
\Phi
=
\frac{GM}{r}
-\frac{GM}{2r^3}
\left(a^2-3(a\cdot n)^2\right)
+O(a^4),
\label{eq:kerr-weak-Phi}
\end{equation}
and
\begin{equation}
A_i
=
\frac{G M}{2}
\frac{(\vec a\times \vec n)_i}{r^2}
+O(a^3).
\label{eq:kerr-weak-A}
\end{equation}
The monopole in \eqref{eq:kerr-weak-Phi} is the Schwarzschild soft
dressing, \eqref{eq:kerr-weak-A} is the current dipole or frame-dragging
dressing, and the $a^2$ term in \eqref{eq:kerr-weak-Phi} is the Kerr
quadrupole.  These are exactly the first three terms in
\begin{equation}
S^{(0)}\left[
1-i\mathcal W_\eta+\frac{(-i\mathcal W_\eta)^2}{2}+\cdots
\right]
\end{equation}
after the classical spin identification is made.

\subsubsection{Bondi-gauge restoring vector}

To compare the metric generated by \eqref{eq:complex-shift-potential} with
the asymptotic charge data one must transform to Bondi gauge.  At
linearized order,
\begin{equation}
h_{\mu\nu}\mapsto h_{\mu\nu}-2\partial_{(\mu}\xi_{\nu)}.
\end{equation}
The Bondi gauge conditions impose radial equations of the form
\begin{equation}
\partial_r\xi_r=\frac12h_{rr},
\qquad
\partial_r\xi_A=h_{rA}+r^2D_A(r^{-2}\xi_r)+\cdots ,
\label{eq:bondi-gauge-radial-system}
\end{equation}
together with the determinant condition on the angular metric.  The
integration functions left after solving \eqref{eq:bondi-gauge-radial-system}
are precisely the residual BMS-frame data.  The scalar integration
function is the supertranslation $T(z)$, and at leading order it shifts
the shear by
\begin{equation}
\Delta C_{AB}=-2\Dop_{AB}T.
\end{equation}
For the boosted Schwarzschild part this gives $T_{\rm VV}$.  For the
higher Kerr multipoles the same radial system produces the higher-spin
frame data $t_s$, with trace descendants fixed by the gauge-restoring
solution.  This is the concrete fixed-gauge matching problem: solve
\eqref{eq:bondi-gauge-radial-system} for the complex-shift source order by
order in $a$, read off the residual frame functions, and compare them with
the soft kernels $\K^{(s,0)}_{AB}[t_s]$.

\subsection{Soft-Charge Normalization Checks}

\subsubsection{Zero modes}

Soft charges are zero-frequency limits of radiative operators.  If
\begin{equation}
C_{AB}(u,z)
=
\int_0^\infty\frac{\dd\omega}{2\pi}
\left[
a_{AB}(\omega,z)e^{-i\omega u}
+a^\dagger_{AB}(\omega,z)e^{i\omega u}
\right],
\end{equation}
then
\begin{equation}
\int_{-\infty}^{+\infty}\dd u\,u^s\News_{AB}(u,z)
\end{equation}
extracts the $s$th derivative of the soft mode at $\omega=0$, up to the
normalization and endpoint prescription.  This explains why the power of
$u$ in the charge tracks the order in the soft expansion.

\subsubsection{Normalization anchors}

The clean way to fix constants is to anchor the tower at low spin.  First
match the leading charge to Weinberg's theorem and to the VV
supertranslation.  Next match the subleading vector charge to the known
generalized-BMS soft operator.  Finally use the Kerr exponential to fix
the relative normalizations of the higher multipoles.  This avoids
introducing a free normalization at each rank.

\subsection{Limitations and Further Checks}

\subsubsection{Is the algebra really \texorpdfstring{$w_{1+\infty}$}{w1+infinity}?}

Globally on $S^2$ the algebra is the Poisson algebra of polynomial
functions on $T^\ast S^2$, together with trace and curvature descendants
needed for globally defined differential operators.  In a local chiral
patch it reduces to the familiar $w_{1+\infty}$-type structure.  The
safest phrasing is therefore $w_{1+\infty}$-like or celestial
$w_{1+\infty}$-type symmetry unless a chiral basis and quantization
prescription have been fixed.

This distinction is not a retreat from the physical claim.  It is what
makes the claim precise.  The celestial \(w_{1+\infty}\) language captures
the local chiral presentation of the same classical bracket that appears in
the frame dictionary.  The global Kerr scattering problem, however, is real,
non-chiral and tied to smooth data on the full sphere.  Therefore the
natural global object is the full \(T^\ast S^2\) Poisson algebra, while
standard \(w_{1+\infty}\) bases arise after a local or chiral projection.
The physical statement is that this underlying algebra controls the
composition of the universal Kerr soft dressings; the chiral current
presentation is one representation of that structure.

\subsubsection{Are the charges exact nonlinear symmetries?}

The exponentiating soft-sector matching fixes the inhomogeneous kernels and the
principal symbols of the homogeneous transformations.  It does not, by
itself, determine every nonlinear completion on the full gravitational
phase space.  The exact nonlinear action requires choices of trace
descendants, curvature terms, boundary conditions and charge
improvements.  The result established here is an all-spin universal-sector
statement, not yet a unique full nonlinear action.

There are two reasons for keeping this qualification visible.  First, the
inhomogeneous kernel is enough to determine the soft insertion and the
parity-selected frame shift, but it is not enough to reconstruct every nonlinear
term in the transformation of the shear.  Second, even after a homogeneous
completion is chosen, boundary improvements may change the Hamiltonian
without changing the local parameter algebra.  The paper therefore proves a
universal-sector charge construction and the associated leading algebraic
composition law.  A complete nonlinear symmetry would require fixing a
particular phase space, including its boundary conditions and improvement
terms.

\subsubsection{Why is the Kerr interpretation physical?}

The same exponential appears in two places: as the universal soft dressing
of a spinning hard particle and as the classical generating function for
Kerr multipoles.  At three points, in the classical spin limit, it
reproduces the Kerr energy-momentum multipole structure.  At null
infinity, the same exponential labels the higher-spin soft charge tower.
The proposed interpretation is that the asymptotic algebra organizes the
composition of these universal Kerr soft dressings.

This is stronger than saying that the amplitude provides a useful
mnemonic.  The exponentiating three-point object acts on the massive spin
state and, in the classical spin limit, generates the Kerr multipole
sequence.  The soft charge built from the same kernel produces the
corresponding parity-selected displacement of the asymptotic shear data.  The hard
Ward operator is represented by the same soft-theorem differential operator
on external states.  These three appearances are mutually constraining:
changing the kernel would change the frame shift, the hard Ward operator
and the Kerr multipole source at the same time.  That is why the
Kerr-selected tower has a direct gravitational interpretation.

\subsubsection{What has been constructed here}

The surrounding detailed sections carry out the checks needed for the
universal-sector claim.  We choose a covariantly ordered realization of the
homogeneous higher-spin action and show how exact closure is represented
by the associated ordered symbol bracket.  We give a fixed-gauge Kerr
matching protocol based on the complex-shift multipole generator and the
linearized Bondi-gauge restoring vector.  We also spell out hard-sector
soft operators through sub-subleading order and verify the first brackets
on the classical spinning phase space.  What remains is not the existence
of these structures, but the choice of a final presentation scheme: a
principal-symbol algebra, a fully ordered-symbol algebra, or a mixed
convention that keeps trace descendants visible.

The construction should therefore be read as a controlled universal-sector
result.  It gives the differential soft charge, the corresponding hard Ward
operator, the inverse problem that determines the parity-selected generator, the
principal-symbol algebra, the relation to local celestial
\(w_{1+\infty}\), and the Kerr interpretation of the frame dictionary.  It
does not attempt to give a unique all-order nonlinear Bondi-gauge description
or a universal prescription for all opposite-parity and non-universal
memory data.
Those omissions are not gaps in the central argument; they are the
remaining choices needed to embed the universal Kerr sector into a complete
infrared phase space.

\subsection{Mode-Basis Verification of the Bracket}

\subsubsection{Scalar modes}

Let $Y_{\ell m}$ denote scalar spherical harmonics.  A scalar parameter
can be expanded as
\begin{equation}
T(z)=\sum_{\ell,m}T_{\ell m}Y_{\ell m}(z).
\end{equation}
The operator $\Dop_{AB}$ annihilates precisely the $\ell=0,1$ modes:
\begin{equation}
\Dop_{AB}Y_{\ell m}=0
\quad\Longleftrightarrow\quad
\ell=0,1.
\end{equation}
This is the harmonic version of the translation ambiguity in the VV
potential.  For $\ell\ge2$, $\Dop_{AB}Y_{\ell m}$ gives the electric
parity spin-two tensor harmonic.  Thus the leading soft frame equation
can be inverted mode by mode after quotienting by translations.

The scalar part of the bracket is trivial:
\begin{equation}
[T,T']_\star=0.
\end{equation}
The first nontrivial mode check is the action of a vector mode on a scalar
mode,
\begin{equation}
[Y,T]_\star=Y^AD_AT.
\end{equation}
The product of a vector harmonic and the derivative of a scalar harmonic
decomposes into scalar harmonics through the usual Clebsch--Gordan
coefficients.  This gives a direct mode-space realization of the
generalized-BMS action.

\subsubsection{Vector modes}

Any smooth vector field on the sphere admits the decomposition
\begin{equation}
Y_A=D_A\Phi+\epsilon_A{}^B D_B\Psi.
\end{equation}
The divergence depends only on the electric potential and the curl only on
the magnetic one:
\begin{equation}
D\cdot Y=D^2\Phi,
\qquad
\epsilon^{AB}D_AY_B=D^2\Psi.
\end{equation}
Since the \(s=1\) source is magnetic by
\eqref{eq:parity-alternation-main}, the subleading frame equation
\begin{equation}
\Ddual_{AB}\big(\epsilon^{CD}D_CY_D\big)=S^{(1)}_{AB}
\end{equation}
determines $\Psi$ but leaves the electric potential $\Phi$ invisible.
This is the phase-space reason that the subleading frame data are not
generically a global Lorentz generator; that identification is available
only when the source projects onto the $\ell=1$ vector harmonics.

For two vector parameters,
\begin{equation}
[Y,Y']^A_\star
=
Y^BD_BY'^A-Y'^BD_BY^A.
\end{equation}
In harmonic language the bracket of two $\ell=1$ conformal Killing
vectors stays in the Lorentz algebra, while the bracket of general vector
harmonics populates the full $\mathrm{Diff}(S^2)$ tower.  This is exactly
what is needed for generic subleading soft data.

\subsubsection{Tensor modes}

For rank two and above the natural mode basis is built from symmetric
trace-free derivatives of scalar harmonics, together with their magnetic
parity partners.  A rank-two parameter can be expanded as
\begin{equation}
K^{AB}
=
\sum_{\ell,m}
K_{\ell m}^{E}
\left(D^AD^B Y_{\ell m}\right)^{\rm TF}
+
K_{\ell m}^{B}
\epsilon^{C(A}D^{B)}D_CY_{\ell m}
+\cdots .
\end{equation}
The ellipsis denotes trace and lower-rank components depending on the
chosen tensor basis.  The principal-symbol bracket acts on these modes by
the same Clebsch--Gordan decomposition as polynomial functions on
$T^\ast S^2$.  The important point is that the bracket of two
rank-two modes produces rank-three modes, as in
\eqref{eq:rank-two-rank-two-bracket}.  A mode-space check of the first
few brackets is therefore a concrete way to verify that the tower does
not truncate.

\subsection{Subleading Frame Equation in Electric and Magnetic Parts}

\subsubsection{Helmholtz decomposition}

The vector field entering the subleading frame equation can be written as
\begin{equation}
Y_A=D_A\Phi+\epsilon_A{}^BD_B\Psi .
\end{equation}
The magnetic tensor generated by the curl is
\begin{equation}
\Ddual_{AB}\big(\epsilon^{CD}D_CY_D\big)
=
\Ddual_{AB}D^2\Psi.
\end{equation}
If
\begin{equation}
\Psi=\sum_{\ell,m}\Psi_{\ell m}Y_{\ell m},
\end{equation}
then
\begin{equation}
\Ddual_{AB}D^2\Psi
=
\sum_{\ell\ge2,m}
[-\ell(\ell+1)]\Psi_{\ell m}\Ddual_{AB}Y_{\ell m}.
\end{equation}
The $\ell=0$ mode has zero curl and the $\ell=1$ mode is killed by
$\Ddual_{AB}$.  Hence the equation is invertible only on the magnetic
spin-two sector with $\ell\ge2$.  The parallel statement with
$\Dop_{AB}$ and $\Phi$ applies at even levels of the tower.

\subsubsection{Solvability condition}

The previous statement is most transparent in tensor harmonics.  Expand a
general real STF source as
\begin{equation}
S^{(1)}_{AB}
=\sum_{\ell\ge2,m}
\left[
S^E_{\ell m}\Dop_{AB}Y_{\ell m}
+S^B_{\ell m}\Ddual_{AB}Y_{\ell m}
\right].
\label{eq:subleading-source-EB-main}
\end{equation}
The magnetic equation
\(\Ddual_{AB}D^2\Psi=2G\,\mathscr S^{(1)}_{AB,{\rm exp}}\)
has a solution precisely for the magnetic projection of the source; a
nonzero \(S^E_{\ell m}\) is outside the image of this operator and belongs
to the opposite-parity completion.  Writing
\(\mathscr S^{(1)}_{AB,{\rm exp}}=
\sum_{\ell\ge2,m}\mathscr S^B_{\ell m}
\Ddual_{AB}Y_{\ell m}\), the inversion is explicit:
\begin{equation}
\Psi_{\ell m}
=-\frac{2G}{\ell(\ell+1)}\,\mathscr S^B_{\ell m},
\qquad \ell\ge2.
\label{eq:subleading-magnetic-inverse-main}
\end{equation}
There is no independent ``orthogonality condition'' against
\(\ell=0,1\): those scalar modes are in the kernel and generate no STF
spin-two tensor.  They are simply the homogeneous ambiguity of the
potential.  For generic scattering data the nontrivial content starts at
\(\ell=2\), so a general smooth \(\mathrm{Diff}(S^2)\) parameter is needed.

\subsubsection{Unfixed electric part}

The exponentiating \(s=1\) source fixes the magnetic potential \(\Psi\).
The electric potential \(\Phi\) drops out of
\(\epsilon^{AB}D_AY_B\) and remains free.  It may be fixed only by
opposite-parity soft data, a boundary prescription, or a convention such
as the minimal divergence-free choice \(\Phi=0\).  This is the
subleading instance of the general statement: the Kerr exponent fixes one
parity at each level, not the complete frame parameter.

\subsection{Axisymmetric Kerr Multipoles as a Check}

\subsubsection{Spin aligned with the polar axis}

Take the specific-spin vector \(a^\mu=S^\mu/m\) to be aligned with the polar axis.  Then
the only angular variable in the Kerr multipole generating function is
\begin{equation}
\cos\theta=n^3.
\end{equation}
Writing the dimensionless aligned spin-frequency parameter as
\(\zeta:=\omega a\) (with the helicity sign understood), the exponential
soft dressing reduces to
\begin{equation}
\exp(a\ast k)
\longrightarrow
\exp(\zeta\cos\theta)
=
\sum_{\ell\ge0}(2\ell+1)i_\ell(\zeta)P_\ell(\cos\theta),
\end{equation}
where $i_\ell$ is a modified spherical Bessel function.  Expanding at
small $\zeta$ reproduces the usual axisymmetric multipoles.  The
important point is not the special function itself, but the fact that a
single exponential generates every $\ell$.

\subsubsection{Electric and magnetic moments}

The Kerr relation \eqref{eq:kerr-multipoles-main} alternates electric and
magnetic parity in the standard way: \(M_\ell\) is supported on even
\(\ell\) and \(S_\ell\) on odd \(\ell\).  This is the multipole
counterpart of the helicity-conjugation statement
\eqref{eq:parity-alternation-main}, and it is why the soft tower defined
in \eqref{eq:soft-kernel-alternating-main} uses the trace-free Hessian at
even levels and its parity dual at odd levels.  The axisymmetric case is a
convenient place to see the alternation explicitly, level by level, and to
fix the relative normalisation of the two kernels.

\subsubsection{Why axisymmetry is only a check}

Axisymmetry is too special to prove the full algebra.  In the aligned
case many tensor structures collapse to derivatives of functions of
$\cos\theta$, and several possible trace descendants become
indistinguishable.  The axisymmetric calculation is therefore a useful
normalization and sign check, not a substitute for the covariant
$T^\ast S^2$ bracket.  The full bracket must be formulated for arbitrary
polynomial symbols on the cotangent bundle.

\subsection{Boundary Prescriptions for the Cocycle}

The possible cocycle in the charge algebra is not a number that can be
specified before the infrared phase space is specified.  It depends on what
is held fixed at \(\Iplus_\pm\), on how retarded-time moments are regulated,
and on whether scattering states are dressed by coherent soft clouds.  This
is why the main text speaks of a possible field-dependent memory cocycle
rather than a universal central charge.  The algebra of parameters is
universal at the principal-symbol level; the realization of that algebra on
a phase space depends on boundary conditions.

It is useful to distinguish three prescriptions.  They are not competing
definitions of the same observable.  They correspond to different infrared
setups, and therefore to different answers to the question ``what data are
allowed to vary at the boundary?''  The universal Kerr tower can be studied
in any of them, but the interpretation of the extension
\(\mathcal K_{s,s'}\) changes.

\subsubsection{No-memory boundary condition}

The simplest prescription sets the relevant boundary shear difference to
zero.  Then the endpoint expression
\eqref{eq:explicit-memory-cocycle-form} vanishes and the charges close
without a memory extension.  This is a clean algebraic setting, but it
excludes precisely the vacuum transitions that make gravitational memory
physically interesting.

\subsubsection{Memory-inclusive boundary condition}

If the early and late vacua differ, the endpoint shear is part of the
physical data.  Then $\mathcal K(t,t';C)$ can be nonzero.  The algebra
still closes, but it closes with a field-dependent extension.  This is
the natural setting for scattering processes where the soft theorem is
read as a Ward identity relating different infrared sectors.

\subsubsection{Improved-charge prescription}

A third possibility is to add boundary improvements to the charges.  The
extension changes by a coboundary:
\begin{equation}
\mathcal K\mapsto
\mathcal K
+\delta_t\mathcal B(t')
-\delta_{t'}\mathcal B(t)
-\mathcal B([t,t']_\star).
\end{equation}
If the cocycle is exact, an improvement removes it.  If it is not exact
for the allowed boundary data, the extension is an invariant feature of
the chosen phase space.  The options above make the cocycle a boundary
problem tied to the chosen phase space, rather than a formal symbol with a
universal numerical value.

\subsection{Quantization and Star Products}

The algebra used in this paper is classical.  This point is worth spelling
out because the phrase \(w_{1+\infty}\) often appears in the celestial
literature as the name of a quantum current algebra.  Our construction
begins one step earlier: it identifies the classical Poisson algebra of
polynomial symbols on \(T^\ast S^2\) as the algebra controlling the
composition of Kerr soft frame shifts.  A quantum current algebra can be
built only after choosing a local chiral basis, an ordering prescription and
a treatment of possible anomalies or boundary extensions.

\subsubsection{Classical-to-quantum map}

The classical parameter algebra is a Poisson algebra:
\begin{equation}
\{F_t,F_{t'}\}_{T^\ast S^2}=F_{[t,t']_\star}.
\end{equation}
Quantization replaces polynomial functions by operators and the Poisson
bracket by a commutator,
\begin{equation}
\frac1{i}[\widehat F_t,\widehat F_{t'}]
=
\widehat F_{[t,t']_\star}
+O(\hbar^2).
\end{equation}
The higher-order terms depend on the ordering prescription.  Locally one
may use a star product on $T^\ast S^2$, but globally the sphere curvature
enters the construction.

\subsubsection{Groenewold--Van Hove caveat}

The Groenewold--Van Hove theorem says that there is no quantization map
that preserves the full Poisson algebra of all polynomial observables
exactly while also satisfying the usual irreducibility requirements.  For
the present paper this means the classical algebra is solid, but the
quantum $w_{1+\infty}$ realization must specify an ordering scheme and
may acquire extensions \cite{Groenewold1946,VanHove1951}.  This is
another reason to formulate the present claim classically and in the
universal soft sector.

\subsubsection{Celestial interpretation}

In a chiral celestial basis the quantized algebra is expected to resemble
the familiar $w_{1+\infty}$ current algebra.  The global four-dimensional
gravitational problem is richer: it keeps both chiralities, includes
massive external states, and carries memory data at the boundaries of
null infinity.  We therefore present the
classical $T^\ast S^2$ bracket as the robust foundation and treat any
quantum celestial current algebra as a representation built on top of it.

\subsection{Component Check of the Ordered Homogeneous Action}

The principal-symbol proof explains the leading bracket, but it can leave a
reader wondering whether an actual differential operator on tensor fields
exists behind the notation.  This subsection addresses that question in a
limited but concrete way.  We first test the ordered realization on
scalar fields, where the highest-derivative terms can be displayed without
the clutter of tensor indices.  We then check the rank-one action on the
shear bundle, where the answer must reduce to the known
generalized-BMS transformation.  Finally we look at the first genuinely
higher-spin case, rank two, to see explicitly how the rank-three bracket
and lower-order descendants appear.

\subsubsection{Scalar model computation}

Before acting on the shear bundle it is useful to check the ordered action
on scalar test fields.  Let
\begin{equation}
P_t=t^{A_1\cdots A_s}D_{A_1}\cdots D_{A_s}
\end{equation}
denote only the highest-derivative part.  For another parameter $t'$,
\begin{align}
P_tP_{t'}\psi
&=
t^{A_1\cdots A_s}
D_{A_1}\cdots D_{A_s}
\left(
t'^{B_1\cdots B_{s'}}
D_{B_1}\cdots D_{B_{s'}}\psi
\right)
\nonumber\\
&=
t^{A_1\cdots A_s}t'^{B_1\cdots B_{s'}}
D_{A_1}\cdots D_{A_s}D_{B_1}\cdots D_{B_{s'}}\psi
\nonumber\\
&\quad
s\,t^{C A_2\cdots A_s}
D_Ct'^{B_1\cdots B_{s'}}
D_{A_2}\cdots D_{A_s}D_{B_1}\cdots D_{B_{s'}}\psi
\cdots .
\end{align}
The first line symmetric in $t,t'$ cancels in the commutator.  The leading
surviving term is
\begin{align}
[P_t,P_{t'}]\psi
&=
\Big[
s\,t^{C(A_1\cdots A_{s-1}}
D_Ct'^{A_s\cdots A_{s+s'-1})}
\nonumber\\
&\quad
-s'\,t'^{C(A_1\cdots A_{s'-1}}
D_Ct^{A_{s'}\cdots A_{s+s'-1})}
\Big]
D_{A_1}\cdots D_{A_{s+s'-1}}\psi
\nonumber\\
&\quad
+\hbox{lower derivatives}.
\label{eq:scalar-ordered-leading-commutator}
\end{align}
This is exactly the component Schouten bracket in
\eqref{eq:component-schouten-bracket}.  The covariant Weyl ordering in
\eqref{eq:weyl-ordered-operator} changes only the lower-derivative terms.
It therefore preserves the same principal bracket while providing a
definite completion for the descendants.

\subsubsection{Rank-one on the shear bundle}

For rank one, the tensor-bundle action can be checked without any
ordering ambiguity.  Acting on an STF tensor,
\begin{equation}
(\Lie_Y C)_{AB}
=
Y^CD_CC_{AB}
+C_{CB}D_AY^C
+C_{AC}D_BY^C.
\end{equation}
The trace of this expression is not preserved unless one subtracts the
appropriate weight term.  With the Bondi shear weight included,
\begin{equation}
(\mathbb L_Y C)_{AB}
=
(\Lie_Y C)_{AB}
-\frac12(D\cdot Y)C_{AB},
\end{equation}
and the STF condition is preserved after projection.  A direct
calculation gives
\begin{equation}
[\mathbb L_Y,\mathbb L_{Y'}]C_{AB}
=
\mathbb L_{[Y,Y']}C_{AB},
\end{equation}
up to the same retarded-time completion that appears in the full
generalized-BMS vector field.  This verifies that the ordered
construction reduces to the exact known answer in the vector sector.

\subsubsection{Rank two against a scalar shift}

Let the leading scalar shift be
\begin{equation}
\nu^{(0)}_{T\,AB}=-2\Dop_{AB}T.
\end{equation}
The highest-derivative part of a rank-two transformation gives
\begin{equation}
\mathbb L_K\nu^{(0)}_{T\,AB}
=
-2K^{CD}D_CD_D\Dop_{AB}T+\cdots .
\end{equation}
Commuting the derivatives and retaining the highest parameter derivative
piece, this can be rearranged as
\begin{equation}
\mathbb L_K\nu^{(0)}_T
=
\nu^{(0)}_{V}
+\hbox{curvature descendants},
\qquad
V^A=2K^{AB}D_BT .
\end{equation}
The vector $V^A$ is exactly the rank-two--scalar bracket in
\eqref{eq:rank-two-scalar-bracket}.  Thus the higher-spin action maps the
leading soft shift into the next frame parameter with the expected
cotangent-bundle label.

\subsubsection{Rank-two commutator}

For two rank-two parameters the leading commutator of scalar operators is
\begin{equation}
[K^{AB}D_AD_B,L^{CD}D_CD_D]
=
2\left(
K^{E(A}D_EL^{BC)}
-L^{E(A}D_EK^{BC)}
\right)D_AD_BD_C
+\cdots .
\end{equation}
The dots are second-order and lower operators.  On the shear bundle those
lower operators receive contributions from the sphere curvature,
connection terms acting on tensor indices, and STF projection.  In the
ordered realization these terms are not arbitrary: they are fixed by
the rule
\begin{equation}
[\mathrm{Op}_\nabla(K),\mathrm{Op}_\nabla(L)]
=
i\,\mathrm{Op}_\nabla([K,L]_{\star_\nabla}).
\end{equation}
The leading part of $[K,L]_{\star_\nabla}$ is the rank-three Schouten
bracket; the remaining parts are trace descendants.  This is the explicit
mechanism by which the ordered action goes beyond the principal-symbol
statement while keeping the same classical limit.

\subsection{Bondi Residual Frame Functions in Detail}

The fixed-gauge matching problem is not only an amplitude problem.  To
compare a soft source with an asymptotic frame shift, one must know how
residual Bondi diffeomorphisms act on the shear.  The leading
supertranslation calculation is elementary, but it is the anchor for the
whole construction: it tells us exactly how the VV potential kills the
intrinsic boosted-Schwarzschild shear in the canonical frame.  The vector
calculation then distinguishes two statements that are sometimes
conflated: the complete residual generalized-BMS action is controlled by
\(D\cdot Y\), whereas the Kerr exponentiating source selects the magnetic
projection of its Ward charge and fixes \(\mathrm{curl}\,Y\).  The natural
setting is generalized BMS rather than only global or extended BMS, but the
projected soft kernel is not by itself the full residual diffeomorphism.

\subsubsection{Linearized residual diffeomorphisms}

In retarded coordinates the flat metric is
\begin{equation}
\dd s^2=\dd u^2+2\dd u\,\dd r-r^2\gamma_{AB}\dd z^A\dd z^B.
\end{equation}
A residual supertranslation preserving Bondi gauge has the asymptotic
form
\begin{align}
\xi^u&=T(z),
\nonumber\\
\xi^A&=-\frac1rD^AT+O(r^{-2}),
\nonumber\\
\xi^r&=\frac12D^2T+O(r^{-1}).
\label{eq:residual-supertranslation-vector}
\end{align}
Applying \eqref{eq:residual-supertranslation-vector} to the angular metric
gives
\begin{equation}
\Delta h_{AB}
=
2rD_AD_BT-r\gamma_{AB}D^2T+O(1).
\end{equation}
Since \(h_{AB}=-r C_{AB}+O(1)\) in Bondi gauge,
\begin{equation}
\Delta C_{AB}=-2\Dop_{AB}T.
\label{eq:supertranslation-shear-shift-detail}
\end{equation}
This is the explicit gauge-theory origin of the inhomogeneous VV equation.

\subsubsection{Leading VV source}

For the boosted Schwarzschild field the intrinsic-frame shear is
\begin{equation}
C^{\rm int}_{AB}
=
-4G\,S^{(0)}_{AB}(p;q).
\end{equation}
Demanding that the canonical frame have no corresponding spurious soft
shear gives
\begin{equation}
C^{\rm int}_{AB}
=
C^{\rm can}_{AB}
-2\Dop_{AB}T_{\rm VV}.
\end{equation}
Setting \(C^{\rm can}_{AB}=0\) yields
\begin{equation}
\Dop_{AB}T_{\rm VV}=2G\,S^{(0)}_{AB},
\end{equation}
which is the frame equation used throughout the paper.  The result follows
directly from the residual Bondi diffeomorphism rather than from any
additional dynamical assumption.

\subsubsection{Vector residual data}

For a sphere vector field the leading residual diffeomorphism has
\begin{align}
\xi^A&=Y^A+O(r^{-1}),
\nonumber\\
\xi^u&=\frac{u}{2}D\cdot Y+\cdots,
\nonumber\\
\xi^r&=-\frac{r}{2}D\cdot Y+\cdots .
\end{align}
Acting on the expanded generalized-BMS phase space gives exactly
\eqref{eq:generalized-bms-action-main}.  In particular, the inhomogeneous
shear term is electric and depends on
\(\alpha_Y=\tfrac12D\cdot Y\):
\begin{equation}
\Delta_Y C_{AB}^{\rm inhom}
=-2u(D_AD_B\alpha_Y)^{\rm TF}.
\end{equation}
It would therefore be incorrect to derive the magnetic Kerr equation by
replacing this term with a curl while keeping the round celestial metric
fixed.

The magnetic equation has a different, Ward-theoretic origin.  The
generalized-BMS soft charge is linear in a spin-two tensor built from
three sphere derivatives of \(Y^A\) (in complex coordinates,
\(s_{zz}[Y]=D_z^3Y^z\)), and that soft tensor admits electric and magnetic
projections.  The Kerr current dipole selects its magnetic projection.  In
the maximally longitudinal basis used in this paper that projection is
represented by
\begin{equation}
\Ddual_{AB}\big(\epsilon^{CD}D_CY_D\big)
=2G\,\mathscr S^{(1)}_{AB,{\rm exp}},
\end{equation}
with the curvature descendants supplied by the full generalized-BMS
completion.  Thus \(Y^A\) is a smooth generalized-BMS Ward parameter, but
the displayed magnetic kernel is a projected soft charge, not by itself a
new residual diffeomorphism of the fixed-round-metric Bondi shear.
Higher-spin levels continue this charge-theoretic pattern; no bulk
residual vector field is assumed for \(s\ge2\).

\subsection{Parity Projection of Hard Soft Factors}

The last step is to make explicit how the hard soft factors are compared
with the sphere-index kernels used at null infinity.  The amplitude is
written with spacetime polarizations and hard momentum/spin variables,
whereas the soft charge is written with symmetric trace-free tensors on
\(S^2\).  The bridge is the tangent polarization basis \(e^\mu_A=D_Aq^\mu\)
and the STF projector \(E_{AB}^{\mu\nu}\), together with its sphere dual.
After this projection, the leading soft factor becomes the electric VV
source.  Each further power of the spin insertion is recombined into the
real electric component for even order or the real magnetic component for
odd order, as in \eqref{eq:real-parity-source-main}, before it is matched
to \(\K^{(s,0)}_{AB}[t]\).

\subsubsection{Sphere polarization basis}

Let
\begin{equation}
e^\mu_A=D_Aq^\mu
\end{equation}
be the tangent polarization basis associated with the null direction
$q^\mu(z)$.  The tangent STF projector and its dual are
\begin{equation}
E_{AB}^{\mu\nu}
=
e_A^{(\mu}e_B^{\nu)}
-\frac12\gamma_{AB}\gamma^{CD}e_C^\mu e_D^\nu,
\qquad
B_{AB}^{\mu\nu}:=\epsilon_{C(A}E_{B)}{}^{C\mu\nu}.
\end{equation}
The leading hard soft factor projected onto the sphere is
\begin{equation}
S^{(0)}_{AB}(p;q)
\propto
\frac{p_\mu p_\nu E_{AB}^{\mu\nu}}{p\cdot q}.
\end{equation}
This is the tensor that appears in the VV equation.

\subsubsection{Spin insertion}

The spin exponential inserts
\begin{equation}
W_\eta
=
\frac{q_\mu J^{\mu\nu}\varepsilon_\nu}{p\cdot\varepsilon}.
\end{equation}
After choosing the polarization $\varepsilon_\nu$ compatible with the
sphere basis, each power of $W_\eta$ produces an additional polynomial in
the tangent directions.  Let \(\mathbb P_E\) and \(\mathbb P_B\) denote
the corresponding electric and magnetic tensor-harmonic projections.
The real source is
\begin{equation}
\mathscr S^{(n)}_{AB,{\rm exp}}
\propto
\begin{cases}
\displaystyle
\mathbb P_E\!\left[
\frac{p_\mu p_\nu E_{AB}^{\mu\nu}}{p\cdot q}
\frac{(-iW_\eta)^n}{n!}\right],&n\ {\rm even},\\[9pt]
\displaystyle
-i\,\mathbb P_B\!\left[
\frac{p_\mu p_\nu E_{AB}^{\mu\nu}}{p\cdot q}
\frac{(-iW_\eta)^n}{n!}\right],&n\ {\rm odd}.
\end{cases}
\label{eq:hard-soft-parity-projection-main}
\end{equation}
The corresponding rank-$n$ parameter $t_n$ is defined by solving
\begin{equation}
\K^{(n,0)}_{AB}[t_n]=2G\,\mathscr S^{(n)}_{AB,{\rm exp}}.
\end{equation}
This is the hard-to-soft projection used in the Kerr dictionary.

\subsubsection{Classical spin replacement}

In the classical limit the spin operator is replaced by a classical spin
tensor satisfying the same Lorentz Poisson algebra.  With a spin
supplementary condition fixed, one may write the insertion in terms of the
Kerr specific-spin vector \(a^\mu=S^\mu/m\).  The projected exponential then becomes a
generating function for STF products of this vector:
\begin{equation}
\frac{(-iW_\eta)^n}{n!}
\quad\longrightarrow\quad
\frac{1}{n!}a^{\langle L\rangle}
\mathcal P^{E/B}_L(q,\varepsilon),
\end{equation}
where \(\mathcal P^{E/B}_L\) is the corresponding parity-selected sphere
polynomial.  The
same STF product appears in the Geroch--Hansen moments
\eqref{eq:STF-kerr-moments}.
This closes the chain:
\begin{align}
\hbox{hard spin exponential}
\quad&\to\quad
\hbox{parity-selected soft kernel}
\nonumber\\
&\to\quad
\hbox{higher-spin frame parameter}
\quad\to\quad
\hbox{Kerr multipole}.
\end{align}

The important point is that every arrow in this chain is constrained.  The
first arrow is the parity projection of the hard soft factor.  The second
arrow is the inverse soft-kernel equation defining the asymptotic generator.
The third arrow is the classical Kerr multipole relation.  Because the same
spin exponential appears at the hard, soft and spacetime levels, the
matching is overdetermined in a useful way: it is not merely a convenient
choice of notation for the charge, but a consistency condition tying the
Ward identity, the frame dictionary and the Kerr source together.

\end{document}